\RequirePackage{fix-cm}
\documentclass[twocolumn,epjc3,coverpage]{svjour3}  
\smartqed
\RequirePackage{graphicx}
\RequirePackage{amsmath}
\RequirePackage{atlasphysics}
\RequirePackage{subfigure}
\RequirePackage{lscape}
\RequirePackage{cite}
\RequirePackage{url}
\RequirePackage[varg]{txfonts}
\RequirePackage[bottom]{footmisc}
\RequirePackage{multirow}
\RequirePackage{afterpage}
\usepackage[hyperindex,breaklinks]{hyperref} 

\journalname{Eur. Phys. J. C}

\newcommand{\antikt}{anti-$k_{t}$}

\newcommand{\heraI}{HERA~}
\newcommand{\cme}{$\sqrt{s}=$}

\newsavebox\verbbox
\begin{lrbox}{\verbbox}
\footnotesize
\verb|doublefsr=1|
\end{lrbox}

\usepackage{preprintcover}
\PreprintCoverPaperTitle{Measurement of the inclusive jet cross-section in \textit{pp}~collisions at $\sqrt{s}$~=~2.76~TeV and comparison to the inclusive jet cross-section at $\sqrt{s}$~=~7~TeV using the ATLAS detector}  
\PreprintIdNumber{CERN-PH-EP-2013-036}
\PreprintCoverAbstract{The inclusive jet cross-section has been measured in proton-proton collisions at 
$\sqrt{s} = 2.76\TeV$ in a dataset corresponding to an integrated luminosity 
of $0.20~\ipb$ collected with the ATLAS detector at the Large Hadron Collider in 2011. 
Jets are identified using the \antikt{} algorithm with two radius parameters of $0.4$ and $0.6$. 
The inclusive jet double-differential cross-section is presented as a function 
of the jet transverse momentum \pT{} and jet rapidity $y$, covering a range of 
$20\leq \pT < 430\GeV$ and $|y| < 4.4$.
The ratio of the cross-section to the inclusive jet cross-section measurement at $\sqrt{s} = 7\TeV$,
published by the ATLAS Collaboration, is calculated as a function of both transverse momentum and the 
dimensionless quantity $\xt = 2 \pT / \sqrt{s}$, in bins of jet rapidity. The systematic uncertainties
on the ratios are significantly reduced due to the cancellation of correlated uncertainties in the two
measurements. Results are compared to the prediction from next-to-leading order perturbative QCD calculations
corrected for non-perturbative effects, and next-to-leading order Monte Carlo simulation. 
Furthermore, the ATLAS jet cross-section measurements at $\sqrt{s} = 2.76\TeV$ and $\sqrt{s}=7\TeV$ are 
analysed within a framework of next-to-leading order perturbative QCD calculations to determine parton
distribution functions of the proton, taking into account the correlations between the measurements. 
\keywords{QCD \and jet \and LHC}}
\PreprintJournalName{Eur. Phys. J. C}

\begin{document}

\title{Measurement of the inclusive jet cross-section in \textit{pp}~collisions at $\sqrt{s}$~=~2.76~TeV and comparison to the inclusive jet cross-section at $\sqrt{s}$~=~7~TeV using the ATLAS detector}
\author{The ATLAS Collaboration}
\date{\today}
\maketitle

\begin{abstract}
The inclusive jet cross-section has been measured in proton-proton collisions at 
$\sqrt{s} = 2.76\TeV$ in a dataset corresponding to an integrated luminosity 
of $0.20~\ipb$ collected with the ATLAS detector at the Large Hadron Collider in 2011.
Jets are identified using the \antikt{} algorithm with two radius parameters of $0.4$ and $0.6$. 
The inclusive jet double-differential cross-section is presented as a function 
of the jet transverse momentum \pT{} and jet rapidity $y$, covering a range of 
$20\leq \pT < 430\GeV$ and $|y| < 4.4$.
The ratio of the cross-section to the inclusive jet cross-section measurement at $\sqrt{s} = 7\TeV$,
published by the ATLAS Collaboration, is calculated as a function of both transverse momentum and the 
dimensionless quantity $\xt = 2 \pT / \sqrt{s}$, in bins of jet rapidity. The systematic uncertainties
on the ratios are significantly reduced due to the cancellation of correlated uncertainties in the two
measurements. Results are compared to the prediction from next-to-leading order perturbative QCD calculations
corrected for non-perturbative effects, and next-to-leading order Monte Carlo simulation. 
Furthermore, the ATLAS jet cross-section measurements at $\sqrt{s} = 2.76\TeV$ and $\sqrt{s}=7\TeV$ are 
analysed within a framework of next-to-leading order perturbative QCD calculations to determine parton
distribution functions of the proton, taking into account the correlations between the measurements. 
\keywords{QCD \and jet \and LHC}
\end{abstract}

\section{Introduction}
\label{sec:intro}
Collimated jets of hadrons
are a dominant feature of high-energy particle interactions. 
In Quantum Chromodynamics (QCD) they can be interpreted in terms of the fragmentation of quarks and gluons
produced in a scattering process.
The inclusive jet production cross-section provides information 
on the strong coupling and the structure of the proton, 
and tests the validity of perturbative QCD (pQCD) down to the shortest accessible distances.

The inclusive jet cross-section has been measured at high energy 
in pro\-ton-anti\-pro\-ton (\ppbar) collisions with \linebreak $\sqrt{s} = 546\GeV{}$ and 630~\GeV{} at the SPS 
\cite{Banner1982203,
Appel1985349,
Arnison:1985rv, 
Arnison1986461, 
Alitti1991232
}, and with 
$\sqrt{s} = 546\GeV$, 630~\GeV, 1.8~\TeV{} and 1.96~\TeV{} at the Tevatron
\cite{Abe:1991ea,Abe1993,Abe:1996wy,Bhatti:1996wx,Abbott:1998ya,Abbott:2000ew, 
Abbott:2000kp,Abbott:2000dh,Affolder:2001hn,Abazov:2001hb,Abulencia:2005jw,
Abulencia:2005yg,Abulencia:2007ez,Aaltonen2008,Abazov2008ae,Abazov:2009nc,
Abazov:2011vi}. 

The Large Hadron Collider (LHC)~\cite{1748-0221-3-08-S08001} at CERN allows the production of jets with transverse momenta in the \TeV{} \linebreak regime, colliding protons on protons (\pp{}) with a centre-of-mass energy of currently up to $\sqrt{s} = 8\TeV{}$. The ATLAS Collaboration has presented early measurements of the inclusive jet cross-section at  $\sqrt{s}=7$ \TeV{} based on a dataset with an integrated luminosity of $17$~\inb{} 
for jets with a transverse momentum of $60 \le \pt < 600$~\GeV{} and a rapidity\footnote{Rapidity is defined as $y = 0.5 \ln [(E + p_\text{z} )/(E - p_\text{z} )]$ where $E$ denotes the energy and $p_\text{z}$ is the component of the momentum along the beam direction.} of $|y|< 2.8$~\cite{Aad:2010wv}, 
as well as for the entire dataset of $37$~\ipb{} taken in $2010$ 
for jets with $20 \le \pt < 1500$~\GeV{} and $|y|<4.4$~\cite{Aad:2011fc}.
The CMS Collaboration has presented results in the kinematic range of $18\leq\pt<1100$~\GeV{} and $|y|<3$ in a dataset of $34$~\ipb{} \cite{CMS:2011ab}, in the range of $35\leq\pt<150$~\GeV{} and $3.2<|y|<4.7$ using $3.1$~\ipb{} \cite{Chatrchyan:2012gwa}, and for $0.1\leq\pt<2$~\TeV{} and $|y|<2.5$ using $5.0$~\ifb{} \cite{2012CMSjets}. 
These data are found to be generally well described by
next-to-leading order (NLO) pQCD calculations, corrected for non-perturbative effects from hadronisation and the underlying event.

At the start of the 2011 data taking period of the LHC, the ATLAS experiment collected \pp{} collision data at \linebreak $\sqrt{s} = 2.76$~\TeV{} corresponding to an integrated luminosity of $0.20~\ipb$. Having a centre-of-mass energy close to the highest energies reached in \ppbar{} collisions, the dataset provides a connection from LHC measurements to previous measurements at the Tevatron. 
Moreover, measurements with the same detector at different centre-of-mass energies provide stringent tests of the theory, since the dominant systematic uncertainties are correlated. These correlations can be explored in a common fit to the measurements at
different $\sqrt{s}$ or in ratios of the inclusive jet double-differential cross-sections.
Hence, uncertainties can be significantly reduced. Such ratios were reported by previous experiments,
UA2~\cite{Appel1985349}, UA1~\cite{Arnison1986461}, CDF~\cite{Abe1993,Bhatti:1996wx} and D0~\cite{Abbott:2000kp}. 

In this paper the inclusive jet double-differential cross-section is measured for $20 \le \pt < 430$~\GeV{} and rapidities of $|y|<4.4$ at $\sqrt{s} = 2.76$~\TeV. Moreover, the ratio to the previously measured cross-section at $\sqrt{s} = 7$~\TeV \cite{Aad:2011fc} is determined
as a function of \pt{} and as a function of the dimensionless quantity $\xt = 2 \pt/\sqrt{s}$~\cite{feynman:PhysRevD.18.3320}.
For the ratio measured as a function of \pt{}, many experimental systematic uncertainties cancel, while for the ratio measured as a function of \xt{}, theoretical uncertainties are reduced.
This allows a precise test of NLO pQCD calculations.

The outline of the paper is as follows.
The definition of the jet cross-section is given in the next section, followed by a brief description of the ATLAS detector in Sect.~\ref{sec:atlas} and the data taking in Sect.~\ref{sec:dataset}. The Monte Carlo simulation, the theoretical predictions and the uncertainties on the predictions are described in Sects.~\ref{sec:mc} and \ref{sec:theory},
followed by the event selection in Sect.~\ref{sec:events} and the jet reconstruction and calibration in Sect.~\ref{sec:jets}. The unfolding of detector effects and the treatment of systematic uncertainties are discussed in Sects.~\ref{sec:unfold} and~\ref{sec:uncert}, followed by the results of the inclusive jet cross-section at $\sqrt{s} = 2.76$~\TeV{} in Sect.~\ref{sec:xsec}. The results of the ratio measurement, including the discussion of its uncertainties, are presented in Sect.~\ref{sec:ratio}. 
In  Sect.~\ref{sec:pdffit} the results of an NLO pQCD fit to these data  are discussed.  
 The conclusion is given in Sect.~\ref{sec:conclusion}.

\section{Definition of the measured variables}
\label{sec:defXsec}
\subsection{Inclusive single-jet cross-section}
Jets are identified using the \antikt{} algorithm~\cite{Cacciari:2008gp} implemented in the {\sc FastJet} \cite{Cacciari200657,Fastjet} software package. Two different values of the radius parameter, $R=0.4$ and $R=0.6$, are used. Inputs to the jet algorithm can be partons in the NLO pQCD calculation, stable particles after the hadronisation process in the Monte Carlo simulation, or energy deposits in the calorimeter in data. 

Throughout this paper, the jet cross-section refers to the cross-section of jets built from stable particles, defined by having a proper mean lifetime of $c\tau>10$ mm.
Muons and neutrinos from decaying hadrons are included in this definition.

The inclusive jet double-differential cross-section, \linebreak $d^{2}\sigma / d\pt dy$, is measured as a function of the jet transverse momentum \pt{} in bins of rapidity $y$. The kinematic range of the measurement is $20\leq\pt<430\GeV$ and $|y| < 4.4$.

The jet cross-section is also measured as a function of the dimensionless quantity \xt{}. For a pure $2 \to 2$ central scattering of the partons, \xt{} gives the momentum fraction of the initial-state partons with respect to the parent proton.

\subsection{Ratio of jet cross-sections at different centre-of-mass energies}
The inclusive jet double-differential cross-section can be related to the invariant cross-section according to
\begin{equation}
   E \frac{d^{3}\sigma}{dp^{3}} = \frac{1}{2\pi \pt} \frac{d^{2}\sigma}{d\pt dy},
\end{equation}
where $E$ and $p$ denote the energy and momentum of the jet, respectively.
The dimensionless scale-invariant cross-section $F(y,\xt)${} can be defined as~\cite{Bjorken:1973kd}:
\begin{equation}
\label{eq:2}
F(y,\xt, \sqrt{s}) = \pt^4  E \frac{d^{3}\sigma}{dp^{3}} = \frac{\pt^3}{2 \pi} \frac{d^{2}\sigma}{d\pt dy} 
=\frac{s}{8 \pi} \xt^3 \frac{d^2\sigma}{d\xt dy}.
\end{equation}
In the simple quark-parton model~\cite{PhysRev.185.1975,PhysRevLett.23.1415}, $F$ does not depend on the centre-of-mass energy, as follows from dimensional analysis.
In QCD, however, several 
effects lead to a violation of the scaling behaviour, introducing a \pt{} (or $\sqrt{s}$) dependence to $F$. The main effects are the scale dependence of the parton distribution functions (PDFs) and the strong coupling constant~$\alphas$.

The cross-section ratio of the invariant jet cross-section measured at $\sqrt{s}=2.76$~\TeV{} 
to the one measured at $\sqrt{s}=7$~\TeV{} is then denoted by:
\begin{equation}
\rho(y,\xt) = \frac{F(y,\xt, 2.76 \TeV)}{F(y,\xt, 7 \TeV)}.
\end{equation} 
The violation of the $\sqrt{s}$~scaling leads to a deviation of $\rho(y, \xt)$ from one.
$\rho(y,\xt)$ is calculated by measuring the bin-averaged inclusive jet double-differential cross-sections at the two centre-of-mass energies in the same \xt{} ranges:
\begin{equation}
\label{eq:4}
\rho(y,\xt) =\left(\frac{2.76\TeV}{7\TeV}\right)^3\cdot \frac{\sigma(y,\xt, 2.76\TeV)}{\sigma(y,\xt, 7\TeV)},
\end{equation} 
where $\sigma(y,\xt, \sqrt{s})$ corresponds to the measured averaged cross-section $d^{2}\sigma / d\pt dy$ in a bin $(y, \pt=\sqrt{s}\cdot\xt/2)$, and $\xt$ is chosen to be at the bin centre.
Here, the \pt{} binning for the inclusive jet cross-section at $\sqrt{s}=2.76$~\TeV{} is 
chosen such that it corresponds to the same \xt{} ranges obtained from the \pt{} bins of the jet cross-section measurement 
at $\sqrt{s}=7$~\TeV{}. The bin boundaries are listed in~\ref{sec:bins}.

The ratio of inclusive double-differential cross-sections
is also measured
as a function of \pt, where the same \pt{} binning is used for both centre-of-mass energies.
This ratio is denoted by
\begin{equation}
\begin{aligned}
\rho(y,\pt) &= \frac{\sigma(y,\pt, 2.76\TeV)}{\sigma(y,\pt, 7\TeV)},
\end{aligned}
\end{equation} 
where $\sigma(y,\pt, \sqrt{s})$ is the measured averaged cross-section $d^{2}\sigma / d\pt dy$ in a bin $(y, \pt)$ at a centre-of-mass energy of $\sqrt{s}$.
Since the uncertainty due to the jet energy scale is the dominant experimental uncertainty at a given \pt{}, the experimental systematic uncertainty is significantly reduced by taking the cross-section ratio in the same \pt{} bins. 

\section{The ATLAS detector}
\label{sec:atlas}
The ATLAS detector consists of a tracking system (inner detector)
in a $2\,\mathrm{T}$ axial magnetic field up to a pseudorapidity\footnote{ATLAS uses 
a right-handed coordinate system with its origin at the nominal 
interaction point (IP) in the centre of the detector and the $z$-axis 
along the beam pipe. The $x$-axis points from the IP to the centre of the
LHC ring, and the $y$-axis points upward. 
The pseudorapidity is defined in terms of the 
polar angle $\theta$ as $\eta=-\ln\tan(\theta/2)$.} 
of $|\eta|=$~2.5, 
sampling electromagnetic and hadronic calorimeters up to $|\eta|=4.9$, 
and muon chambers in an azimuthal magnetic field provided by a system of toroidal magnets. 
A detailed description of the ATLAS detector
can be found elsewhere~\cite{Aad:2008zzm}. 

The inner detector 
consists of layers of
silicon pixel detectors, silicon microstrip detectors and transition
radiation tracking detectors. It is used in this analysis to identify
candidate collision events by constructing vertices from tracks.
Jets are reconstructed using the energy deposits in the calo\-ri\-me\-ter, whose granularity
and material varies as a function of~$\eta$.
The electromagnetic calorimeter uses lead as an absorber, liquid argon (LAr) 
as the active medium and has a fine gra\-nu\-la\-ri\-ty.
It consists of a barrel ($|\eta|<1.475$) and an endcap ($1.375<|\eta|<3.2$) region. 
The hadronic calorimeter is divided into
three distinct regions: a barrel region ($|\eta|<0.8$) and
an extended barrel region ($0.8<|\eta|<1.7$) instrumented with
a steel/scintillating-tile modules, and an endcap
region ($1.5<|\eta|<3.2$) using copper/LAr modules.
Finally, the forward calorimeter ($3.1<|\eta|<4.9$) is instrumented with 
copper/LAr and tungsten/LAr modules
to provide electromagnetic and hadronic energy measurements,
respectively. 

The ATLAS trigger system is composed of three consecutive levels: level 1, level 2 and the event filter,
with progressively increasing computing time per event, finer granularity and access to more detector systems.
For jet triggering, the relevant systems are the minimum bias trigger scintillators (MBTS), located in front of the endcap cryostats covering $2.1 < |\eta| < 3.8$, as well as calorimeter triggers for central jets, covering $|\eta| < 3.2$, and for forward jets, covering $3.1 < |\eta| < 4.9$, respectively.

\section{Data taking}
\label{sec:dataset}
The  proton-proton collision data at $\sqrt{s} = 2.76$ TeV 
were collected 
at the start of the 2011 data taking period of the LHC. 
The total integrated luminosity of the collected data is $0.20~\ipb$. 
The proton bunches were grouped in nine bunch trains. 
The time interval between two consecutive bunches was $525\,\mathrm{ns}$. 
The average number of interactions per bunch crossing is found to be  $\mu = 0.24$.
All events used in this analysis were collected with good operational status
of the relevant detector components for jet measurements.

The data at $\sqrt{s} = 7$ TeV have a total integrated luminosity of $37~\ipb$. 
Further details are given in Ref.~\cite{Aad:2011fc}.

\section{Monte Carlo simulation}
\label{sec:mc}
Events used in the simulation of the detector response are produced
by the \pythia{} 6.423 generator~\cite{Sjostrand:2006za}, 
using the MRST 2007 LO* PDFs~\cite{Sherstnev:2007nd}. 
The generator utilises leading-order (LO)
pQCD matrix elements for $2 \rarrow 2$ processes, 
along with a leading-logarithmic \pt-ordered parton shower~\cite{Corke:2010zj}, 
an underlying event simulation with multiple parton interactions~\cite{multInter}, 
and the Lund string model for 
hadronisation~\cite{lundString}. 
The event generation uses the ATLAS Minimum Bias Tune~1 (AMBT1) set of parameters~\cite{Aad:2010ac}. 
Additional proton-proton collisions occurring in the same bunch crossing have not been simulated
because the average number of interactions per beam crossing is so small.

The {\sc Geant} software toolkit~\cite{Agostinelli:2002hh} within the ATLAS 
simulation framework~\cite{Aad:2010ah} simulates the propagation of the generated 
particles through the ATLAS detector and their interactions with the detector material.

The \herwigpp{} 2.5.1~\cite{Bahr:2008pv,Gieseke:2011na} generator is used in addition to \pythia{} in the evaluation of non-perturbative effects in the theory prediction. 
It is based on the  $2 \rarrow 2$ LO pQCD matrix elements
and a lead\-ing-logarithmic angular-ordered parton shower~\cite{Gieseke:2003rz}. The cluster model~\cite{Webber:1983if} is used for the hadronisation, and an underlying event simulation is based on the eikonal model~\cite{Bahr:2008dy}.

\section{Theoretical predictions}
\label{sec:theory}

\subsection{NLO pQCD prediction}
\label{sec:NLOpQCD}
The NLO pQCD predictions are calculated using the \nlojetpp{}~4.1.2~\cite{Nagy:2003tz} program. 
For fast and flexible calculations with various PDFs and factorisation and renormalisation scales, the {\sc APPLgrid} software~\cite{applgrid:2009} is interfaced with \nlojetpp.
The renormalisation scale, $\mu_R$, and the factorisation scale, $\mu_F$, are chosen for each event as $\mu_R=\mu_F=\pt^\mathrm{max}(y_i)$, where $\pt^\mathrm{max}(y_i)$ is the maximum jet transverse momentum found in a rapidity bin $y_i$. 
If jets are present in different rapidity bins, several scales within the event are used.

The default calculation uses the CT10~\cite{Lai:2010vv} PDF set. 
Predictions using the PDF sets MSTW 2008~\cite{Martin:2009iq}, NNPDF 2.1 (100)~\cite{Ball:2010de, Forte:2010ta}, HERAPDF 1.5~\cite{HERAPDF15} and ABM 11 NLO ($n_f=5$)~\cite{Alekhin:2012ig} are also made for comparison.
The value for $\alphas$ is taken from the corresponding PDF set.

Three sources of uncertainty in the NLO pQCD calculation are considered, namely the uncertainty on the PDF sets, the choice of factorisation and renormalisation scales, and the uncertainty on the value of the strong coupling constant, $\alphas$. 
The PDF uncertainty is defined at $68 \%$ confidence level (CL) and evaluated following the prescriptions given for each PDF set and the PDF4LHC recommendations~\cite{Botje:arXiv1101.0538}.
The uncertainty on the scale choice is evaluated by varying the renormalisation scale and the factorisation scale by a factor of two with respect to the original choice in the calculation.  
The considered variations are
\begin{equation}
\begin{split}
 (f_{\mu_R}, f_{\mu_F})= &(0.5,0.5),\ (0.5,1),\ (1,0.5),\\
                         &(1,2),\ (2,1),\ (2,2),
\end{split}
\end{equation}
where $f_{\mu_R}$ and $f_{\mu_F}$ are factors for the variation of renormalisation and factorisation scales, hence $\mu_R = f_{\mu_R}\cdot \pt^\mathrm{max}$ and $\mu_F = f_{\mu_F} \cdot \pt^\mathrm{max}$.
The envelope of the resulting variations is taken as the scale uncertainty. 
The uncertainty reflecting the $\alphas$ measurement precision is evaluated following the recommendation of the 
CTEQ group~\cite{Lai:2010nw}, by calculating the cross-section using a series of PDFs 
which are derived with various fixed $\alphas$ values.
Electroweak corrections are not included in the theory predictions. The effect is found to be $O(10\%)$ at high \pt{}, and negligible at small \pt{} for $\sqrt{s}=7\TeV{}$~\cite{Dittmaier:2012kx}.

The theoretical predictions for the cross-section ratios at the two different energies, $\rho(y,\xt)$ or $\rho(y,\pt)$, are also obtained from the NLO pQCD calculations. The evaluation of the prediction at $\sqrt{s}=7$ \TeV{} is given in Ref.~\cite{Aad:2011fc}, and the procedure is identical to the one used for $\sqrt{s}=2.76$ \TeV{} in the present analysis. Hence, the uncertainty on the ratio is determined using the same variation in each component of the considered uncertainties simultaneously for both $\sqrt{s}=2.76$ \TeV{} and $\sqrt{s}=7$ \TeV{} cross-section predictions. 

\subsection{Non-perturbative corrections}
The fixed-order NLO pQCD calculations, described in \linebreak Sect.~\ref{sec:NLOpQCD}, predict the parton-level cross-section, which should be corrected for non-perturbative effects before comparison with the measurement at particle level. 
The corrections are derived using LO Monte Carlo event generators complemented by the leading-logarithmic parton shower 
by evaluating the bin-wise ratio of the cross-section with and without hadronisation and the underlying event. 
Each bin of the NLO pQCD cross-section is then multiplied by the corresponding correction for non-perturbative effects. 
The baseline correction factors are obtained from \pythia{} 6.425~\cite{Sjostrand:2006za} with the AUET2B CTEQ6L1 tune~\cite{ATL-PHYS-PUB-2011-009}.
The uncertainty is estimated as the envelope of the correction factors obtained from a series of different generators and tunes:
\pythia{} 6.425 using the tunes
AUET2B LO** \cite{ATL-PHYS-PUB-2011-009}, 
AUET2 LO** \cite{ATL-PHYS-PUB-2011-008}, 
AMBT2B CTEQ6L1 \cite{ATL-PHYS-PUB-2011-009}, 
AMBT1 \cite{Aad:2010ac}, 
\Perugia{}~2010 \cite{Skands:2010akv4}  
and \Perugia{}~2011 \cite{Skands:2010akv4};
\pythia{} 8.150 \cite{Sjostrand2008852} with tune 4C \cite{ATL-PHYS-PUB-2011-009}; 
and \herwigpp{} 2.5.1 \cite{Bahr:2008pv} with tune UE7000-2 \cite{ATL-PHYS-PUB-2011-009}.
The AMBT2B CTEQ6L1 and AMBT1 tunes, which are \linebreak based on observables sensitive to the modelling of minimum bias interactions, 
are included to provide a systematically different estimate of the underlying event activity. 

The NLO pQCD prediction for the cross-section ratio also needs corrections for non-perturbative effects. 
The same procedure is used to evaluate non-perturbative corrections for the cross-section at $\sqrt{s}=7$ \TeV{} using the same series of generator tunes. 
A ratio of corrections at $\sqrt{s}=2.76$ \TeV{} and $\sqrt{s}=7$ \TeV{} is calculated for each generator tune. 
As for the cross-section, \pythia{} 6.425 with the AUET2B CTEQ6L1 tune is used as the central value of the correction factor for the cross-section ratio and the envelope of the correction factors from the other tunes is taken as an uncertainty.

\subsection{Predictions from NLO matrix elements with parton-shower Monte Carlo simulation}
\label{sec:powheg}
The measured jet cross-section is also compared to predictions from \powheg{} jet pair production, revision 2169~\cite{Alioli:2010xa,POWHEGspikes}. 
\powheg{} is an NLO generator that uses the \powhegbox{} 1.0 package~\cite{Nason:2004rx,Frixione:2007vw,Alioli:2010xd}, which can be interfaced 
to different Monte Carlo programs to simulate the parton shower, the hadronisation and the underlying event. 
This simulation using a matched parton shower is expected to produce a more accurate theoretical prediction. However, ambiguities in the matching procedure, non-optimal tuning of parton \linebreak shower-parameters, and the fact that it is a hybrid between an NLO matrix element calculation and the currently available LO parton-shower generators may introduce additional theoretical uncertainties.

In the \powheg{} algorithm, each event is built by first producing a QCD $2 \rarrow 2$ partonic scattering. The renormalisation and factorisation scales for the fixed-order NLO prediction are set to be equal to the transverse momentum of the outgoing partons, $\pt^\mathrm{Born}$. In addition to the hard scatter, \powheg{} also generates the hardest partonic emission in the event. The event is evolved to the particle level using a parton-shower event generator, where the radiative emissions in the parton showers are limited by the  matching \linebreak scale~$\mu_M$ provided by \powheg{}. The simulation of parton showers uses \pythia{} with the ATLAS underlying event tunes, AUET2B~\cite{ATL-PHYS-PUB-2011-009} and \Perugia{}~2011~\cite{Skands:2010akv4}. The tunes are derived from the standalone versions of these event generators, with no optimisation for the \powheg{} predictions. The CT10 PDF set is used in both \powheg{} and \pythia{}.

To avoid fluctuations in the final observables after the showering process, the \powheg{} event generation is per\-for\-med using a new option\footnote{The origin of these fluctuations are rare event topologies in gluon emissions $q \rightarrow qg$ and gluon splittings $g \rightarrow q \bar{q}$, related to the fact that by default \powheg{} \textsc{Box} 1.0 does not consider the corresponding configurations with opposite ordering of the \pt{} for the final state parton: $q \rightarrow gq$ and $g \rightarrow \bar{q}q$. These processes can be activated in revision~2169 using the \powheg{} option \usebox{\verbbox}, which offers an improved handling of the suppression of these events. More details are given in Ref.~\cite{POWHEGspikes}.} that became available recently~\cite{POWHEGspikes}. 
For $\pt{} < 100\GeV$, this new prediction differs by $O(10\%)$ from the \powheg{} prediction at $\sqrt{s}=7\TeV$ from the previous analysis, which followed a different approach~\cite{Aad:2011fc}. 
The uncertainty from the renormalisation and factorisation scales for the \powheg{} prediction is expected to be similar to that obtained with \nlojetpp{}. The matching scale can potentially have a large impact on the cross-section prediction at particle level, affecting the parton shower, initial-state radiation and multiple interactions, but a procedure to estimate this uncertainty is currently not well defined. Therefore no uncertainties are shown for the \powheg{} curves.

\subsection{Prediction for the inclusive jet cross-section at $\sqrt{s}=2.76$ TeV}
\begin{figure*}[htb]
\begin{center}
\subfigure[$|y|<0.3$]{\includegraphics[width=5.2cm]{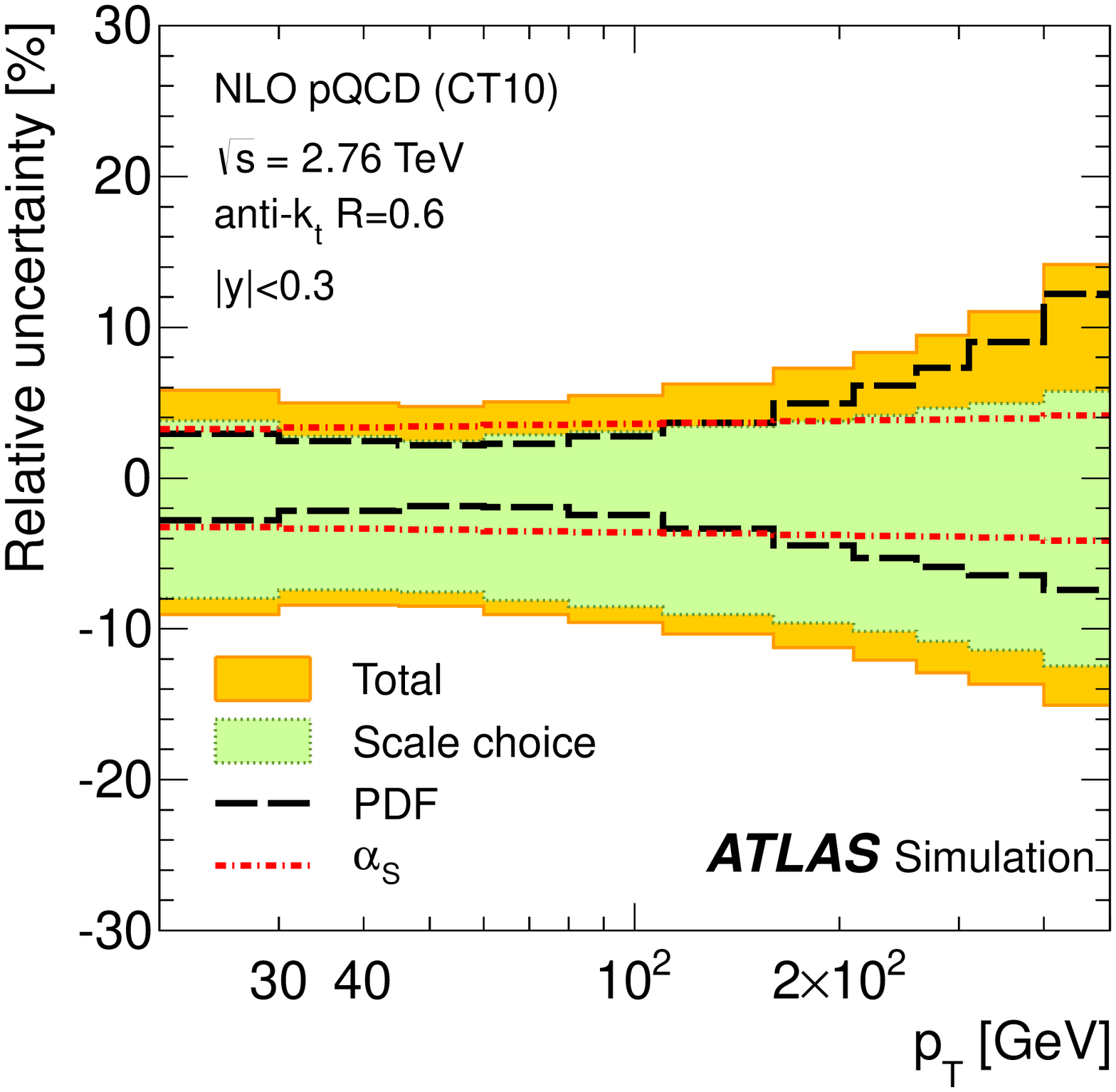}}
\subfigure[$2.1\leq|y|<2.8$]{\includegraphics[width=5.2cm]{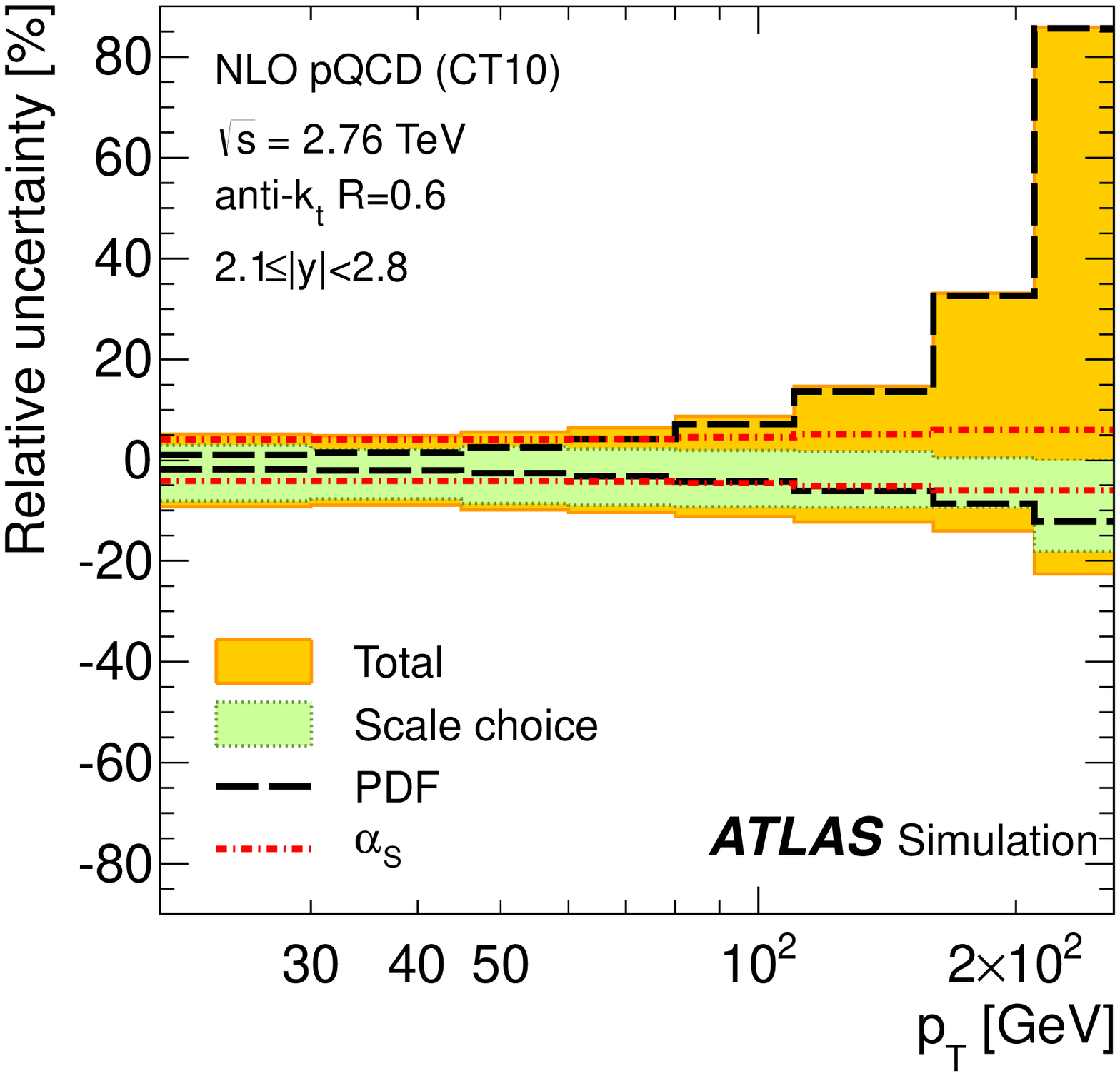}}
\subfigure[$3.6\leq|y|<4.4$]{\includegraphics[width=5.2cm]{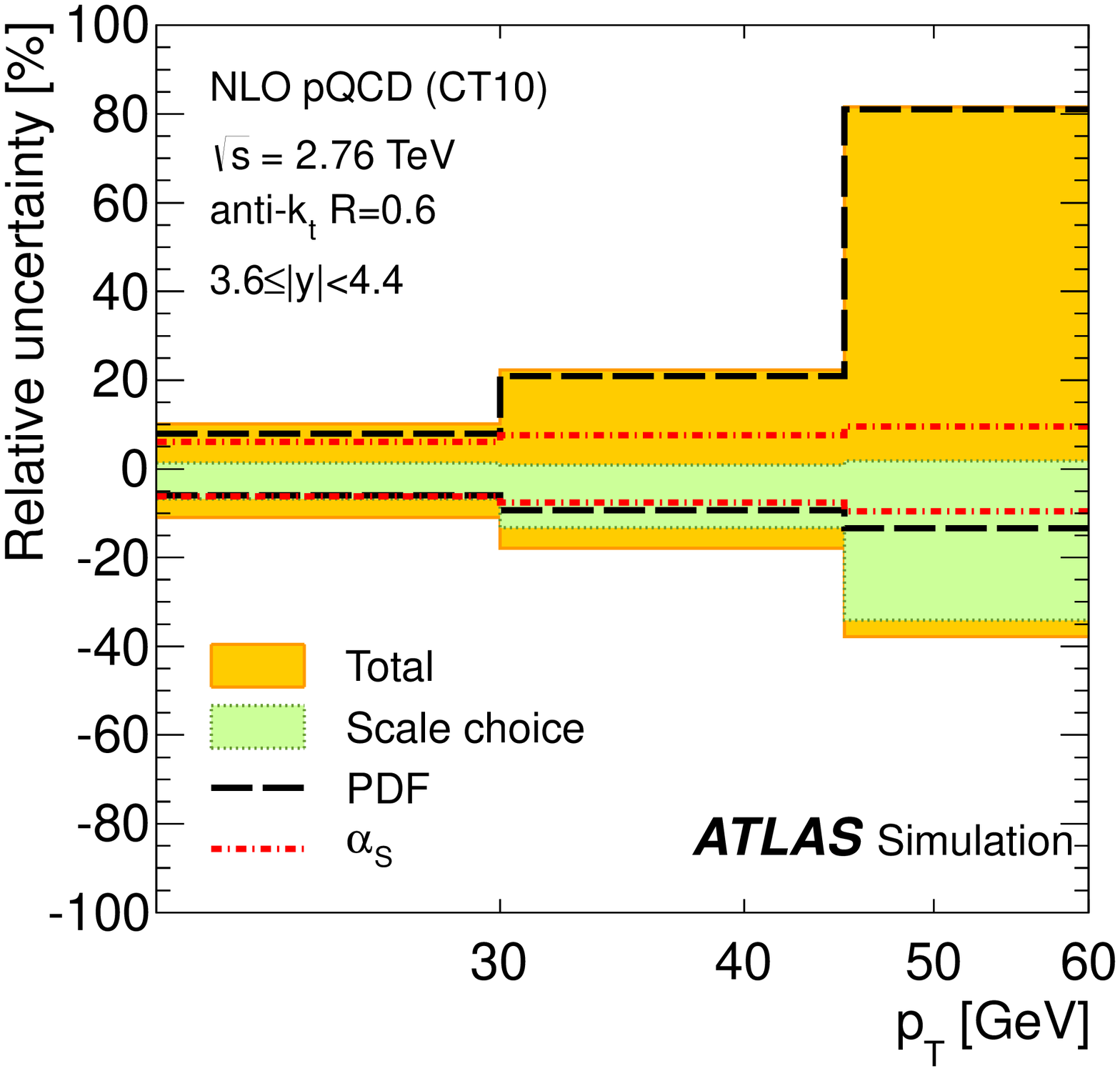}}
\caption{The uncertainty in the NLO pQCD prediction of the inclusive jet cross-section at $\sqrt{s}=2.76$ TeV, calculated using 
\nlojetpp{} with the CT10 PDF set, for \antikt{} jets 
with $R=0.6$ shown in three representative rapidity bins as a function of the jet \pt. 
In addition to the total uncertainty, the uncertainties from the scale choice, 
the PDF set and the strong coupling constant, $\alphas$, are shown separately.}
\label{fig:nloxs_06}
\end{center}
\end{figure*}

The evaluated relative uncertainties of the NLO pQCD calculation for the inclusive jet cross-section at $\sqrt{s}=2.76$ TeV are shown in Fig.~\ref{fig:nloxs_06} as a function of the jet \pt{} for representative rapidity bins and $R=0.6$. 
In the central rapidity region, the uncertainties are about $5\%$ for $\pt\lesssim100$ GeV, increasing to about $15\%$ in the highest jet \pt{} bin.
In the most forward region, they are $10\%$ in the lowest \pt{} bin and up to $80\%$ in the highest \pt{} bin. 
In the higher \pt{} region, the upper bound on the uncertainty is driven by the PDF uncertainty, while the lower bound and the uncertainty at low \pt{} are dominated by the scale choice. The uncertainties for $R=0.4$ are similar.

\begin{figure*}[htb]
\begin{center}
\subfigure[$R=0.4$]{\includegraphics[width=7.5cm]{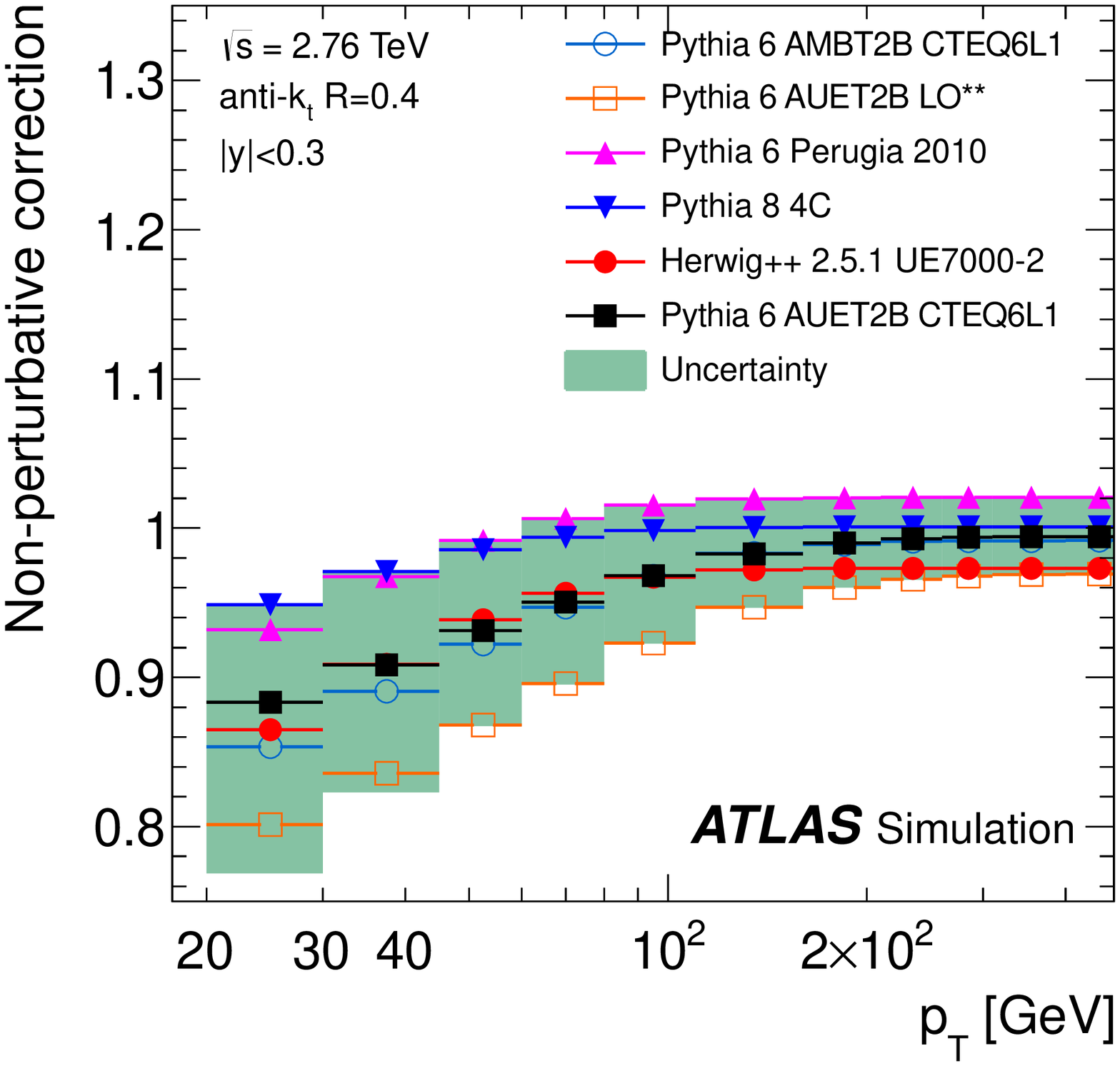}}
\subfigure[$R=0.6$]{\includegraphics[width=7.5cm]{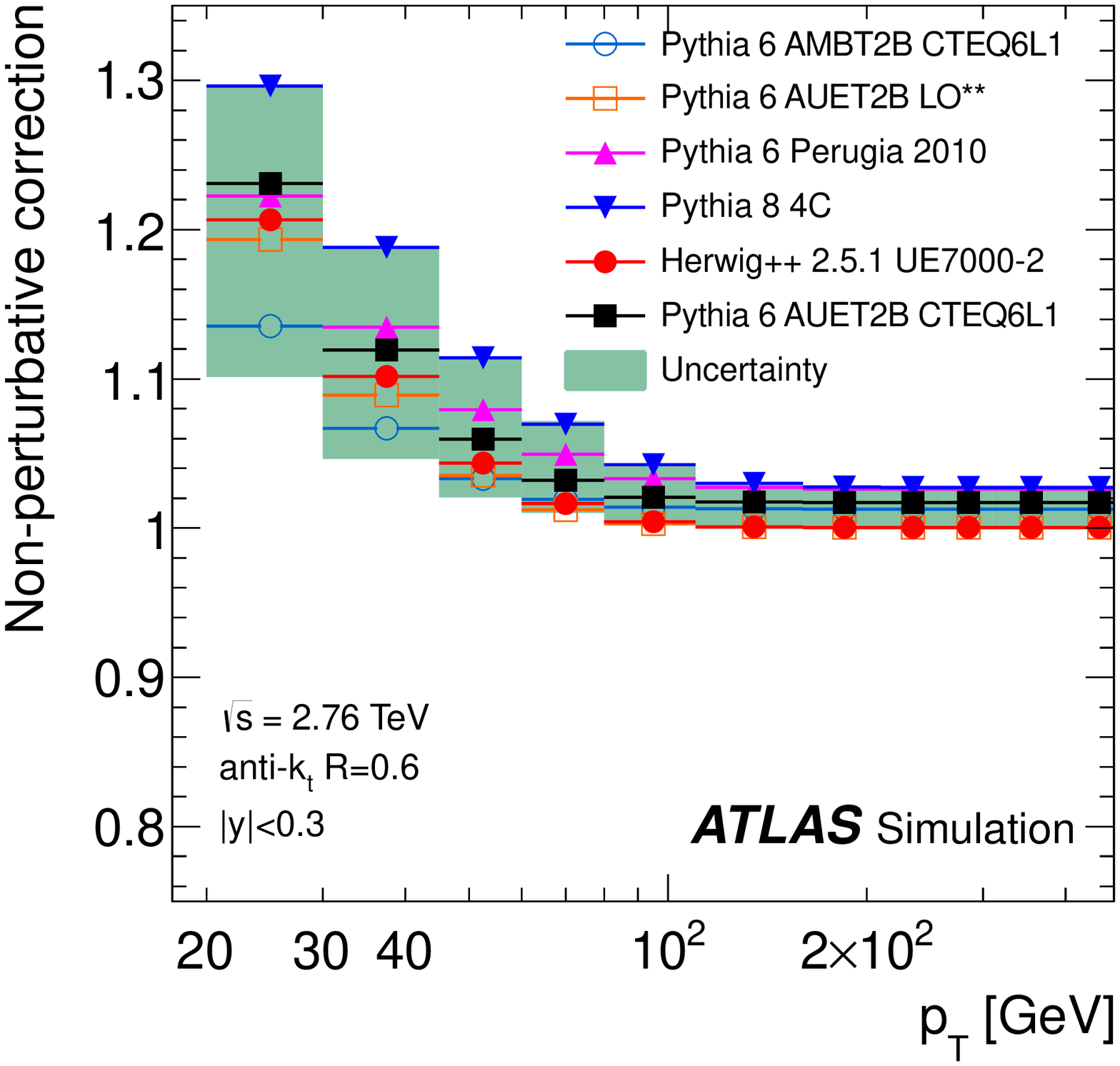}}
\caption{Non-perturbative correction factors for the inclusive jet cross-section for \antikt{} jets with (a) $R=0.4$ and (b) $R=0.6$ in the jet rapidity region $|y|<0.3$ as a function of the jet \pt{} for Monte Carlo simulations with various tunes. The correction factors derived from \pythia{} 6 with the AUET2B CTEQ6L1 tune (full-square) are used for the NLO pQCD prediction in this measurement, with the uncertainty indicated by the shaded area. For better visibility, some tunes used in the uncertainty determination are not shown.}
\label{fig:npcxs}
\end{center}
\end{figure*}

The correction factors for non-perturbative effects and their uncertainties are shown in Fig.~\ref{fig:npcxs} for the inclusive jet cross-section at $\sqrt{s}=2.76$ TeV in the central rapidity bin. 
For jets with $R=0.4$, the correction is about $-10\%$ in the lowest \pt{} bin, while for jets with $R=0.6$, it is about $+20 \%$ as a result of the interplay of the hadronisation and the underlying event for the different jet sizes.
In the high-\pt{} region, the corrections are almost unity for both jet radius parameters, and the uncertainty is at the level of~$\pm2\%$.
 
\subsection{Prediction for the cross-section ratio}
Figs.~\ref{fig:nlorxtrpt_06_a}--\subref{fig:nlorxtrpt_06_c} show the uncertainty on the NLO pQCD calculation of $\rho(y,\xt)$ in representative rapidity bins for $R=0.6$. 
They are significantly reduced to a level of a few percent in the central rapidity region compared to the uncertainties on the cross-sections shown in Fig.~\ref{fig:nloxs_06}. 
The dominant uncertainty at low \pt{} is the uncertainty on the renormalisation and factorisation scale choice, while at high \pt{} the uncertainty due to the PDF contributes again significantly. 
The NLO pQCD calculation of $\rho(y,p_T)$ has an uncertainty of less than $\pm 5\%$ for \pt{} up to 200~\GeV{} 
in the central rapidity region, as shown in Fig.~\ref{fig:nlorxtrpt_06_d}.
The uncertainty increases for higher \pt{} of the jet due mostly to the uncertainties on the PDFs, which are below $10 \%$ for central jets. In the forward region, it reaches up to $80 \%$ in the highest \pt{} bins, as shown in Figs.~\ref{fig:nlorxtrpt_06_e} and~\ref{fig:nlorxtrpt_06_f}. 
The corresponding uncertainties for jets with $R=0.4$ are similar, except for a larger contribution due to the scale choice in the uncertainty on $\rho(y,p_T)$.

\begin{figure*}[htb]
\begin{center}
\subfigure[$\rho(y,\xt),\ |y|<0.3$]{\label{fig:nlorxtrpt_06_a}\includegraphics[width=5.2cm]{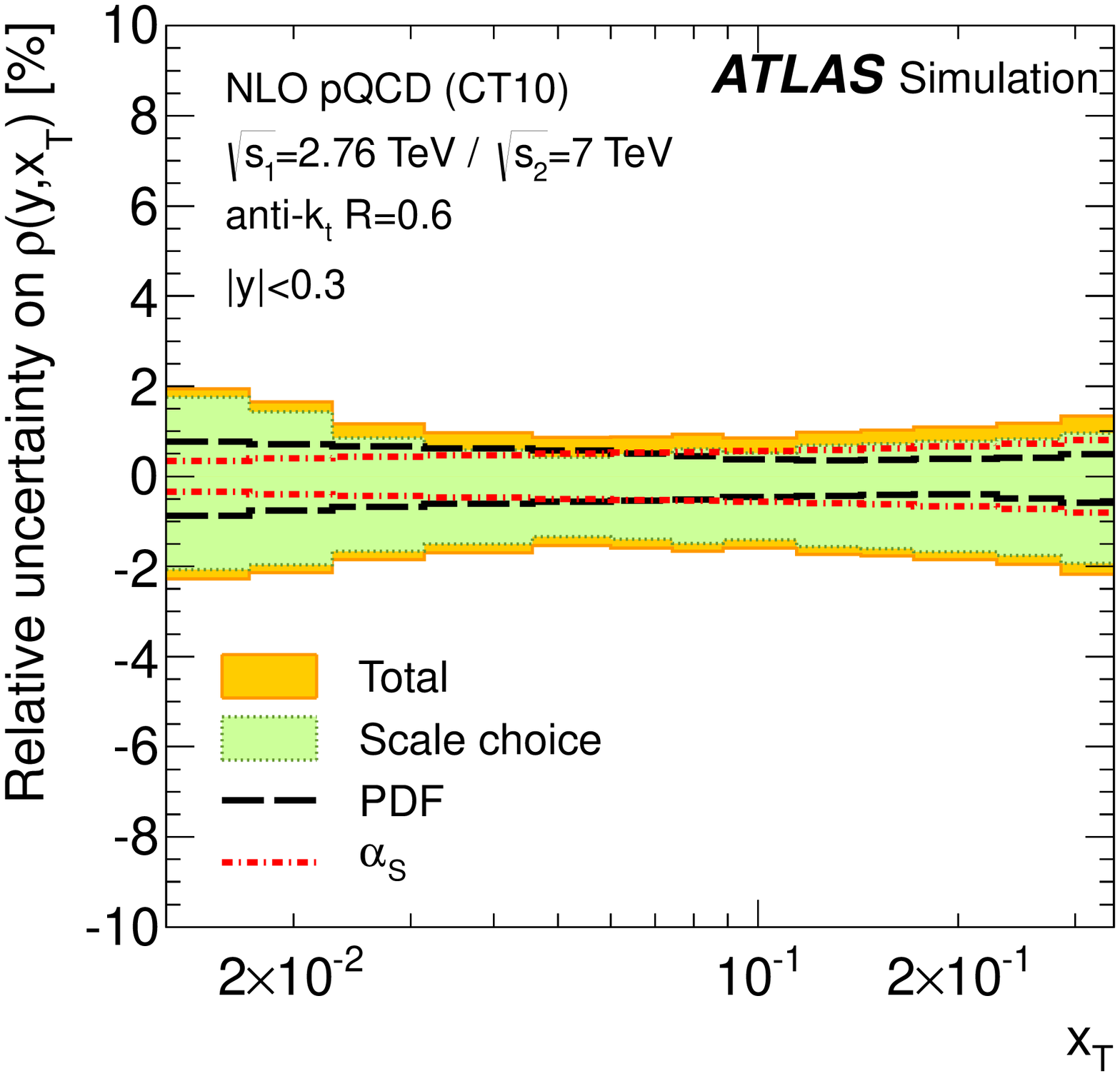}}
\subfigure[$\rho(y,\xt),\ 2.1\leq|y|<2.8$]{\label{fig:nlorxtrpt_06_b}\includegraphics[width=5.2cm]{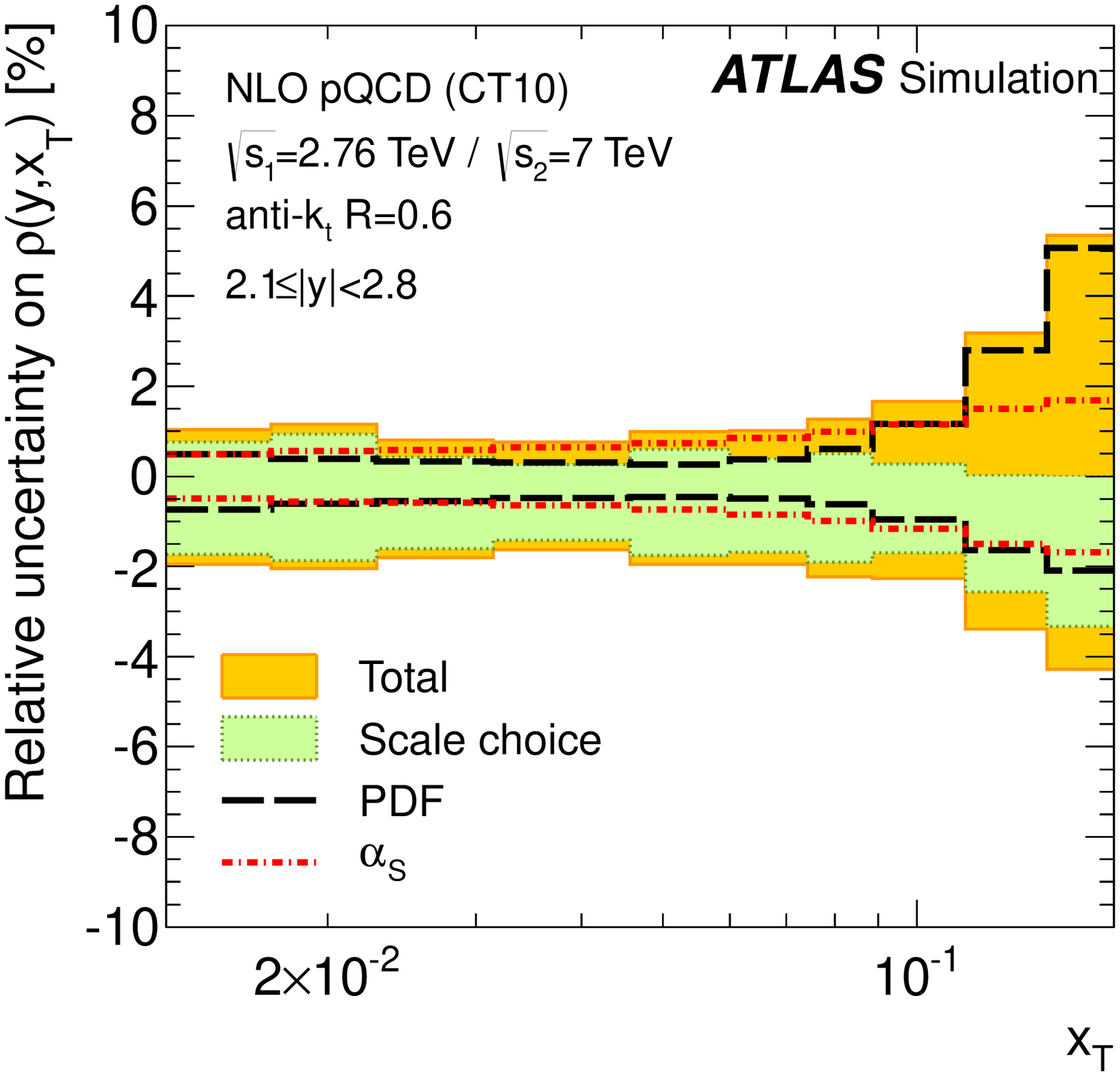}}
\subfigure[$\rho(y,\xt),\ 3.6\leq|y|<4.4$]{\label{fig:nlorxtrpt_06_c}\includegraphics[width=5.2cm]{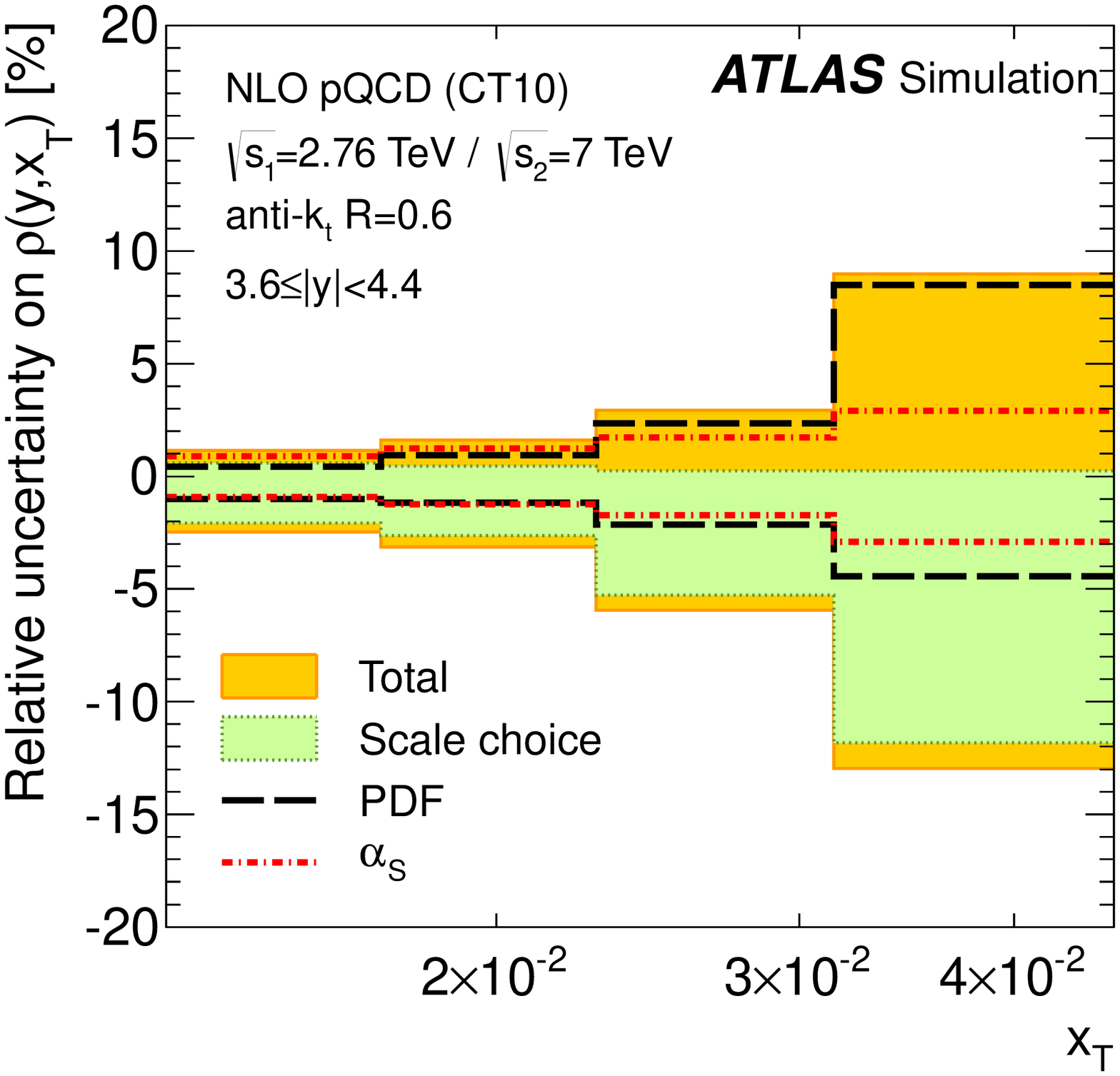}}\\
\subfigure[$\rho(y,\pt),\ |y|<0.3$]{\label{fig:nlorxtrpt_06_d}\includegraphics[width=5.2cm]{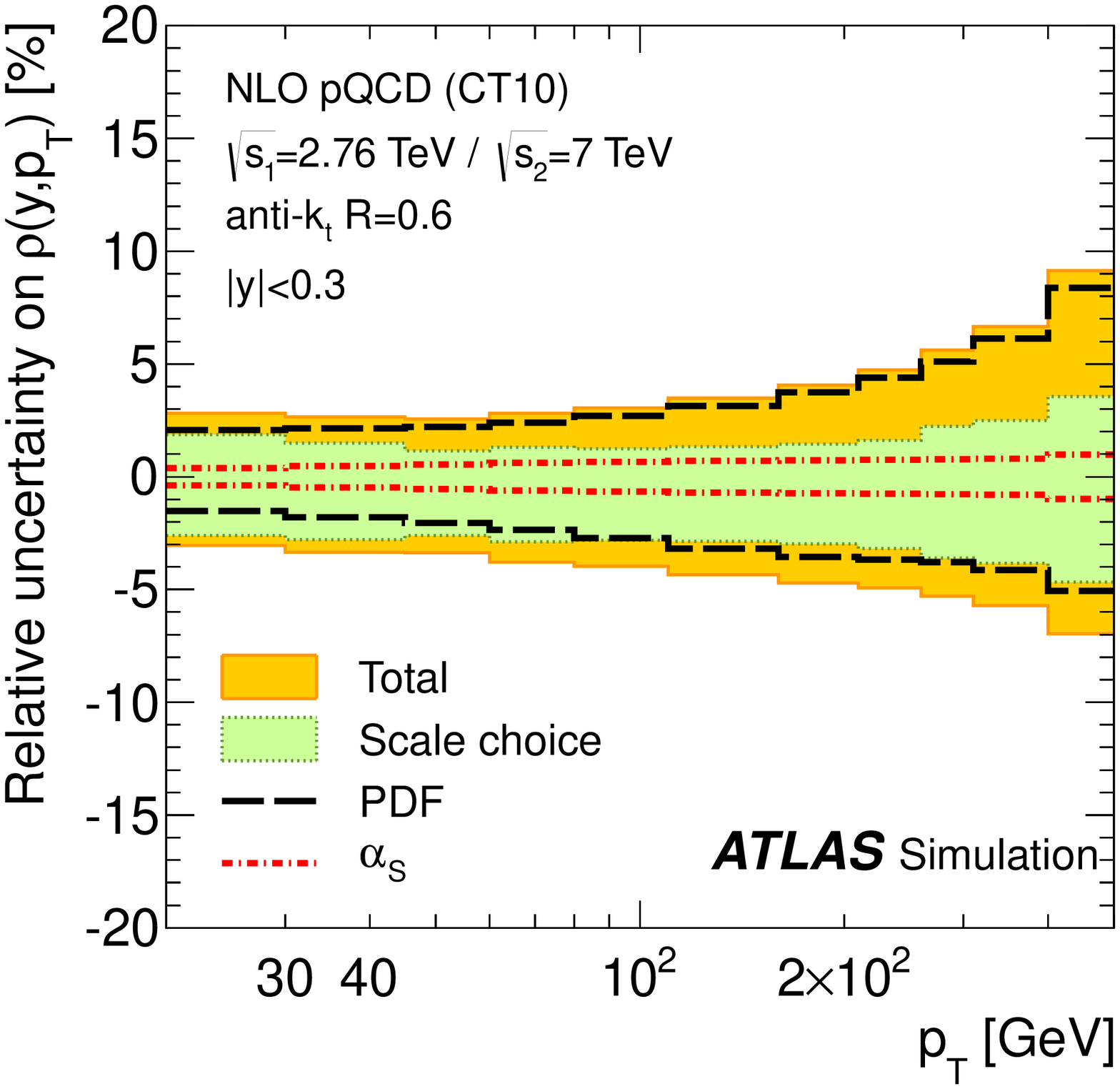}}
\subfigure[$\rho(y,\pt),\ 2.1\leq|y|<2.8$]{\label{fig:nlorxtrpt_06_e}\includegraphics[width=5.2cm]{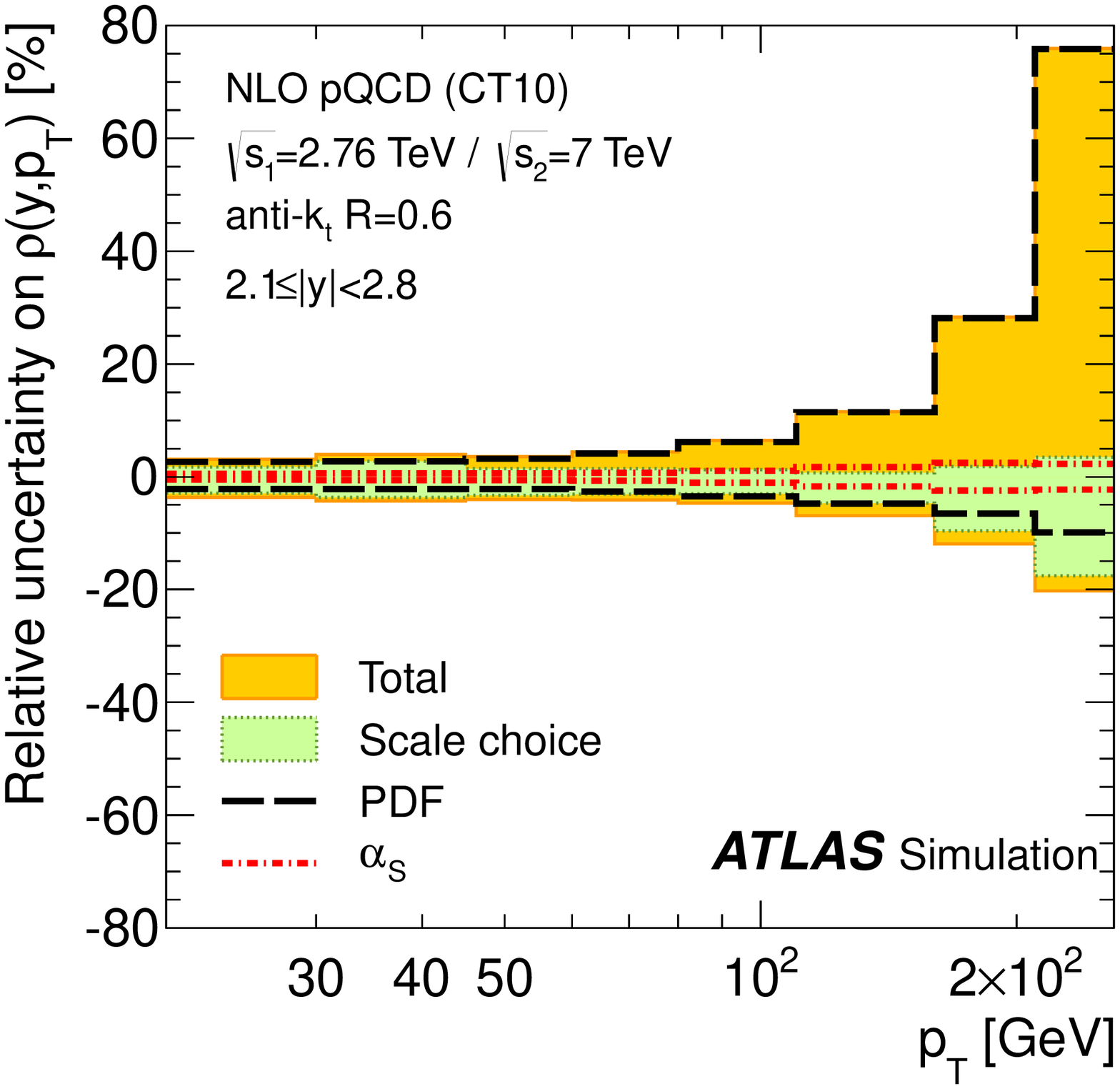}}
\subfigure[$\rho(y,\pt),\ 3.6\leq|y|<4.4$]{\label{fig:nlorxtrpt_06_f}\includegraphics[width=5.2cm]{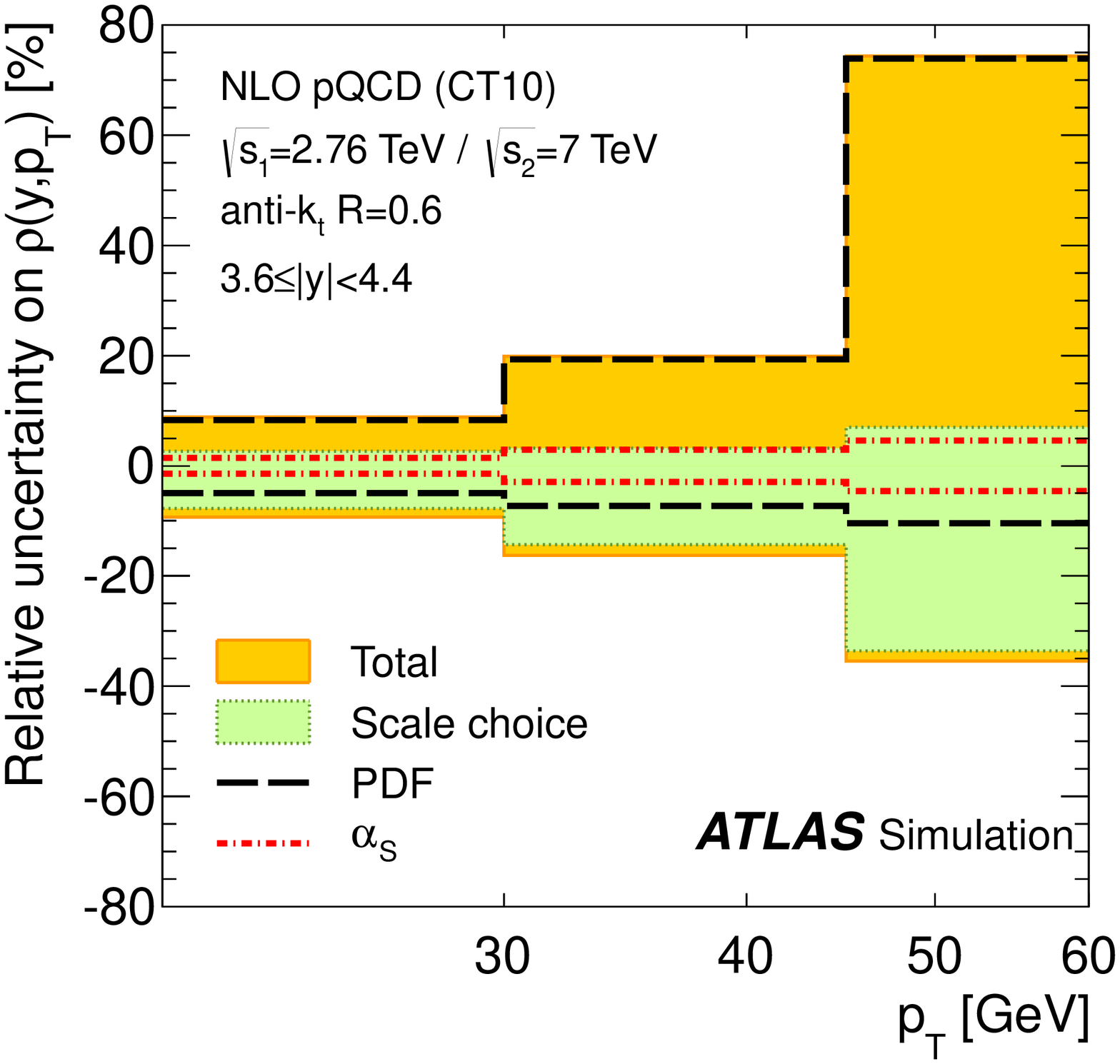}}
\caption{The uncertainty in the NLO pQCD prediction of the cross-section ratio $\rho(y,\xt)$ ((a)--(c)) 
and $\rho(y,\pt)$ ((d)--(f)), 
calculated using \nlojetpp{} with the CT10 PDF set, for \antikt{} jets 
with $R=0.6$ shown in three representative rapidity bins as a function of the jet \xt{} and of the jet \pt{}, respectively. 
In addition to the total uncertainty, the uncertainties from the scale choice, 
the PDF set and the strong coupling constant, $\alphas$, are shown separately.}
\label{fig:nlorxtrpt_06}
\end{center}
\end{figure*}

Non-perturbative corrections to $\rho(y,x_T)$ have a different \xt{} dependence for jets 
with $R=0.4$ and $R=0.6$, as shown in Figs.~\ref{fig:npcrxtrpt_a} and~\ref{fig:npcrxtrpt_b}.
The behaviour of $\rho(y,\xt)$ is driven by the corrections for the cross-section at $\sqrt{s}=2.76$~\TeV{}
since $\pt^{7 \TeV}=(7/2.76)\cdot\pt^{2.76 \TeV}$ in the same \xt{} bins (see\linebreak \ref{sec:bins})
and since the non-per\-tur\-ba\-tive correction is almost flat in the high-\pt{} region.
For jets with $R=0.4$, the correction is $-10\%$ in the lowest \xt{} bin. For $R=0.6$, the correction in this region is in the opposite direction, increasing the prediction by $+10 \%$. The uncertainty in the lowest \xt{} bin for both radius parameters is $\sim \pm10 \%$. 
The non-perturbative corrections to $\rho(y,p_T)$ are shown in Figs.~\ref{fig:npcrxtrpt_c} and~\ref{fig:npcrxtrpt_d}, where a similar \pt{} dependence for $R=0.4$ and $R=0.6$ is found. 
They amount to  $-10 \%$ for jets with $R=0.4$ and $-25 \%$ for jets with $R=0.6$ in the lowest \pt{} bins.
This is due to the correction factors for the NLO pQCD prediction 
at $\sqrt{s}=7$~\TeV{}~\cite{Aad:2011fc} being larger than those at $\sqrt{s}=2.76$~\TeV{}.
Corrections obtained from \pythia{} with various tunes generally agree within $5 \%$ for central jets, while the non-perturbative corrections 
from \herwigpp{} deviate from the ones of the \pythia{} tunes by more than $10 \%$ in the lowest \pt{} bin.

\begin{figure*}[htb]
\begin{center}
\subfigure[$\rho(y,\xt),\ R=0.4$]{\label{fig:npcrxtrpt_a}\includegraphics[width=7.5cm]{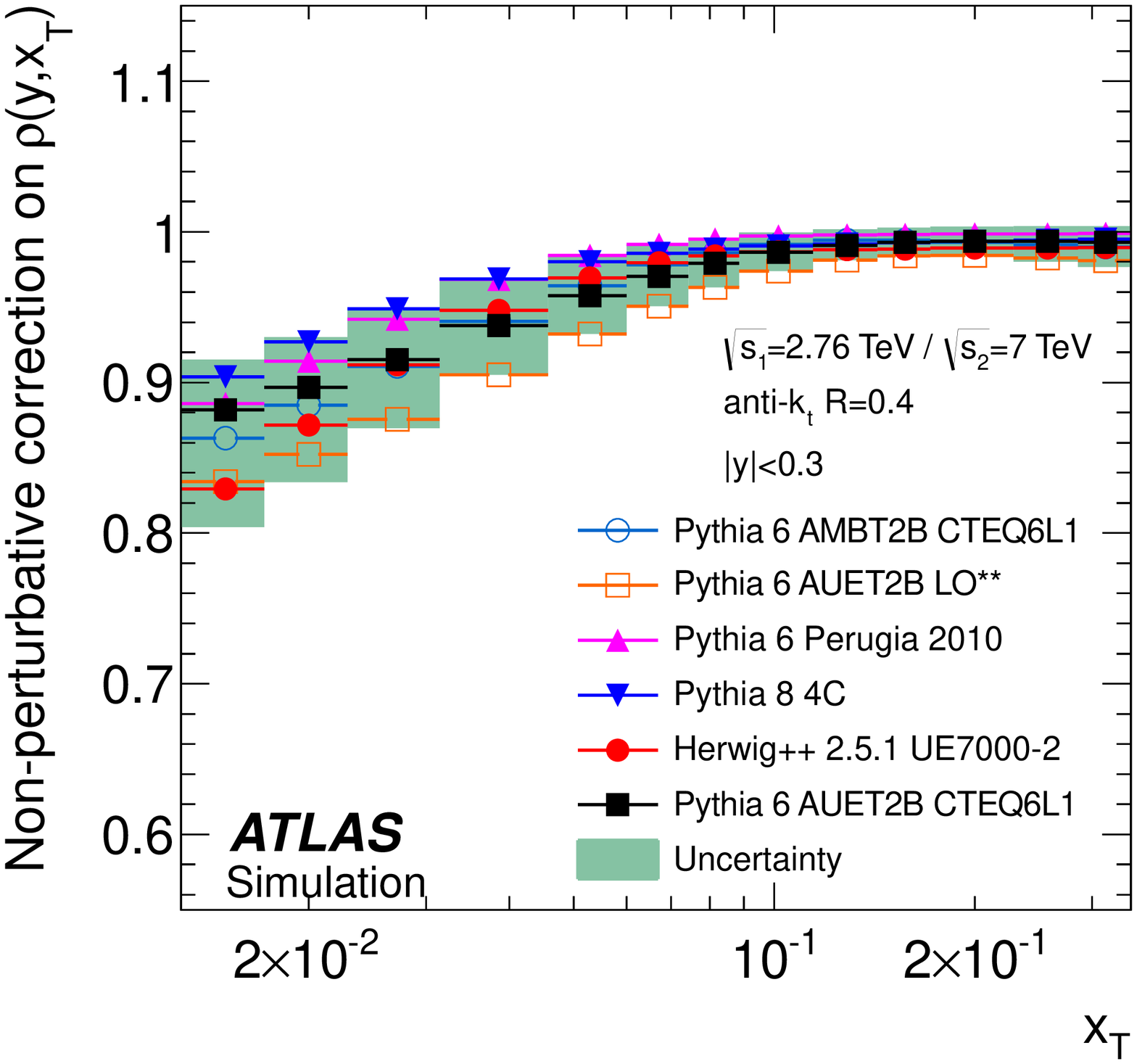}}
\subfigure[$\rho(y,\xt),\ R=0.6$]{\label{fig:npcrxtrpt_b}\includegraphics[width=7.5cm]{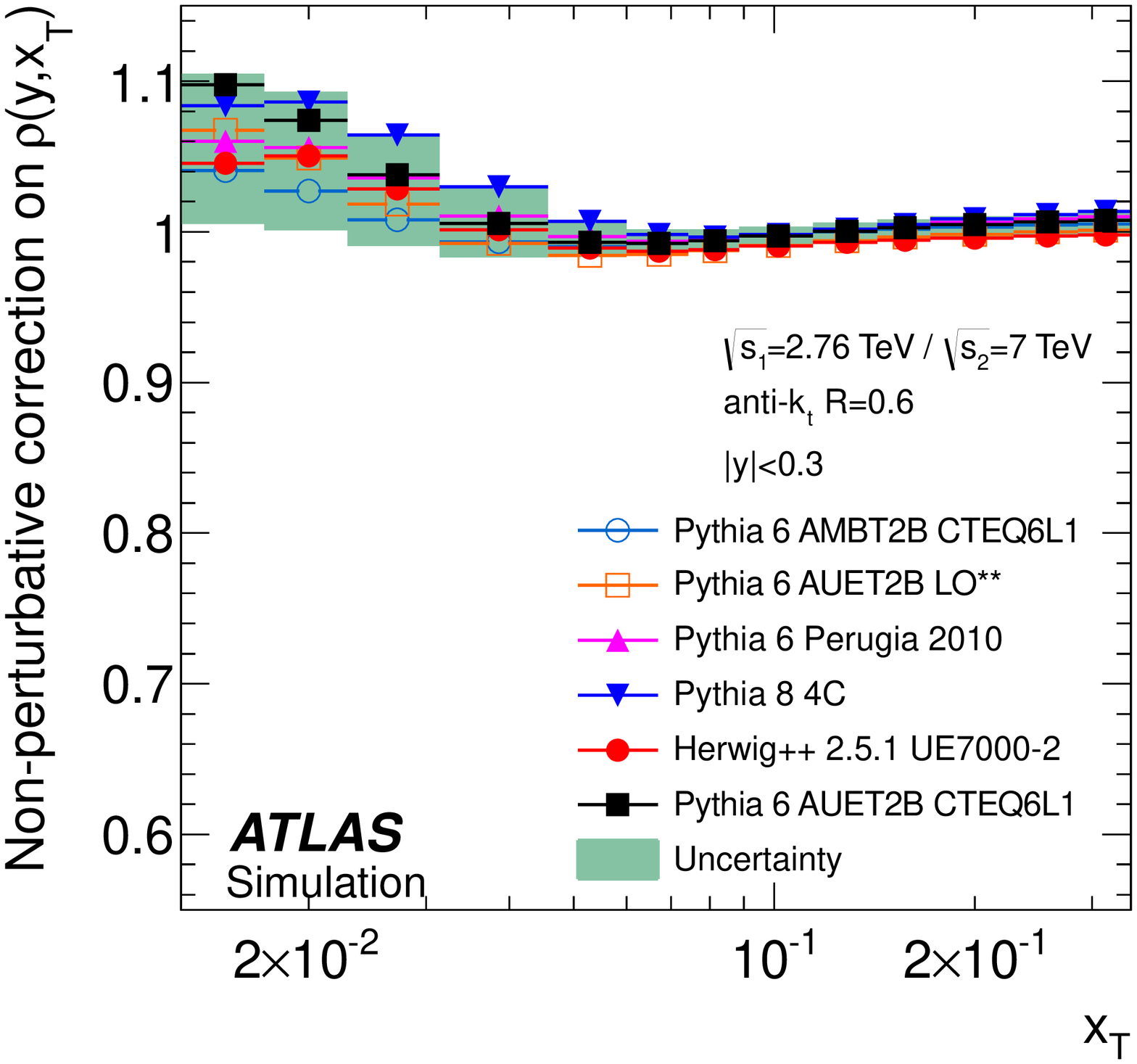}}\\
\subfigure[$\rho(y,\pt),\ R=0.4$]{\label{fig:npcrxtrpt_c}\includegraphics[width=7.5cm]{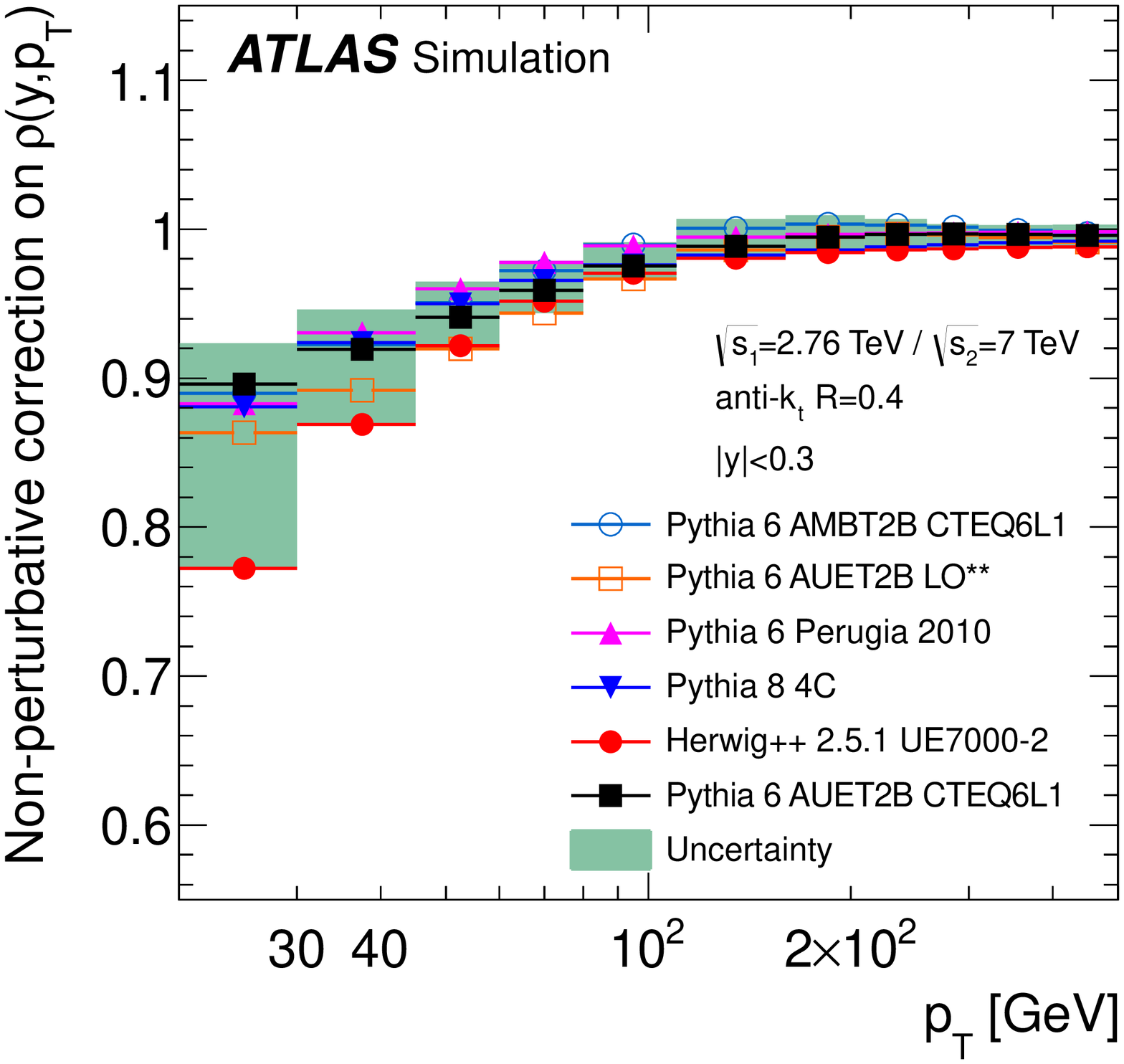}}
\subfigure[$\rho(y,\pt),\ R=0.6$]{\label{fig:npcrxtrpt_d}\includegraphics[width=7.5cm]{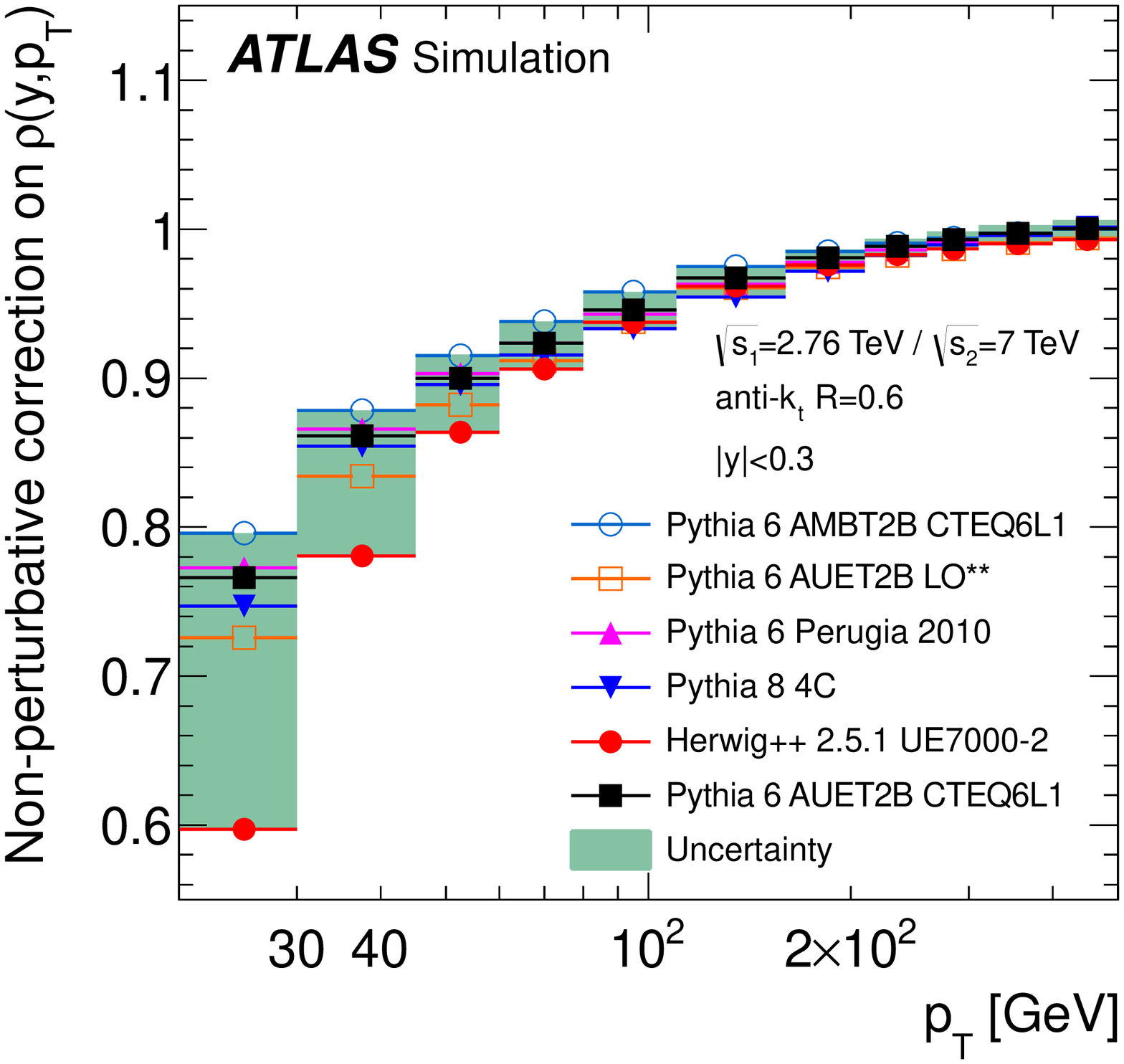}}
\caption{Non-perturbative correction factors for the cross-section ratios, $\rho(y,\xt)$ and $\rho(y,\pt)$, 
for \antikt{} jets with $R=0.4$ or $R=0.6$ shown for a jet rapidity of $|y|<0.3$ for Monte Carlo simulations with various tunes  
as a function of the jet \xt{} and of the jet \pt{}, respectively. 
The correction factors derived from \pythia{} 6 with the AUET2B CTEQ6L1 tune (full-square) are used for the NLO pQCD prediction in this measurement, with the uncertainty indicated by the shaded area. For better visibility, some tunes used in the uncertainty determination are not shown.}
\label{fig:npcrxtrpt}
\end{center}
\end{figure*}

\section{Event selection}
\label{sec:events}
Events are selected online using various trigger definitions according to the \pt{} and the rapidity $y$ of the jets \cite{Aad:2012xs}. 
In the lowest \pt{} region ($\pt < 35$~\GeV{} for $|y| < 2.1$, $\pt < 30$~\GeV{} for $2.1 \leq |y| <2.8$, $\pt < 28$~\GeV{}  for $2.8 \leq |y| <3.6$, and $\pt < 26$~\GeV{} for $3.6 \leq |y| < 4.4$), a trigger requiring at least two hits in the MBTS is used. For the higher \pt{} region, jet-based triggers are used, which select events that contain a jet with sufficient transverse energy at the electromagnetic scale\footnote{The electromagnetic scale is the basic calorimeter signal scale for the ATLAS calorimeter. It has been established using test-beam measurements for electrons and muons to give the correct response for the energy deposited in electromagnetic showers, but it does not correct for the lower response of the calorimeter to hadrons.}. 
The efficiency of the jet-based triggers is determined using the MBTS, and the one for MBTS using the independent trigger
from the Zero Degree Calorimeter~\cite{Jenni:1009649}.
Only triggers that are $>99\%$ efficient for a given jet \pt{} value are used. 
In the region $2.8 < |y| < 3.6$, both a central and a forward jet trigger are used in combination to reach an efficiency of $>99\%$.   
Events are required to have at least one well-re\-con\-structed event vertex, 
which must have at least three associated tracks with a minimum \pt{} of $150\MeV$.

\section{Jet reconstruction and calibration}
\label{sec:jets}
The reconstruction procedure and the calibration factors for the jet cross-section measurement at $\sqrt{s}=2.76$~\TeV{} are nearly identical to those used for the measurement at $\sqrt{s}=7$~\TeV{} with $2010$ data~\cite{Aad:2011fc}; the few exceptions are explicitly specified below.

Jets are reconstructed with the \antikt{} algorithm using as input objects
topological clusters~\cite{topoclusters,Aad:2011he} of energy deposits in the ca\-lo\-ri\-meter, 
calibrated at the electromagnetic scale. 
The four-momenta of the reconstructed jets are corrected event-by-event using the actual vertex position. 
A jet energy scale (JES) correction is then applied to correct for detector effects 
such as energy loss in dead material in front of the calorimeter or between calorimeter segments,
and to compensate for the lower calorimeter response to hadrons than to electrons or photons
\cite{topoclusters,Aad:2011he}. 
Due to the low number of interactions per bunch crossing, an offset
correction accounting for additional energy depositions from multiple
interactions in the same bunch crossing, so-called pile-up, is not applied in
this measurement.

The estimation of the uncertainty in the jet energy measurement  
uses single-hadron calori\-meter response measurements \cite{Aad:2012vm} and systematic
Monte Carlo simulation variations.
An uncertainty of about $2.5 \%$ in the central calorimeter region 
over a wide momentum range of  $60 \leq \pt < 800\GeV$ 
is obtained ~\cite{Aad:2011he}. For jets with lower \pt{} and for forward jets the uncertainties are larger.

All reconstructed jets with $\pt > 20$~\GeV{}, $|y| < 4.4$ and a positive decision from the trigger that is used 
in the corresponding jet kinematic region are considered in this analysis. Jets are furthermore required to pass jet quality selections to reject fake jets reconstructed from non-collision signals, such as beam-related background, cosmic rays or detector noise. The applied selections were established with the $\sqrt{s}=7$~\TeV{} data in $2010$~\cite{Aad:2011fc,Aad:2011he} and are validated in the $\sqrt{s}=2.76$ \TeV{} data by studying distributions of the selection variables with techniques similar to those in Ref.~\cite{Aad:2011he}. The rate of fake jets after the jet selection is negligible.

The efficiency of the jet quality selection is measured using a tag-and-probe method~\cite{Aad:2011he}.
The largest inefficiency is found to be below $4 \%$ for jets with $\pt=20$~\GeV.
Within the statistical uncertainty, the measured efficiency is in good agreement 
with the efficiency previously measured for $\sqrt{s}=7$~\TeV{} data in 2010.
Because of the larger number of events in the $2010$ data at $\sqrt{s}=7$~\TeV, 
the jet selection efficiency from the $2010$ data is taken. 

Various types of validity and consistency checks have been performed on the data, such as testing the expected invariance of the jet cross-section as a function of $\phi$, or the stability of the jet yield over time. No statistically significant variations are detected. The basic kinematic variables are described by the Monte Carlo simulation within the systematic uncertainties.

\section{Unfolding of detector effects}
\label{sec:unfold}
Corrections for the detector inefficiencies and resolutions are performed to extract the particle-level cross-section, \linebreak
based on a transfer matrix that relates the \pt{} of the jet at particle-level and the reconstruction-level.

For the unfolding, the Iterative, Dynamically Stabilised (IDS) Bayesian unfolding method~\cite{Malaescu:2009dm} is used. The method takes into account the migrations of events across the bins and uses data-driven regularisation. It is performed separately for each rapidity bin, since migrations across \pt{} bins are significant. The migrations across rapidity bins, which are much smaller, are taken into account using the bin-by-bin unfolding.

The Monte Carlo simulation to derive the transfer matrix is described in Sect.~\ref{sec:mc}. 
The Monte Carlo samples are reweighted on a jet-by-jet basis as a function of jet \pt{} 
and rapidity. The reweighting factors are obtained from the ratio of calculated cross-sections using the MSTW 2008 NLO PDF set~\cite{Martin:2009iq} with respect to the MRST 2007 LO* PDF set~\cite{Sherstnev:2007nd}. This improves the description of the jet~\pt{} distribution in data.
Additionally, a jet selection similar to the jet quality criteria in data is applied to jets with low \pt{} in the Monte Carlo simulation at $|\eta|\sim1$.

The transfer matrix for the jet \pt{} is derived by matching a particle-level jet to a reconstructed jet based on a geometrical criterion, in which a particle-level jet and a reconstructed jet should be closest to each other within a radius of $R'=0.3$ in the $(\eta, \phi)$-plane.
The spectra of unmatched particle-level and reconstructed jets are used to provide the matching efficiencies, obtained from the 
number of the matched jets divided by the number of all jets including unmatched jets, both for particle-level jets, $\epsilon^\mathrm{part}$, and for reconstructed jets, $\epsilon^\mathrm{reco}$.

The data are unfolded to particle level using a three-step procedure, namely, correction for matching inefficiency at reconstructed level, 
unfolding for detector effects and then correction for matching inefficiency at particle level. 
The final result is given by the equation:
\begin{equation}
N^\mathrm{part}_i=\sum_jN^\mathrm{reco}_j\cdot\epsilon^\mathrm{reco}_j\ A_{ij}\ /\epsilon^\mathrm{part}_i,
\end{equation}
where $i$ and $j$ are the particle-level and reconstructed bin indices, respectively, and $N^\mathrm{part}_k$ and $N^\mathrm{reco}_k$ are the number of particle-level jets and the number of reconstructed jets in bin $k$. $A_{ij}$~is an unfolding matrix, which gives the probability for a re\-con\-structed-level jet with a certain re\-con\-structed-level \pt{} to have a given particle-level \pt{}. It is determined using the IDS meth\-od.
The number of iterations is chosen such that the bias in the closure test (see below) is small and at most at the percent level.
In this measurement, this is achieved after one iteration. 

The precision of the unfolding technique has been studied using a data-driven closure test~\cite{Malaescu:2009dm}. 
In this study the particle-level \pt{} spectrum in the Monte Carlo simulation is re\-weighted and convolved through the folding matrix, which gives the probability for a particle-level jet with a certain particle-level \pt{} to have a given reconstructed-level \pt{}.
The weights are chosen such that significantly improved agreement between the resulting reconstructed spectrum and data is attained. 
The reconstructed spectrum in this reweighted Monte Carlo simulation is then unfolded using the same procedure as for the data.
Comparison of the spectrum obtained from the unfolding with the original reweighted particle-level spectrum provides an estimate of the bias, which is interpreted as the systematic uncertainty. 

As an estimate of further systematic uncertainties, the unfolding procedure is repeated using different transfer matrices created with tighter and looser matching criteria of $R'=0.2$ and $R'=0.4$. 
The deviations of the results from the nominal unfolding result are considered as an additional uncertainty on the unfolding procedure. 

The statistical uncertainties are propagated through the unfolding by performing pseudo-experiments.
An ensemble of pseudo-experiments is created in which each bin of the transfer matrix is varied according to its statistical uncertainty from the Monte Carlo samples. 
A separate set of pseudo-\-experiments is performed in which the data spectrum is fluctuated according to the statistical uncertainty taking the correlation between jets produced in the same event into account.
The unfolding is then applied to each pseudo-\-experi\-ment, and the resulting ensembles are used to calculate the covariance matrix of the corrected spectrum, from which the uncertainties are obtained.

The unfolding procedure is repeated for the propagation of the uncertainties on the jet energy and angle measurements, as described in the next section. 

\section{Systematic uncertainties on the cross-section measurement}
\label{sec:uncert}
The following sources of systematic uncertainty are considered in this measurement:
the trigger efficiency, jet reconstruction and calibration, 
the unfolding procedure and the luminosity measurement. 

An uncertainty on the trigger efficiency of $1 \%$ is conservatively chosen 
for most of the kinematic region ($|y|<2.8$; $\pt \geq 45 \GeV{}$ in $2.8\leq|y|<3.6$; and $\pt \geq 30 \GeV{}$ in $3.6\leq|y|<4.4$).
A $2 \%$ systematic uncertainty is assigned 
for jets with $\pt < 45$ \GeV{} in the region $2.8 \leq |y| < 3.6$ 
or with $\pt < 30$ \GeV{} in the region $3.6 \leq |y| < 4.4$, as the triggers are used for \pt{} 
close to the lowest \pt{} point with $99 \%$ efficiency for these jets.

The uncertainty on the jet reconstruction efficiency is the same as in the previous measurement 
at $\sqrt{s}=7$ \TeV~\cite{Aad:2011fc} and is $2 \%$ for $\pt<30$ \GeV{} 
and $1 \%$ for $\pt>30$ \GeV{}. It is evaluated using jets reconstructed from tracks~\cite{Aad:2011he}. 
The uncertainty on the jet selection efficiency from the measurement
at $\sqrt{s}=7\TeV{}$ is applied in this measurement, but a minimal uncertainty of $0.5\%$
is retained. The latter accounts for the level of agreement of the central value
in the comparison between the used jet selection efficiency and the measured jet
selection efficiency at $\sqrt{s} = 2.76\TeV{}$.

The uncertainty due to the jet energy calibration is evaluated using the
same uncertainties on the sources as in the previous measurement
at $\sqrt{s}=7\TeV$~\cite{Aad:2011fc}.
Effects from the systematic uncertainty sources are propagated through the unfolding procedure to provide 
the uncertainties on the measured cross-sections. 
The JES uncertainty 
and its sources are described in detail in Ref.~\cite{Aad:2011he},
where the total JES uncertainty is found to be less than $2.5\%$ in the central calorimeter region 
for jets with $60 < \pt{} < 800\GeV{}$, and maximally $14\%$ for $\pt{} < 30\GeV{}$ in the most forward region. 
The JES applied to the reconstructed jets in the Monte Carlo simulation 
is varied separately for each JES uncertainty source both up and down by one standard deviation. 
The resulting \pt{} spectra are unfolded using the nominal unfolding matrix. 
The relative shifts with respect to the nominal unfolded spectrum are taken as uncertainties 
on the cross-section.

The uncertainty on the jet energy resolution (JER) is assigned by considering the difference between data and Monte Carlo 
simulation in the estimated JER using \insitu{} techniques~\cite{Aad:2012ag}. 
The measured resolution uncertainty ranges from $20\%$ to $10\%$ for jets within $|y| < 2.8$ and with transverse momenta from $30\GeV{}$ to $500\GeV{}$.
The difference between data and MC is found to be within $10\%$.
The effect of this uncertainty on the cross-section measurement is evaluated by smearing the energy of reconstructed jets in the Monte Carlo simulation such that the resolution is worsened by the one-standard-deviation uncertainty. 
Then a new transfer matrix is constructed and used to unfold the data spectra. 
The relative difference between the cross-sections unfolded with the modified transfer matrix and with the nominal one is taken as the uncertainty in the measurement. 

The jet angular resolution is estimated in Monte Carlo simulation from the polar angle between the reconstructed jet and its matched jet at particle level. A new transfer matrix with angular resolution degraded by $10\%$ is used for the data unfolding, and the relative difference from the nominal unfolded result is assigned as the resulting uncertainty. 

The uncertainties in the unfolding procedure are de\-scri\-bed in Sect.~\ref{sec:unfold}.
The closure test and the variation of the matching criterion used to construct the transfer matrix are examined. 
The impact of a possible mis-modelling of the jet \pt{} spectrum in the Monte Carlo simulation 
is assessed in the closure test of the unfolding procedure. 

The integrated luminosity is calculated by measuring \pp{} interaction rates with several
ATLAS devices. The absolute calibration is derived from van der Meer scans ~\cite{vanderMeer:1968zz,Aad2011a}.
In total, four scan sessions were performed during the collection of 
the dataset used in the jet cross-section measurements reported here. The 
uncertainty in the luminosity determination arises from three main contributions: 
bunch-population measurements, beam conditions during the luminosity calibration 
scans, and long-term consistency of the different algorithms used to measure the 
instantaneous luminosity during data collection. 
The uncertainty on the luminosity for the $2.76\TeV$ dataset is $\pm 2.7\%$,  
dominated by the irreproducibility of beam 
conditions during the calibration scans. The total systematic uncertainty for 
the 2010 dataset at $\sqrt{s} = 7 \TeV$ is $\pm 3.4\%$~\cite{Aad:2013ucp}, dominated by
bunch-population measurement uncertainties. Because of significant improvements 
to the beam instrumentation im\-ple\-men\-ted between the two 
running periods, and because the dominant systematic uncertainties are of 
independent origins in the two datasets, these luminosity uncertainties are 
treated as uncorrelated.

The evaluated systematic uncertainties on the measured cross-section are added in quadrature and shown in Fig.~\ref{fig:syst_06} for representative rapidity bins and $R=0.6$. Results for jets with $R=0.4$ are similar.
The systematic uncertainty on this measurement is driven by the uncertainties on the JES. 
The very steeply falling jet \pt{} spectrum, especially for large rapidity, transforms even relatively modest uncertainties on the transverse momentum into large changes in the measured differential cross-section. 
The uncertainty on the jet energy resolution also has a sizable effect on the total systematic uncertainty of the measurement in the low \pt{} bins. 
Other sources of uncertainty are found to have a smaller impact on the results. 

\begin{figure*}[htb]
\begin{center}
\subfigure[$|y|<0.3$]{\includegraphics[width=5.2cm]{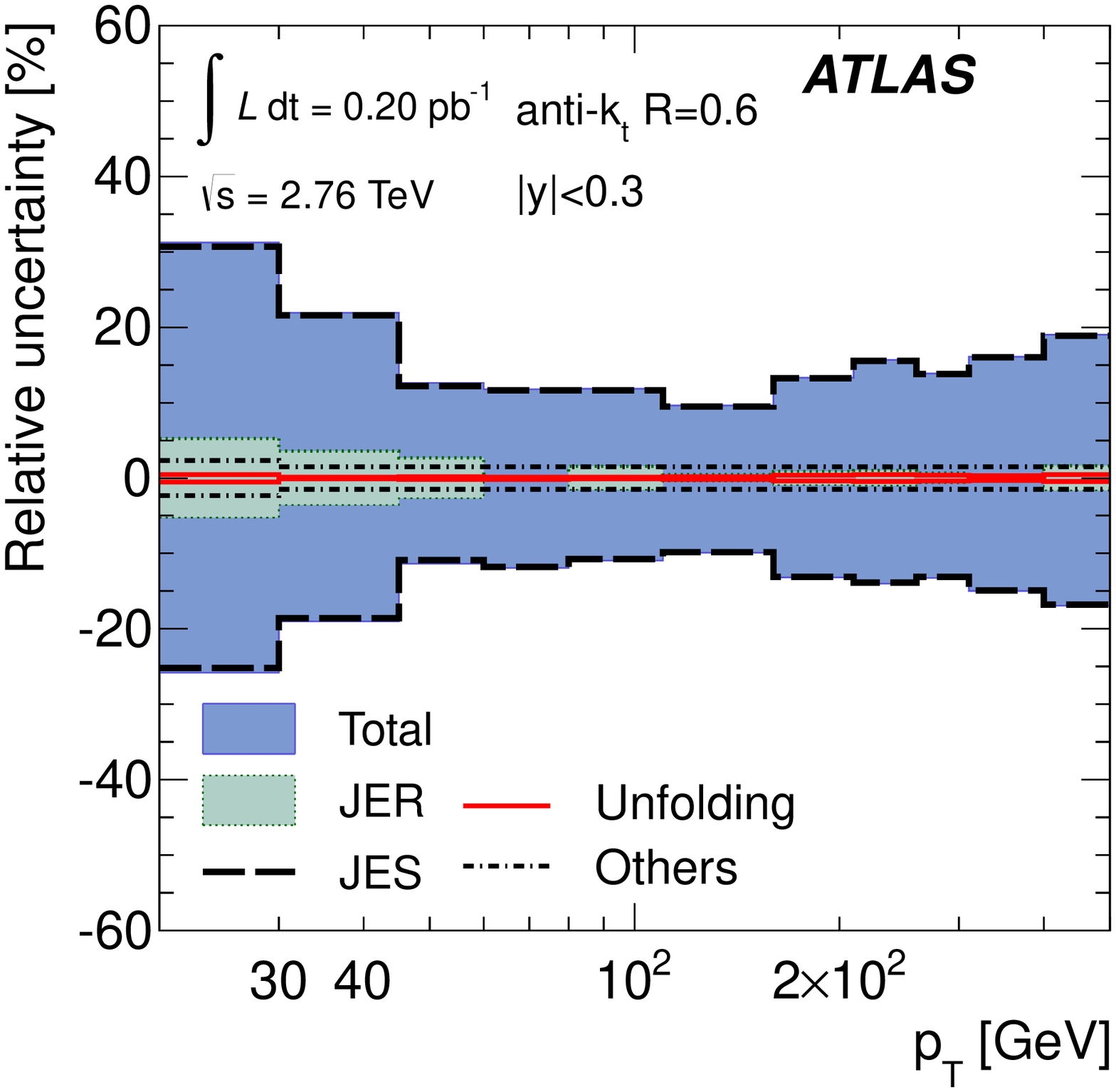}}
\subfigure[$2.1\leq|y|<2.8$]{\includegraphics[width=5.2cm]{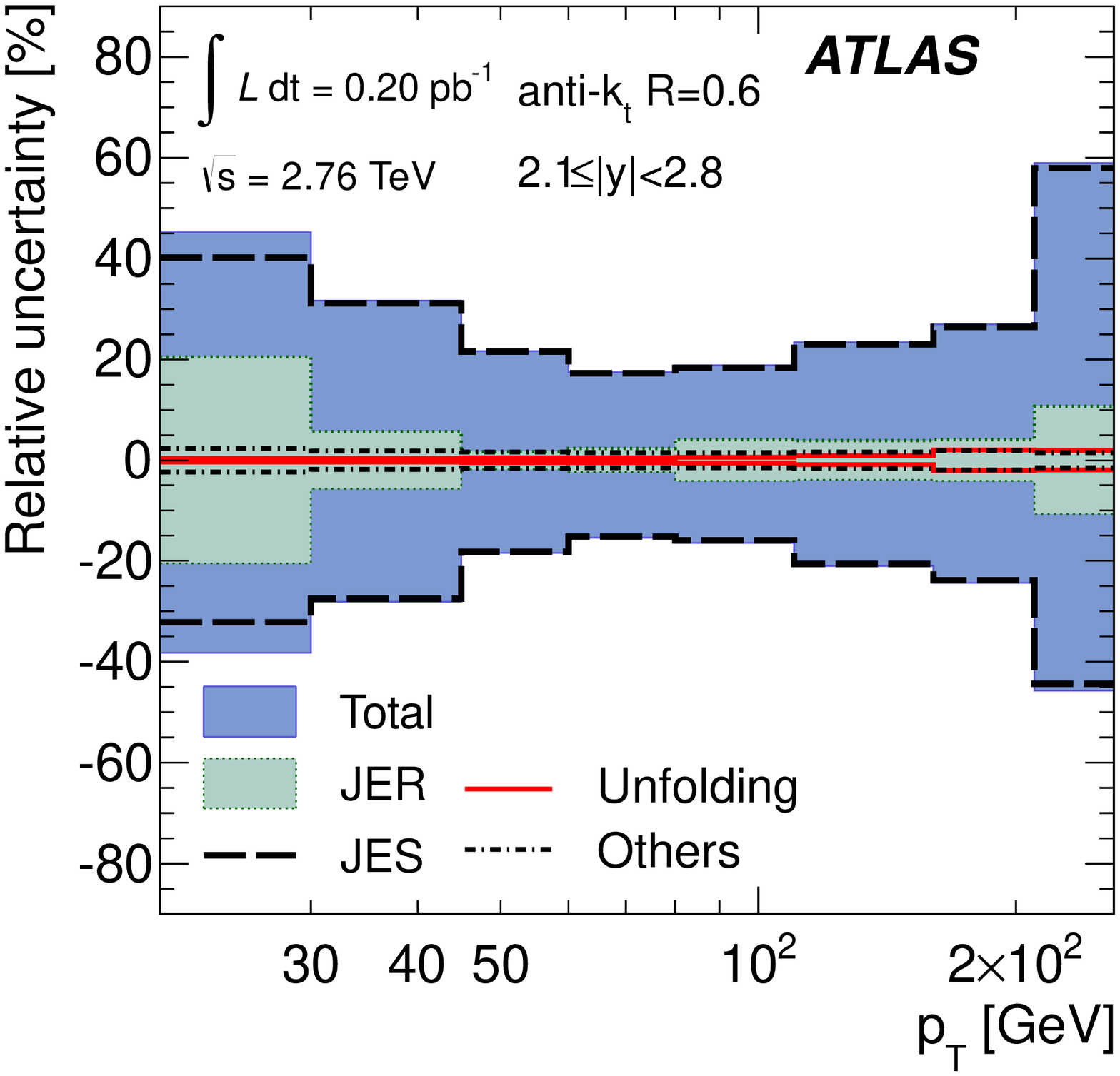}}
\subfigure[$3.6\leq|y|<4.4$]{\includegraphics[width=5.2cm]{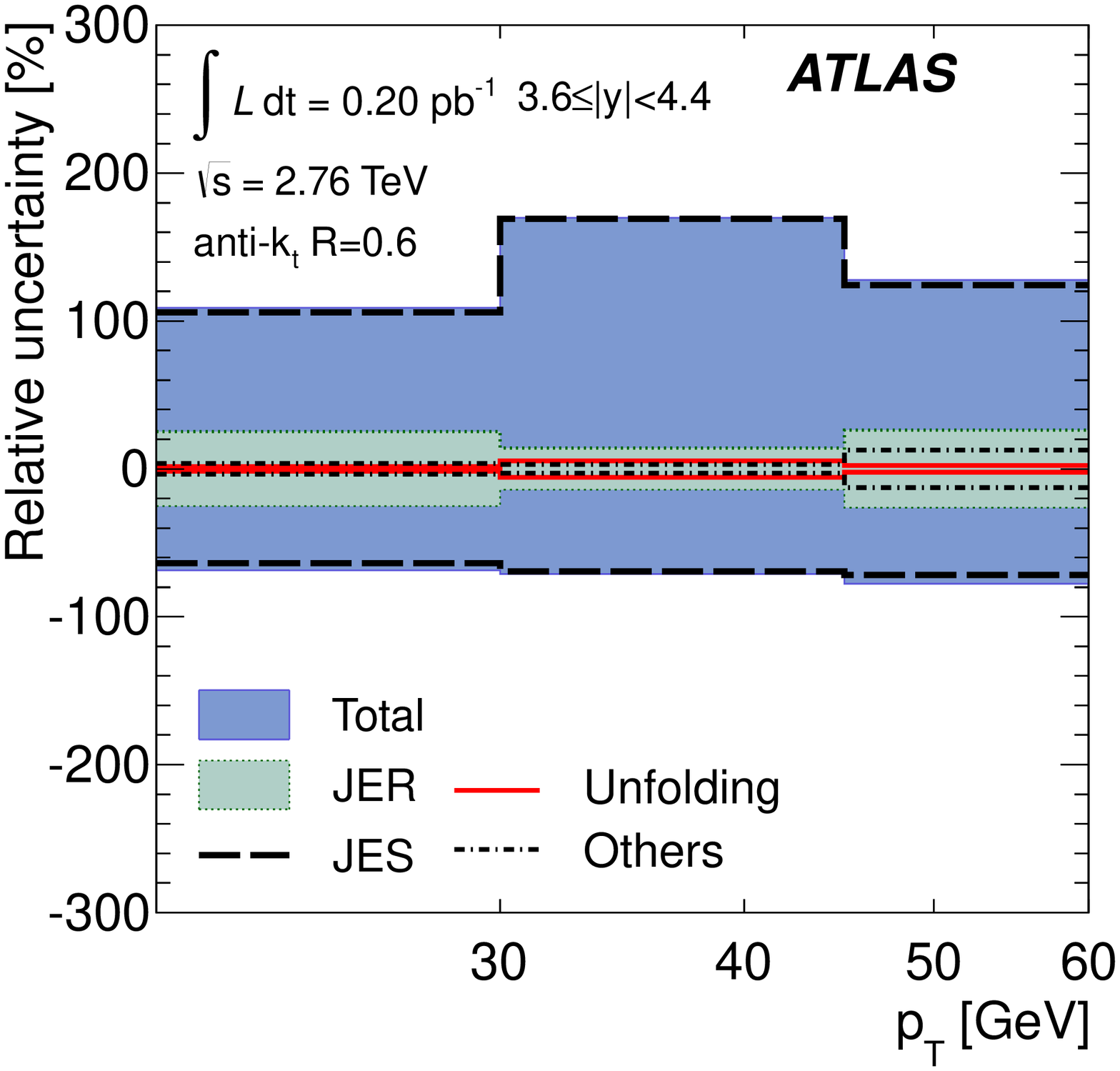}}
\caption{The systematic uncertainty on the inclusive jet cross-section measurement for \antikt{} jets 
with $R=0.6$ in three representative rapidity bins, as a function of the jet \pt. 
In addition to the total uncertainty, the uncertainties from the jet energy scale (JES), 
the jet energy resolution (JER), the unfolding procedure and the other systematic sources are shown separately.
The $2.7\%$ uncertainty from the luminosity measurement and the statistical uncertainty are not shown.}
\label{fig:syst_06}
\end{center}
\end{figure*}

A total of $22$ independent sources of systematic uncertainty have been considered. 
The correlations of the systematic uncertainties across \pt{} and $y$ are examined and summarised in Table~\ref{tab:uncert}. 
In the table, 88 independent nuisance parameters
describe the correlations of systematic uncertainties over the whole phase space. 
The systematic effect on the cross-section measurement associated with each nuisance parameter is treated as completely correlated in \pt{} and $y$. 
The table also shows the correlation with respect to the previous $\sqrt{s}=7$ \TeV{} measurement using $2010$ data, which is used in the extraction of the cross-section ratio in Sect.~\ref{sec:ratio}.

\begin{table*}
 \caption{Description of the bin-to-bin uncertainty correlation in the measurement of the inclusive jet cross-section at $\sqrt{s}=2.76$ \TeV. 
Each number corresponds to a nuisance parameter for which the corresponding uncertainty is fully correlated in the \pt{} of the jet. 
Bins with the same nuisance parameter are treated as fully correlated, while bins with different nuisance parameters are uncorrelated. 
Numbers are assigned to be the same as in the previous publication~\cite{Aad:2011fc}.
The sources labelled by $u_i$ are sources uncorrelated in \pt{} and $y$ of the jet. 
The correlation with the previous cross-section measurement at $\sqrt{s}=7$ \TeV~\cite{Aad:2011fc} is indicated in the last column, where full correlation is indicated by a Y and no correlation by a N.
The description of the JES uncertainty sources can be found in Refs.~\cite{Aad:2011he,Aad:2012vm}.
JES14 is a source due to the pile-up correction and is not considered in this measurement. The sources JES6 and JES15 were merged together in the previous measurement and the sum of the two uncertainties added in quadrature is fully correlated with the JES6 in the previous measurement, indicated by the symbol ``*" in the table.
The nuisance parameter label 31 is skipped in order to be able to keep the same numbers for corresponding nuisance parameters in the two jet cross-section measurements. The values for the nuisance parameters are given in Tables~\ref{tab:xs_r04_y0}-\ref{tab:rpt_r06_y6}.}
  \label{tab:uncert} 
    \small
    \centering
    \begin{tabular}{l|rrrrrrr|c}
    \hline \hline
       Uncertainty source & \multicolumn{7}{c|}{$|y|$ bins} & Correlation\\
                          & 0-0.3 & 0.3-0.8 & 0.8-1.2 & 1.2-2.1 & 2.1-2.8 & 2.8-3.6 & 3.6-4.4 & to 7 \TeV  \\ 
       \hline
       Trigger efficiency          & $u_1$ & $u_1$ & $u_1$ & $u_1$ & $u_1$ & $u_1$ & $u_1$ & N \\  
       Jet reconstruction eff.     & 83 & 83 & 83 & 83 & 84 & 85 & 86 & Y\\
       Jet selection eff.          &  $u_2$  & $u_2$   & $u_2$   & $u_2$   & $u_2$   & $u_2$   & $u_2$   & N\\
       JES1: Noise thresholds      &  1 &  1 &  2 &  3 &  4 &  5 &  6 & Y\\
       JES2: Theory UE             &  7 &  7 &  8 &  9 & 10 & 11 & 12 & Y\\
       JES3: Theory showering      & 13 & 13 & 14 & 15 & 16 & 17 & 18 & Y\\
       JES4: Non-closure           & 19 & 19 & 20 & 21 & 22 & 23 & 24 & Y\\
       JES5: Dead material         & 25 & 25 & 26 & 27 & 28 & 29 & 30 & Y\\
       JES6: Forward JES generators& 88 & 88 & 88 & 88 & 88 & 88 & 88 & *\\
       JES7: $E/p$ response        & 32 & 32 & 33 & 34 & 35 & 36 & 37 & Y\\
       JES8: $E/p$ selection       & 38 & 38 & 39 & 40 & 41 & 42 & 43 & Y\\
       JES9: EM + neutrals         & 44 & 44 & 45 & 46 & 47 & 48 & 49 & Y\\
       JES10: HAD $E$-scale        & 50 & 50 & 51 & 52 & 53 & 54 & 55 & Y\\
       JES11: High $p_T$           & 56 & 56 & 57 & 58 & 59 & 60 & 61 & Y\\
       JES12: $E/p$ bias           & 62 & 62 & 63 & 64 & 65 & 66 & 67 & Y\\
       JES13: Test-beam bias       & 68 & 68 & 69 & 70 & 71 & 72 & 73 & Y\\
       JES15: Forward JES detector & 89 & 89 & 89 & 89 & 89 & 89 & 89 & *\\
       Jet energy resolution       & 76 & 76 & 77 & 78 & 79 & 80 & 81 & Y\\
       Jet angle resolution        & 82 & 82 & 82 & 82 & 82 & 82 & 82 & Y\\
       Unfolding: Closure test     & 74 & 74 & 74 & 74 & 74 & 74 & 74 & N\\
       Unfolding: Jet matching     & 75 & 75 & 75 & 75 & 75 & 75 & 75 & N\\
       Luminosity                  & 87 & 87 & 87 & 87 & 87 & 87 & 87 & N\\
    \hline \hline
    \end{tabular}
\end{table*}

\section[Inclusive jet cross-section at sqrt(s) = 2.76 TeV]{Inclusive jet cross-section at $\sqrt{s}$ = 2.76 TeV}
\label{sec:xsec}
The inclusive jet double-differential cross-section is shown in Figs.~\ref{fig:xsec2p76_04} and \ref{fig:xsec2p76_06} for jets reconstructed with the \antikt{} algorithm with $R=0.4$ and $R=0.6$, respectively. 
The measurement spans jet transverse momenta from $20$ \GeV{} to $430$ \GeV{} 
in the rapidity region of $|y|<4.4$, covering seven orders of magnitude in cross-section. 
The results are compared to NLO pQCD predictions calculated with \nlojetpp{}
using the CT10 PDF set. Corrections for non-perturbative effects are applied.

\begin{figure*}
  \centering
  \includegraphics[width=1.5\columnwidth]{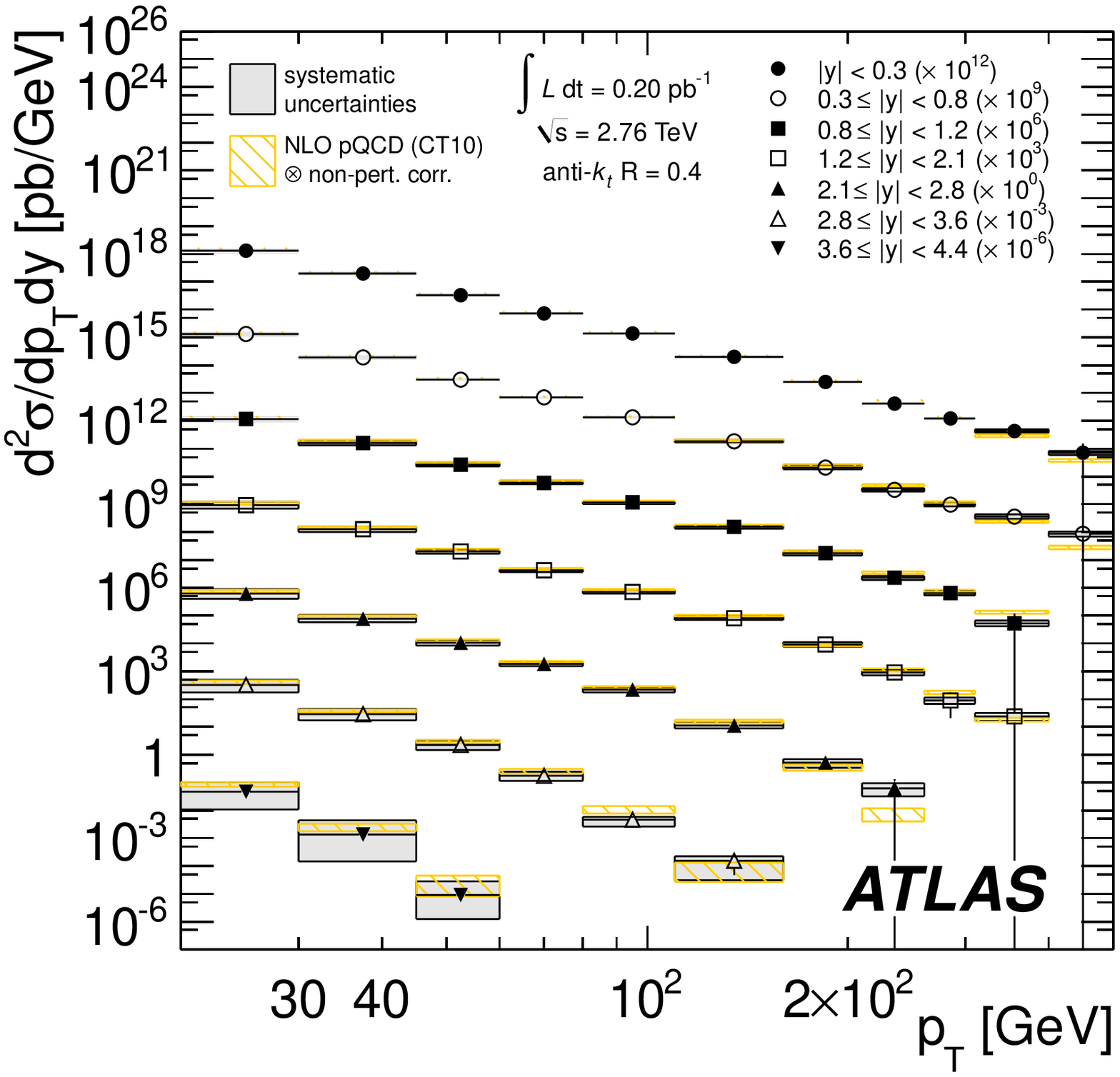}
  \caption{Inclusive jet double-differential cross-section as a function of the jet \pt{} in bins of rapidity, for \antikt{} jets with $R=0.4$.
For presentation, the cross-section is multiplied by the factors indicated in the legend. 
The shaded area indicates the experimental systematic uncertainties. 
The data are compared to NLO pQCD predictions calculated using \nlojetpp{} 
 with the CT10 PDF set, to which non-perturbative corrections have been applied. 
The hashed area indicates the predictions with their uncertainties. The 2.7\% uncertainty from the luminosity measurements is not shown.}
  \label{fig:xsec2p76_04}
\end{figure*}

\begin{figure*}
  \centering
  \includegraphics[width=1.5\columnwidth]{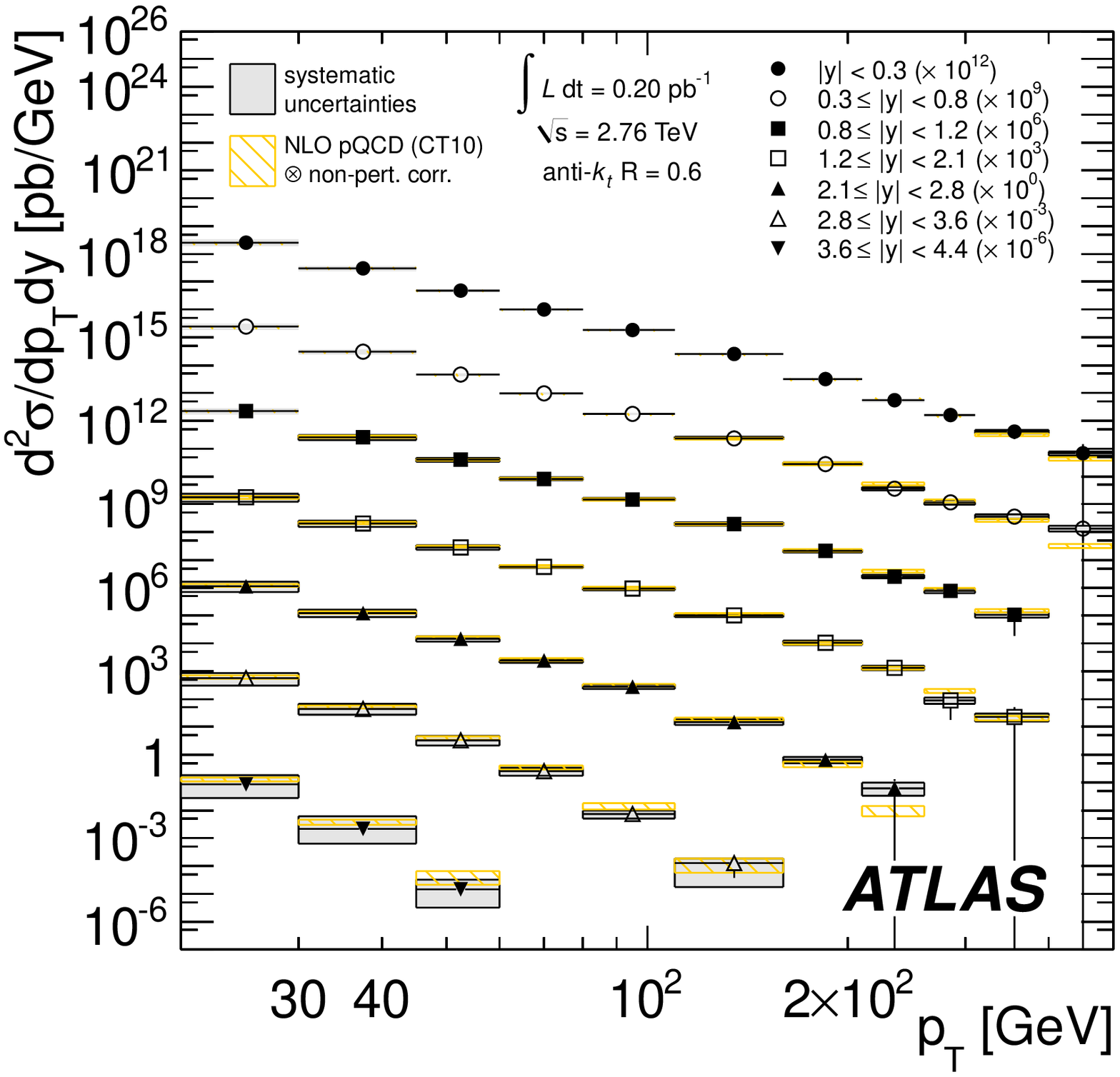}
  \caption{Inclusive jet double-differential cross-section as a function of the jet $p_T$ in bins of rapidity, for \antikt{} jets with $R=0.6$.
For presentation, the cross-section is multiplied by the factors indicated in the legend. 
The shaded area indicates the experimental systematic uncertainties. 
The data are compared to NLO pQCD predictions calculated using \nlojetpp{} with the CT10 PDF set, to which non-perturbative corrections have been applied.
The hashed area indicates the predictions with their uncertainties. The 2.7\% uncertainty from the luminosity measurements is not shown.}  
  \label{fig:xsec2p76_06}
\end{figure*}

The ratio of the measured cross-sections to the NLO pQCD predictions
using the CT10 PDF set is presented 
in Figs.~\ref{fig:xsecband2p76_04} and \ref{fig:xsecband2p76_06} for jets with $R=0.4$ and $R=0.6$, respectively. 
The results are also compared to the predictions obtained using the PDF sets MSTW 2008, NNPDF 2.1, HERAPDF 1.5 and ABM 11. 
The measurement is consistent with all the theory predictions using different PDF sets within their systematic uncertainties for jets with both radius parameters. However, the data for jets with $R=0.4$ have a systematically lower cross-section than any of the theory predictions, while such a tendency is seen only in the forward rapidity regions in the measurement for jets with $R=0.6$.  

\begin{figure*}
  \centering
  \includegraphics[width=0.9\textwidth]{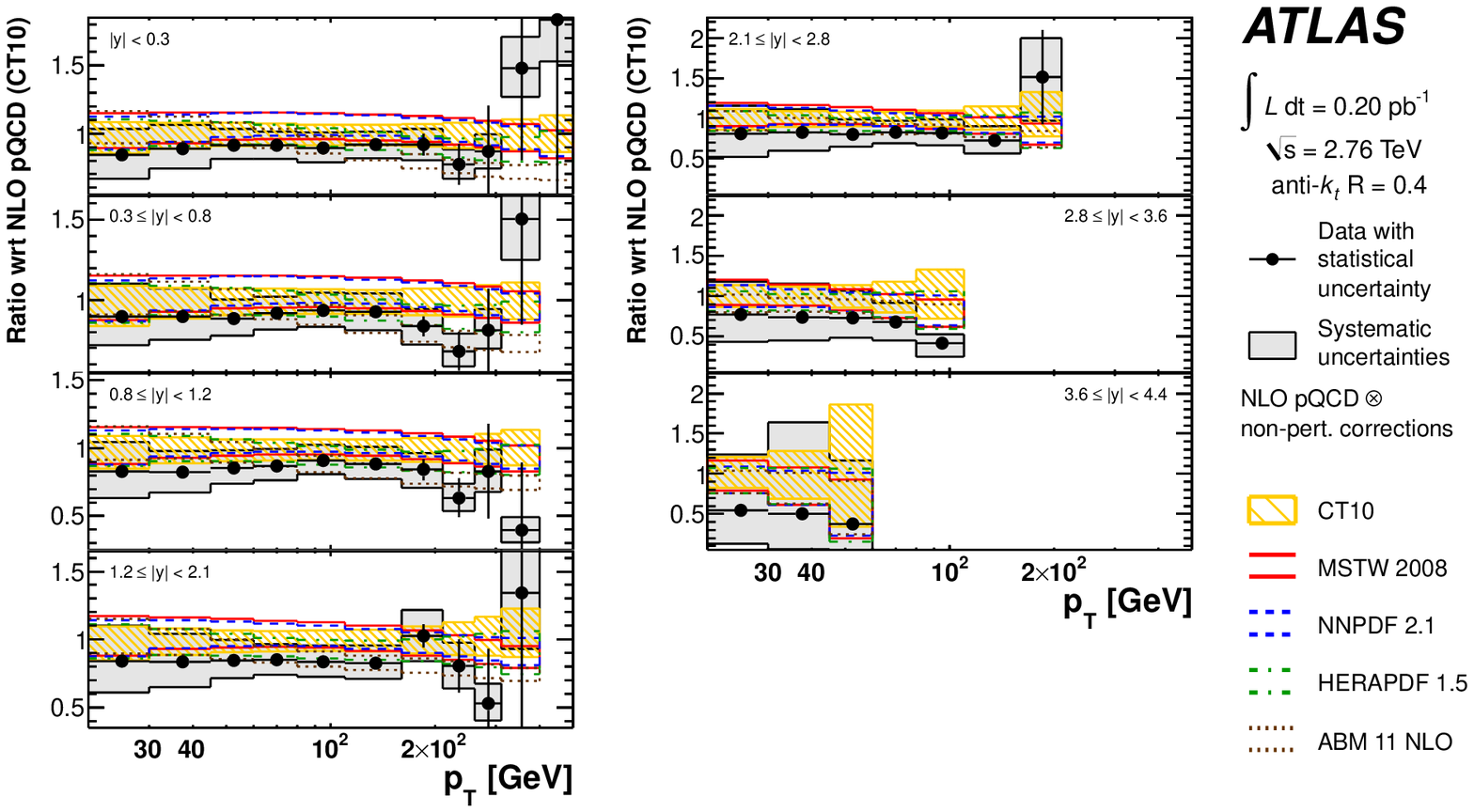}
  \caption{Ratio of the measured inclusive jet double-differential cross-section to the NLO pQCD prediction 
calculated with \nlojetpp{} with the CT10 PDF set corrected for non-perturbative effects. 
The ratio is shown as a function of the jet \pt{} in bins of jet rapidity, for \antikt{} jets with $R=0.4$.
The figure also shows NLO pQCD predictions obtained with different PDF sets, namely  ABM~11, NNPDF~2.1, HERAPDF~1.5 and MSTW2008. Statistically insignificant data points at large \pt{} are omitted. The 2.7\% uncertainty from the luminosity measurements is not shown.}
  \label{fig:xsecband2p76_04}
\end{figure*}

\begin{figure*}
  \centering
  \includegraphics[width=0.9\textwidth]{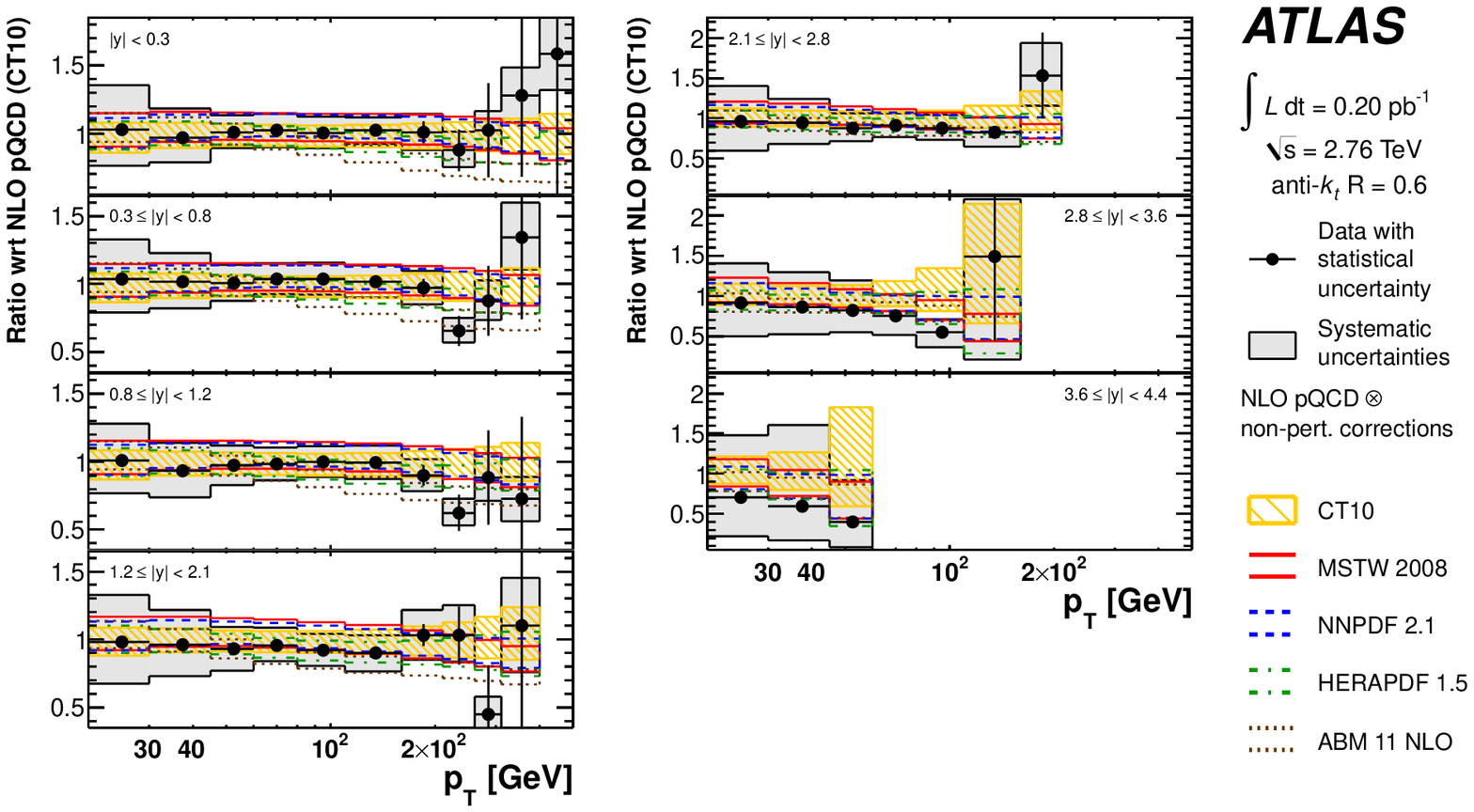}
  \caption{Ratio of the measured inclusive jet double-differential cross-section to the NLO pQCD prediction 
calculated with \nlojetpp{} with the CT10 PDF set corrected for non-perturbative effects. 
The ratio is shown as a function of the jet \pt{} in bins of jet rapidity, for \antikt{} jets with $R=0.6$.
The figure also shows NLO pQCD predictions obtained with different PDF sets, namely  ABM~11, NNPDF~2.1, HERAPDF~1.5 and MSTW2008. Statistically insignificant data points at large \pt{} are omitted. The 2.7\% uncertainty from the luminosity measurements is not shown.}
  \label{fig:xsecband2p76_06}
\end{figure*}

The comparison of the data with the \powheg{} prediction for \antikt{} jets with $R=0.4$ and $R=0.6$ is shown in Figs.~\ref{fig:xsecpowheg2p76_R04} and~\ref{fig:xsecpowheg2p76_R06} as a function of the jet \pt{} in bins of rapidity. 
In general, the \powheg{} prediction is found to be in good agreement with the data. Especially in the forward region, the shape of the data is very well reproduced by the \powheg{} prediction, while small differences are observed in the central region. As seen in the previous measurement at $\sqrt{s}=7\TeV$~\cite{Aad:2011fc}, the \Perugia{}~2011 tune gives a consistently larger prediction than the default \pythia{} tune AUET2B, which is generally in closer agreement with data. 
In contrast to the NLO pQCD prediction with corrections for non-per\-tur\-bative effects, 
the \powheg{} prediction agrees well with data for both
radius parameters $R=0.4$ and $R=0.6$. This might be attributed to the matched parton shower approach from \powheg{}~(see Sect.~\ref{sec:powheg}).

\begin{figure*}
  \centering
  \includegraphics[width=0.9\textwidth]{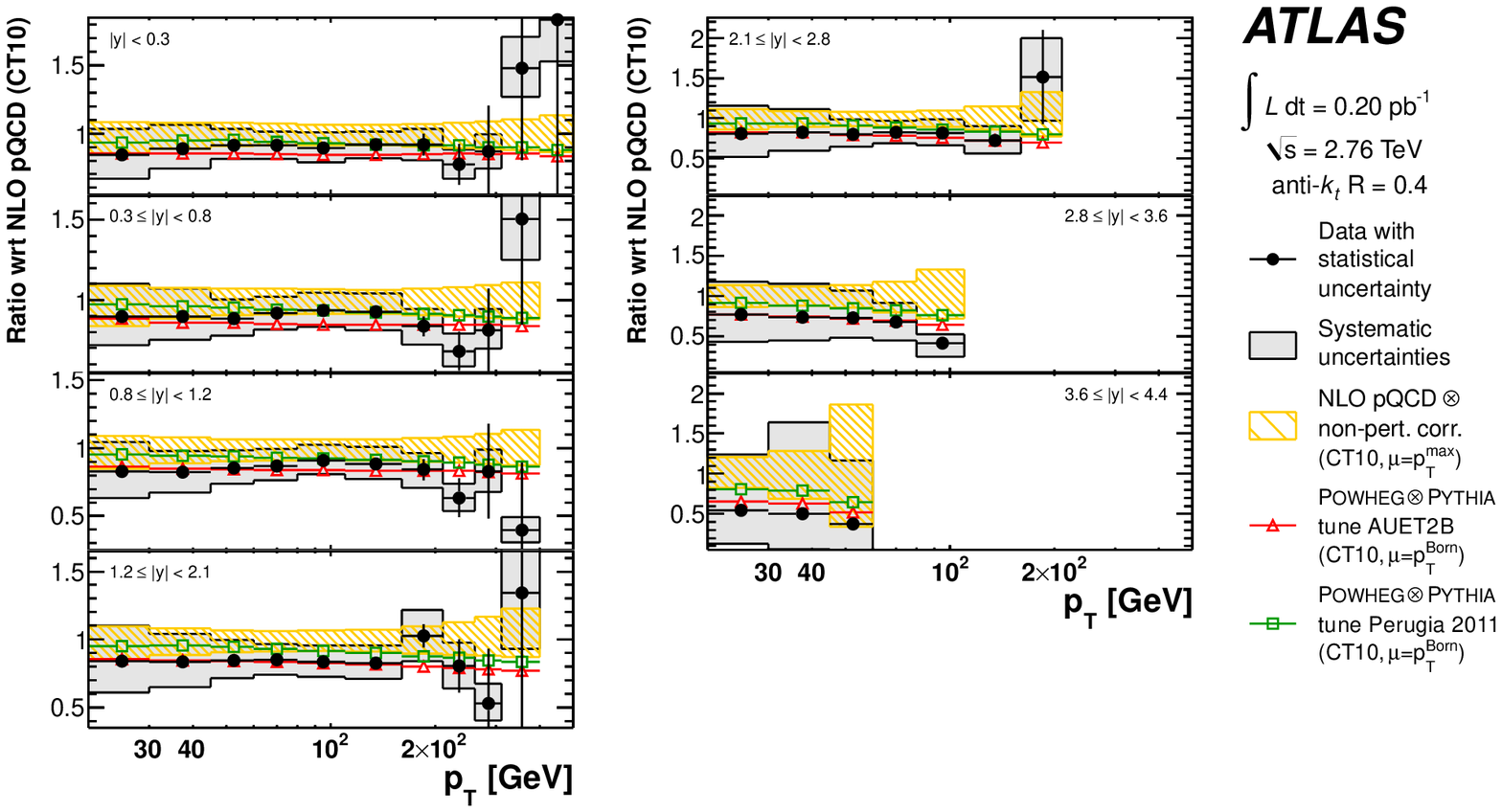}
  \caption{Ratio of the measured inclusive jet double-differential cross-section to the NLO pQCD prediction 
calculated with \nlojetpp{} with the CT10 PDF set corrected for non-perturbative effects. 
The ratio is shown as a function of the jet \pt{} in bins of jet rapidity, for \antikt{} jets with $R=0.4$.
The figure also shows predictions from \powheg{} using \pythia{} for the simulation of the parton shower and hadronisation with the AUET2B tune and the \Perugia{}~2011 tune. Only the statistical uncertainty is shown on the \powheg{} predictions. Statistically insignificant data points at large \pt{} are omitted. The 2.7\% uncertainty from the luminosity measurements is not shown.}
  \label{fig:xsecpowheg2p76_R04}
\end{figure*}

\begin{figure*}
  \centering
  \includegraphics[width=0.9\textwidth]{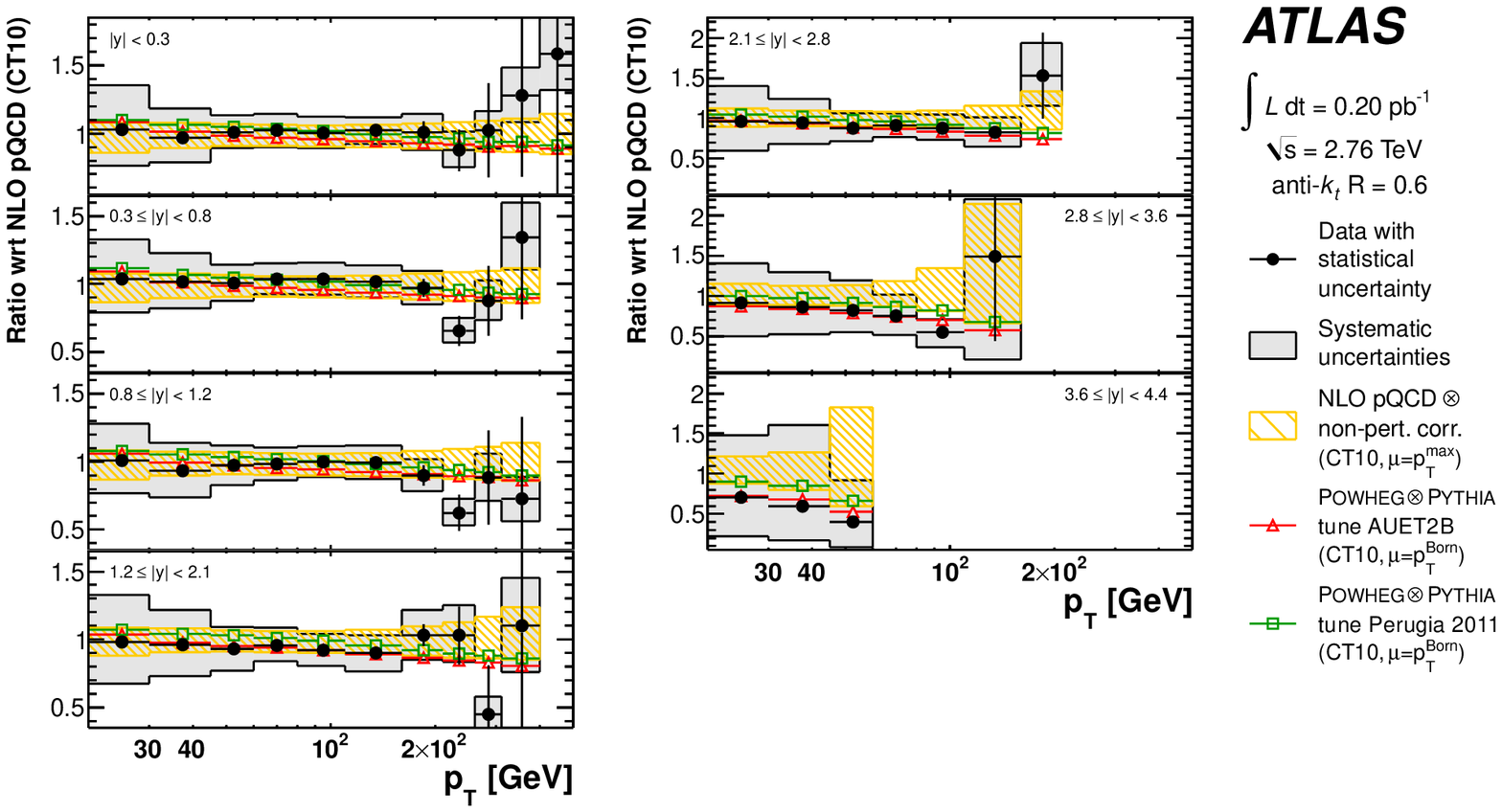}
  \caption{Ratio of the measured inclusive jet double-differential cross-section to the NLO pQCD prediction 
calculated with \nlojetpp{} with the CT10 PDF set corrected for non-perturbative effects. 
The ratio is shown as a function of the jet \pt{} in bins of jet rapidity, for \antikt{} jets with $R=0.6$.
The figure also shows predictions from \powheg{} using \pythia{} for the simulation of the parton shower and hadronisation with the AUET2B tune and the \Perugia{}~2011 tune. Only the statistical uncertainty is shown on the \powheg{} predictions. Statistically insignificant data points at large \pt{} are omitted. The 2.7\% uncertainty from the luminosity measurements is not shown.}
  \label{fig:xsecpowheg2p76_R06}
\end{figure*}

\section[cross-section ratio of sqrt(s)=2.76 TeV to sqrt(s)=7 TeV]{Cross-section ratio of $\sqrt{s}=2.76$ TeV to $\sqrt{s}=7$ TeV}
\label{sec:ratio}

\subsection{Experimental systematic uncertainty}
As indicated in Table~\ref{tab:uncert}, the systematic uncertainties on the measurement due to jet 
reconstruction and calibration are considered as fully  correlated 
between the measurements at $\sqrt{s}=2.76$ \TeV{} and $\sqrt{s}=7$ \TeV{}.
For each correlated systematic source $s_i$, the relative uncertainty $\Delta \rho_{s_i}/\rho$ on the cross-section ratio 
is calculated as 
\begin{equation}
\frac{\Delta \rho_{s_i}}{\rho}=\frac{1+\delta^{2.76 \TeV}_{s_i}}{1+\delta^{7 \TeV}_{s_i}}-1,
\end{equation}
where $\delta^{2.76 \TeV}_{s_i}$ and $\delta^{7 \TeV}_{s_i}$ are relative uncertainties caused by a source $s_i$ 
in the cross-section measurements at  $\sqrt{s}=2.76$ \TeV{} and $\sqrt{s}=7$ \TeV{}, respectively. 
Systematic uncertainties that are uncorrelated between the two centre-of-mass energies are added in quadrature. 
The uncertainties on the trigger efficiency and the jet selection efficiency, and the ones from the unfolding procedure 
are conservatively considered as uncorrelated between the two measurements at the different energies. 
The measurement at $\sqrt{s}=7$ \TeV{} has an additional uncertainty due to pile-up effects in the jet energy calibration. It is added to the uncertainty in the cross-section ratio.
The uncertainties in the luminosity measurements are also treated as uncorrelated (see Sect.~\ref{sec:uncert}), resulting in a luminosity uncertainty of $4.3\%$.
The uncertainty on the momentum of the proton beam, based on the LHC magnetic model, is at the level of $0.1\%$~\cite{Wenninger:EnergyCalibration} and highly correlated between different centre-of-mass energies; hence, it is negligible for the ratio.

The experimental systematic uncertainties on both \linebreak$\rho(y,\xt)$ and $\rho(y,\pt)$ 
are shown in Fig.~\ref{fig:R_syst_06} for representative rapidity bins for jets with $R=0.6$. 
For $\rho(y,\xt)$ the uncertainties are 5\%--20\% for the central jets and $^{+160\%}_{-60\%}$ for the forward jets. 
For jets with $R=0.4$, uncertainties are similar,
except for central jets with low \pt{} where the uncertainty is within $\pm15\%$.
A significant reduction of the uncertainty is obtained for $\rho(y,\pt)$, being well below $5\%$
in the central region. In the forward region, the uncertainty is $\pm70\%$ for jets 
with $R=0.6$, and $^{+100\%}_{-70\%}$ for jets with $R=0.4$. 

\begin{figure*}[htb]
\begin{center}
\subfigure[$\rho(y,\xt), |y|<0.3$]{\includegraphics[width=5.2cm]{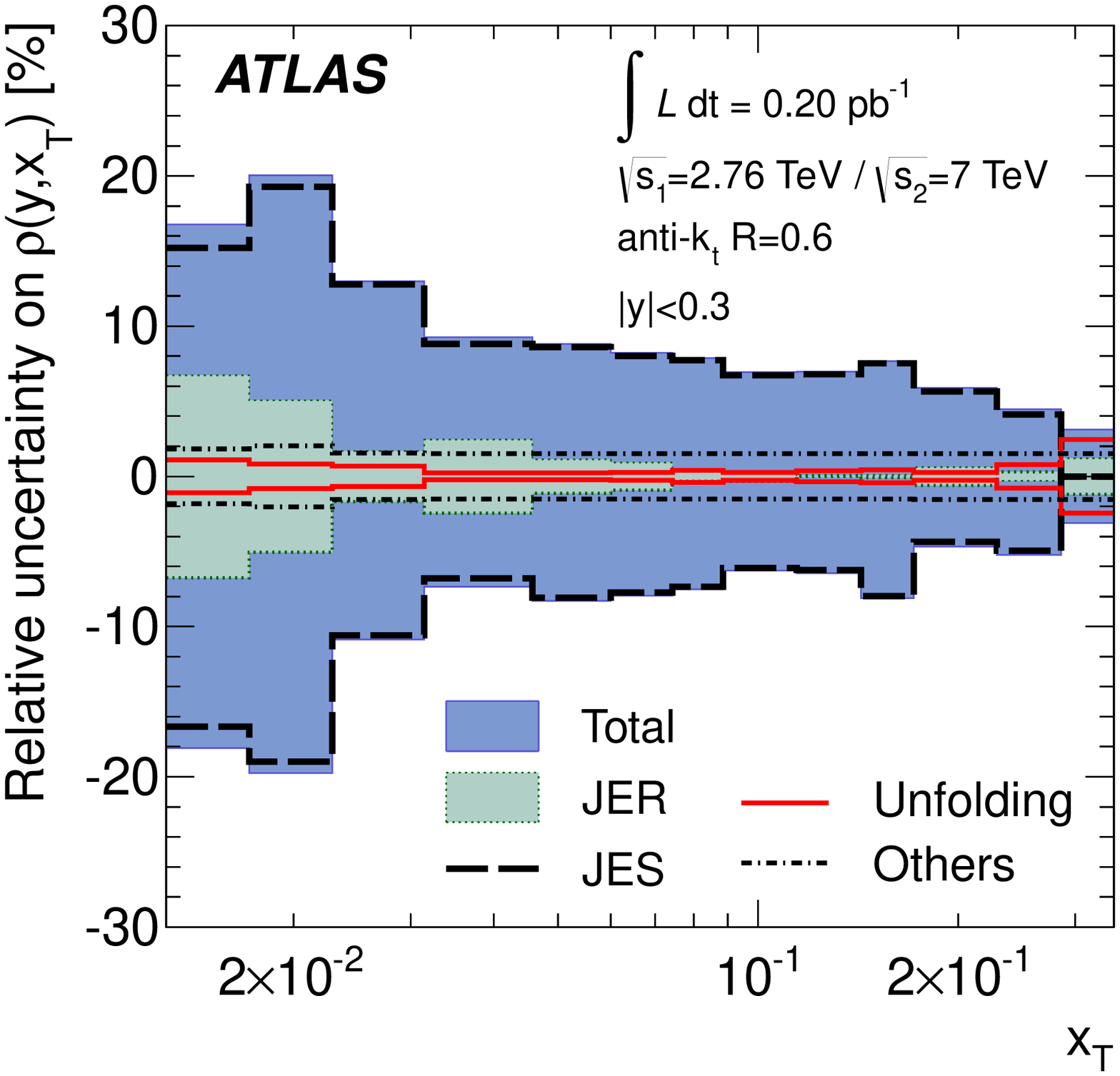}}
\subfigure[$\rho(y,\xt), 2.1\leq|y|<2.8$]{\includegraphics[width=5.2cm]{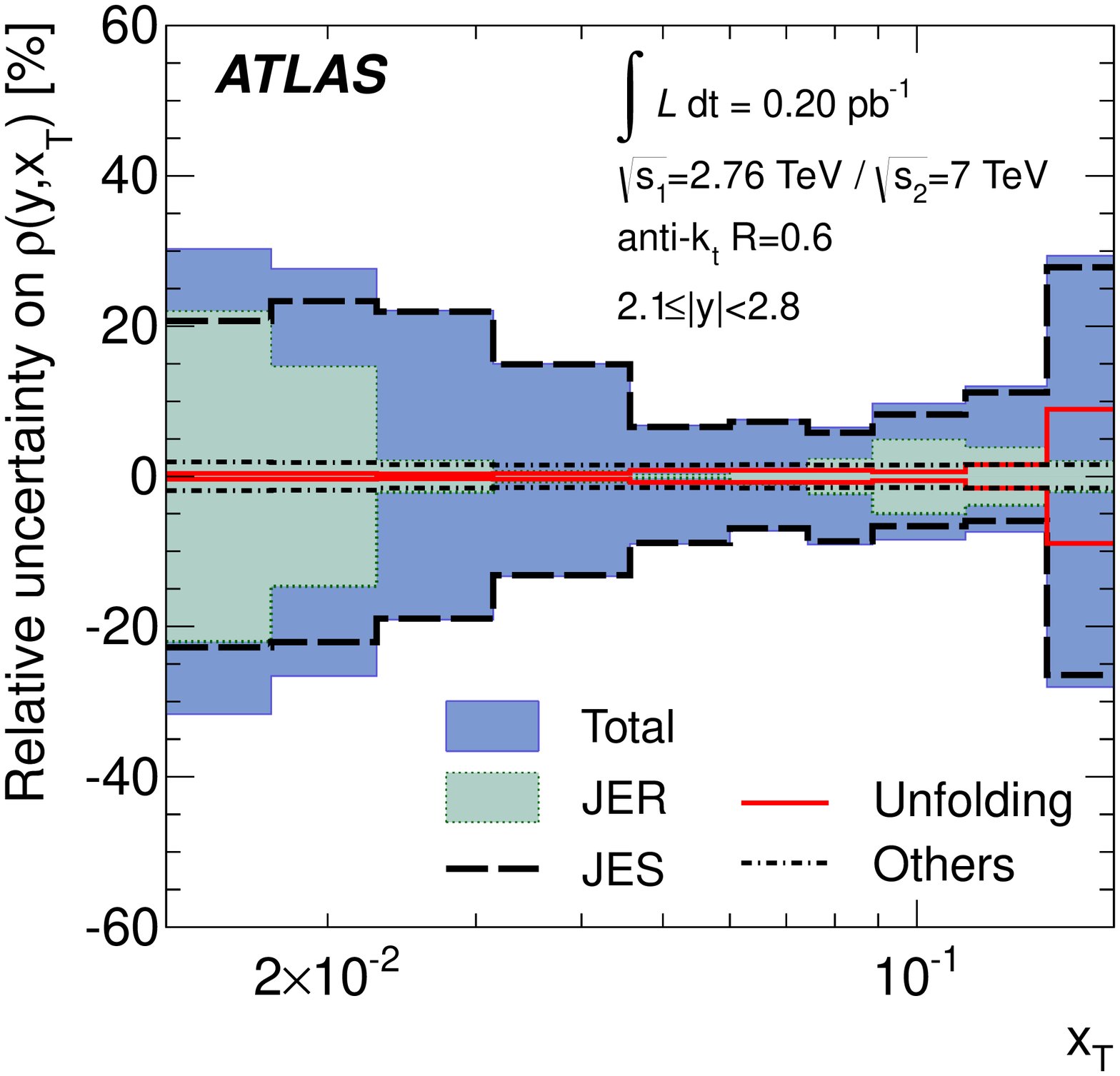}}
\subfigure[$\rho(y,\xt), 3.6\leq|y|<4.4$]{\includegraphics[width=5.2cm]{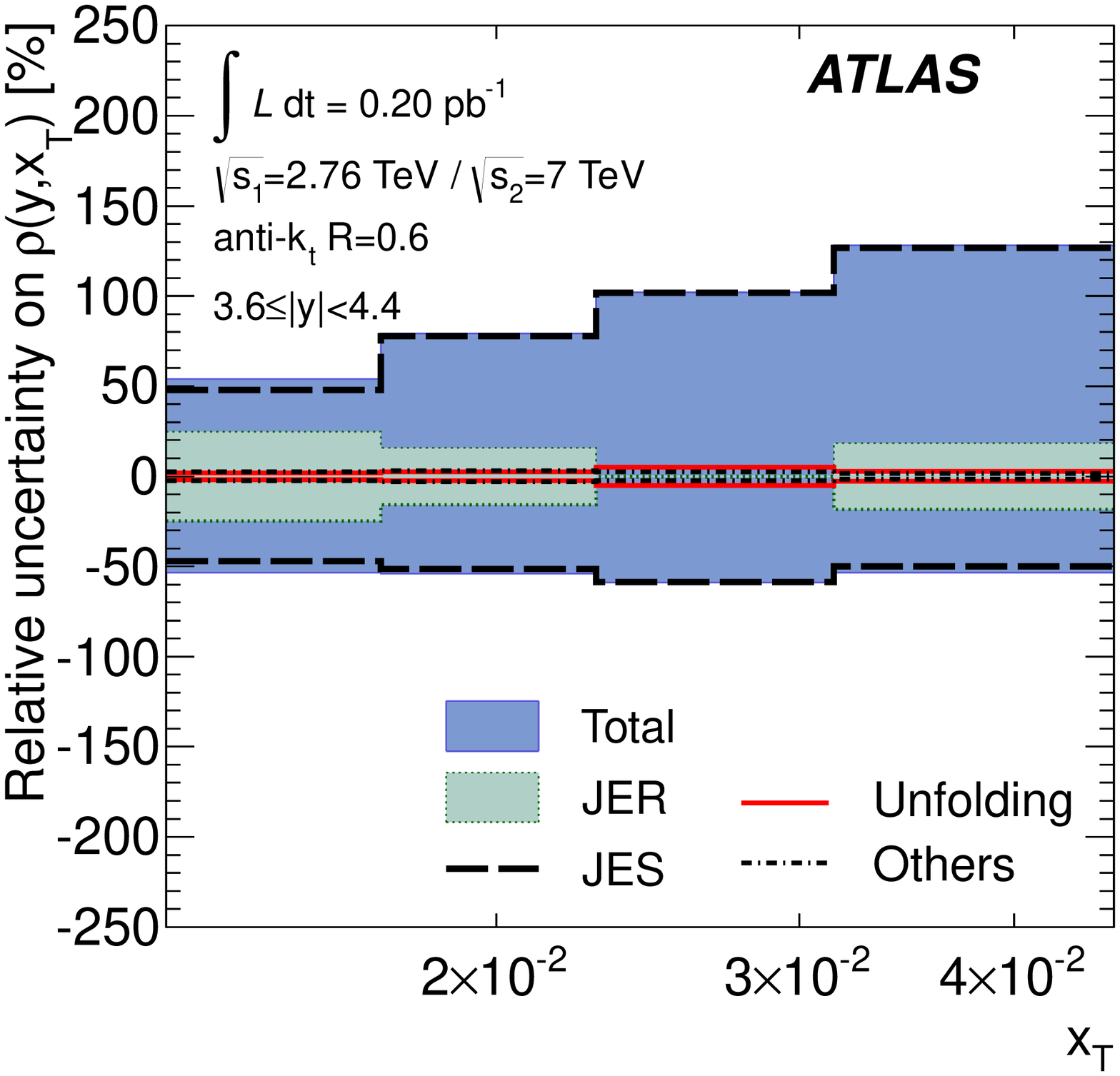}}\\
\subfigure[$\rho(y,\pt), |y|<0.3$]{\includegraphics[width=5.2cm]{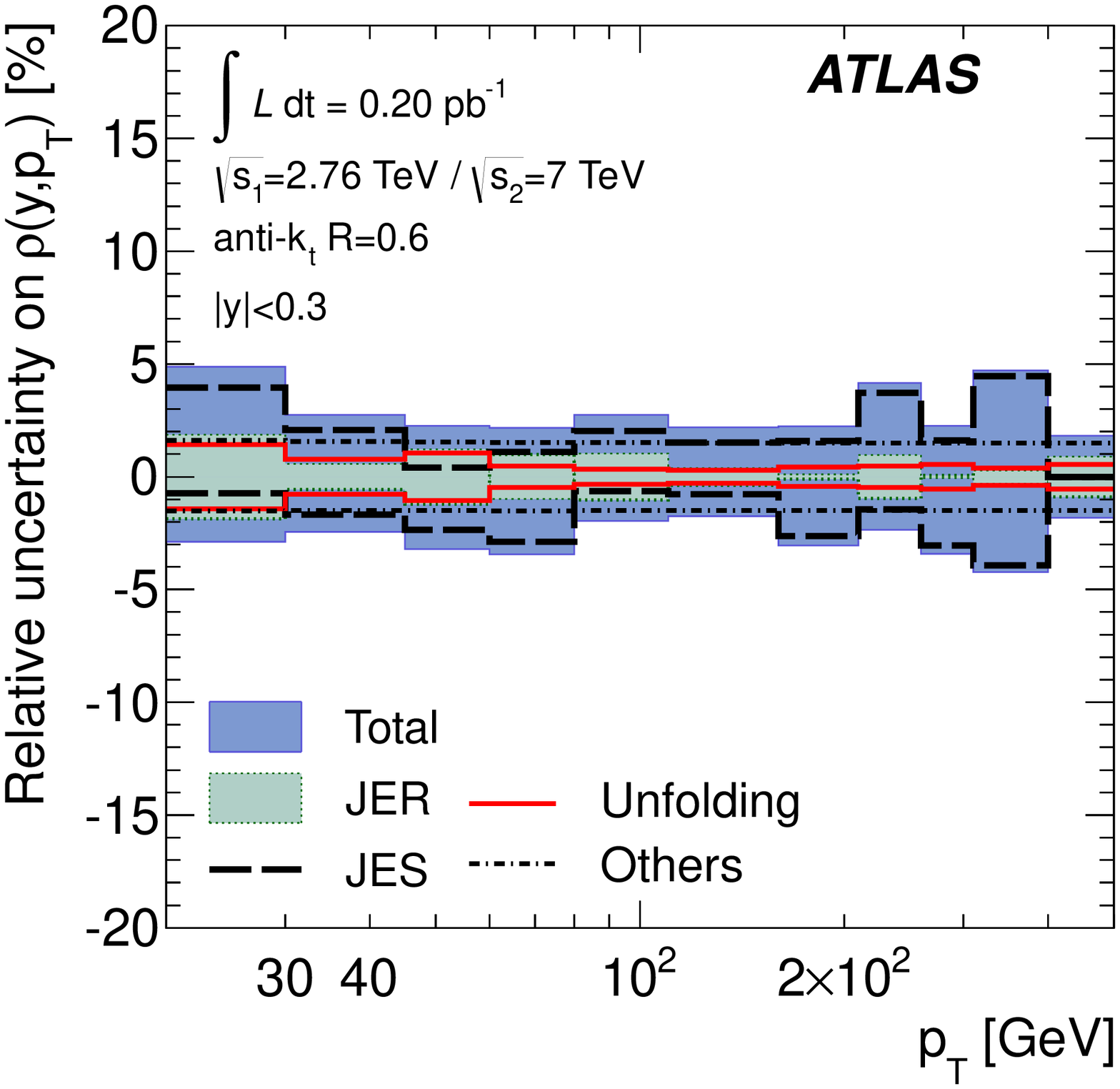}}
\subfigure[$\rho(y,\pt), 2.1\leq|y|<2.8$]{\includegraphics[width=5.2cm]{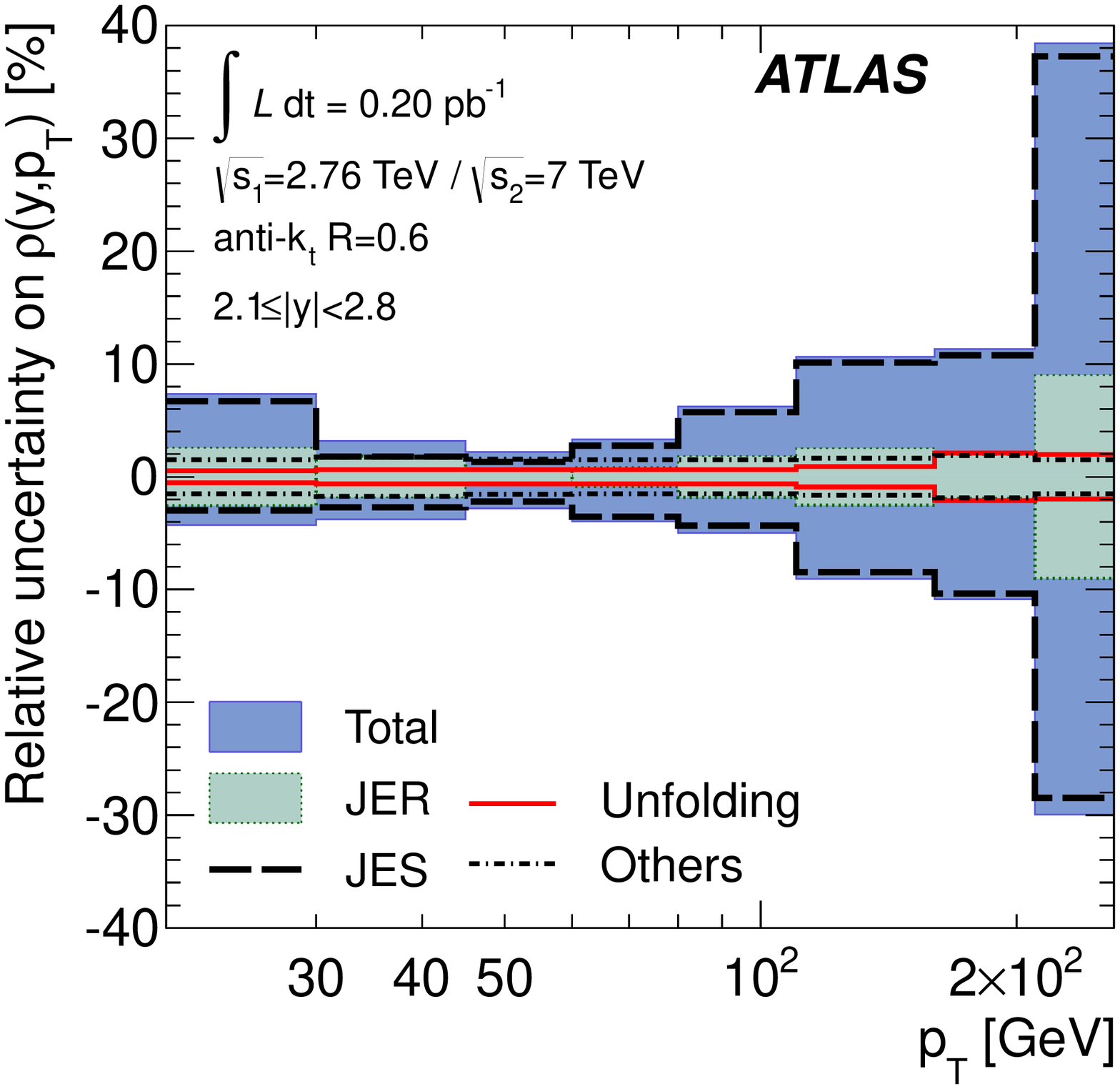}}
\subfigure[$\rho(y,\pt), 3.6\leq|y|<4.4$]{\includegraphics[width=5.2cm]{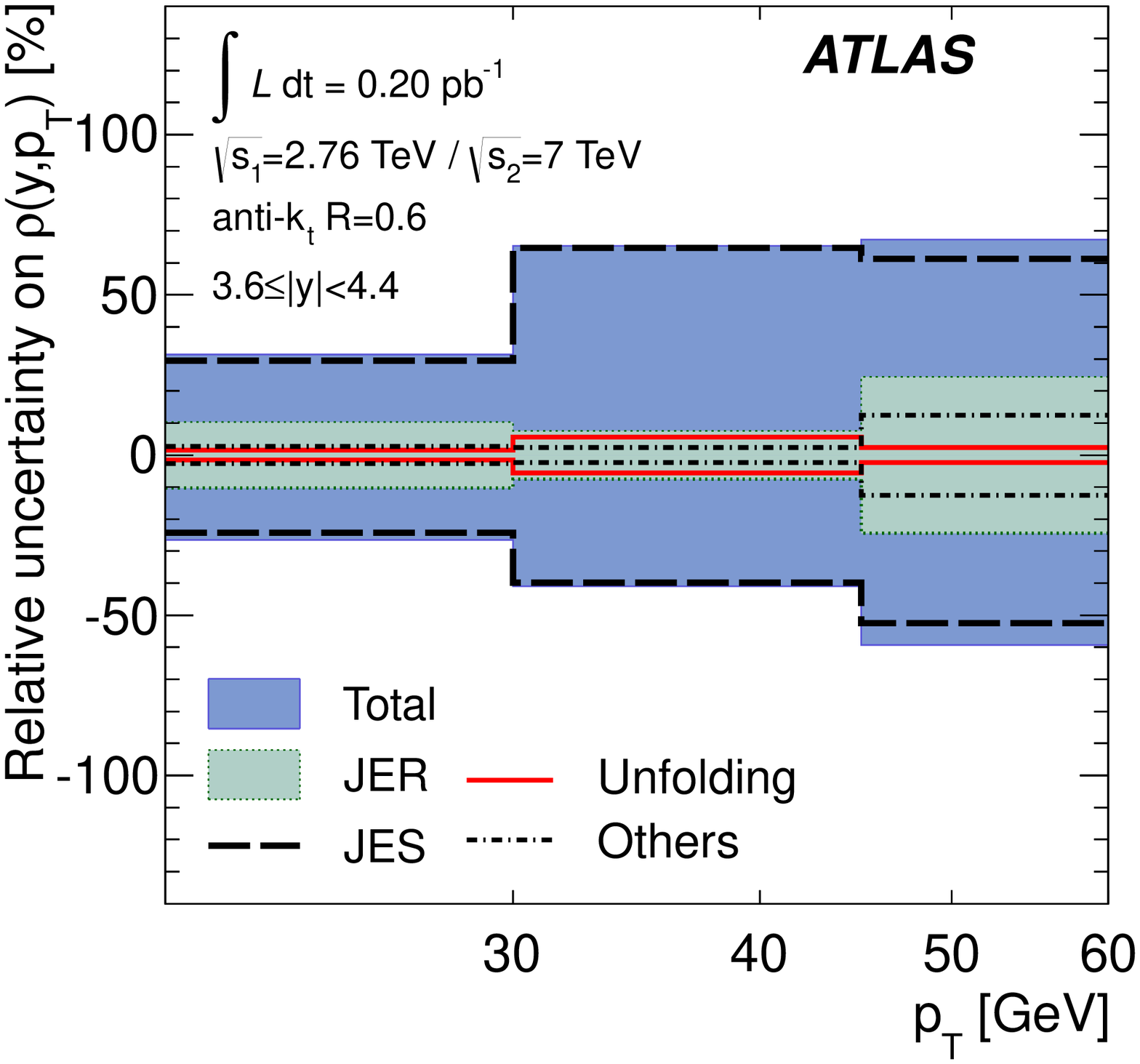}}
\caption{The systematic uncertainty on the cross-section ratios, $\rho(y,\xt)$ and $\rho(y,\pt)$, for \antikt{} 
jets with $R=0.6$ in three representative rapidity bins, as a function of the jet \xt{} and of the jet \pt{}, respectively. 
In addition to the total uncertainty, the uncertainties from the jet energy scale (JES), 
the jet energy resolution (JER), the unfolding procedure and other systematic sources are shown separately.
The 4.3\% uncertainty from the luminosity measurements and the statistical uncertainty are not shown. }

\label{fig:R_syst_06}
\end{center}
\end{figure*}

\subsection{Results}
\label{subsec:ratio}
Figures~\ref{fig:xsratio_xt_r04} and \ref{fig:xsratio_xt_r06} show the extracted 
cross-section ratio of the inclusive jet cross-section measured at $\sqrt{s}=2.76$ \TeV{} 
to the one measured at $\sqrt{s}=7$ \TeV, as a function of \xt{}, for jets with $R=0.4$ and $R=0.6$, respectively.
The measured cross-section ratio is found to be $1.1<\rho(y,\xt)<1.5$ for both radius parameters.
This approximately constant behaviour reflects 
both the asymptotic freedom of QCD and evolution of the gluon distribution in the proton as a function of the QCD scale.  
The measurement shows a slightly different \xt{} dependence for jets with $R=0.4$ and $R=0.6$, which may be 
attributed to different \xt{} dependencies of non-perturbative corrections for the two radius parameters, 
already seen in Figs.~\ref{fig:npcrxtrpt} (a) and.~\ref{fig:npcrxtrpt}~(b).
The measurement is then compared to the NLO pQCD prediction, to which corrections for non-per\-tur\-ba\-tive effects are applied, obtained using the CT10 PDF set. It is in good agreement with the prediction. 
\begin{figure*}
  \centering
  \includegraphics[width=0.9\textwidth]{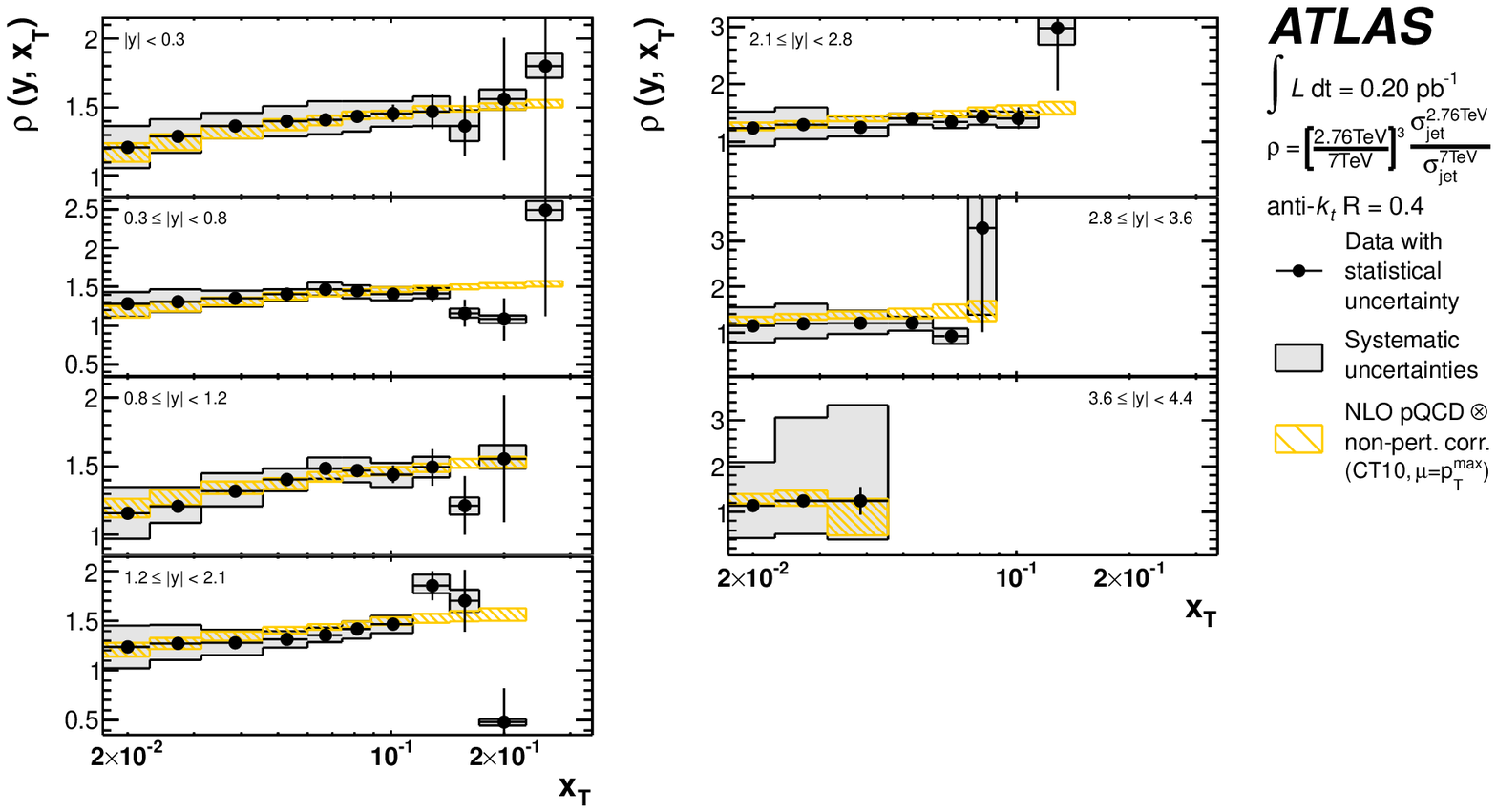}  
  \caption{Ratio of the inclusive jet cross-section at $\sqrt{s}=2.76$ \TeV{} to the one at $\sqrt{s}=7$ \TeV{} 
as a function of \xt{} in bins of jet rapidity, for \antikt{} jets with $R=0.4$. 
The theoretical prediction is calculated at next-to-leading order with the CT10 PDF set 
and corrected for non-perturbative effects.  Statistically insignificant data points at large \xt{} are omitted. The 4.3\% uncertainty from the luminosity measurements is not shown.}
  \label{fig:xsratio_xt_r04}
\end{figure*}
\begin{figure*}
  \centering
  \includegraphics[width=0.9\textwidth]{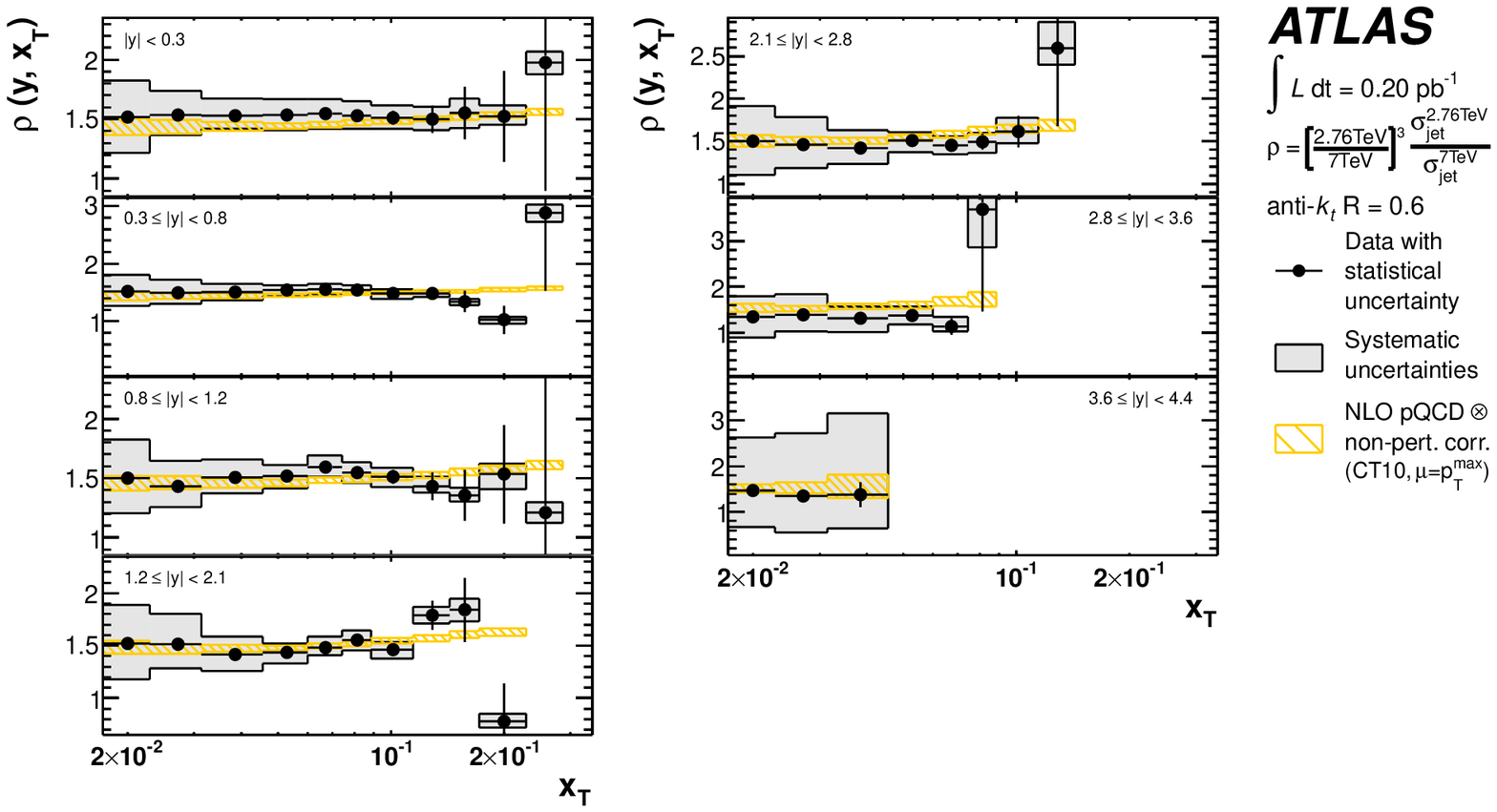}  
  \caption{Ratio of the inclusive jet cross-section at $\sqrt{s}=2.76$ \TeV{} to the one at $\sqrt{s}=7$ \TeV{} 
as a function of \xt{} in bins of jet rapidity, for \antikt{} jets with $R=0.6$. 
The theoretical prediction is calculated at next-to-leading order with the CT10 PDF set and corrected for non-perturbative effects. Statistically insignificant data points at large \xt{} are omitted. The 4.3\% uncertainty from the luminosity measurements is not shown.}
  \label{fig:xsratio_xt_r06}
\end{figure*}

Figures~\ref{fig:ratioxt_pwg_04} and~\ref{fig:ratioxt_pwg_06} show the same cross-section ratio compared to predictions from \powheg{} with the CT10 PDF set.
The tunes AUET2B and \Perugia{}~2011 give very similar predictions in general, and also agree well with the pQCD prediction with non-perturbative corrections applied.

\begin{figure*}
  \centering
  \includegraphics[width=0.9\textwidth]{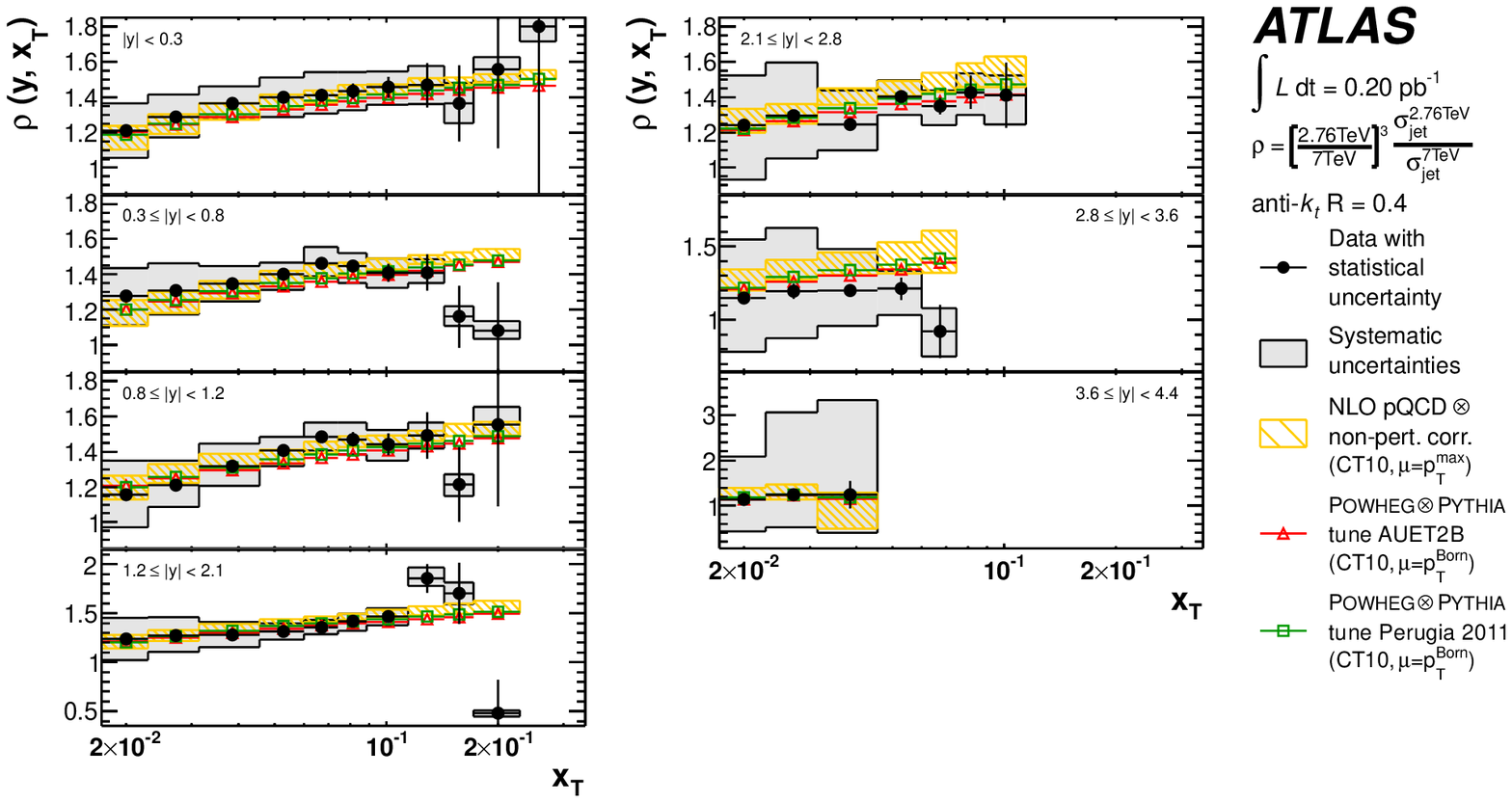}
  \caption{Ratio of the measured inclusive jet double-differential cross-section at $\sqrt{s}=2.76$ TeV to the one at $\sqrt{s}=7$ TeV as a function of the jet \xt{} in bins of jet rapidity, for \antikt{} jet with $R=0.4$.
The theoretical prediction from \nlojetpp{} is calculated using the CT10 PDF set with corrections for non-perturbative effects applied.
Also shown are \powheg{} predictions using \pythia{} for the simulation of the parton shower and hadronisation with the AUET2B tune and the \Perugia{}~2011 tune. Only the statistical uncertainty is shown on the \powheg{} predictions. Statistically insignificant data points at large \xt{} are omitted. The 4.3\% uncertainty from the luminosity measurements is not shown.}
  \label{fig:ratioxt_pwg_04}
\end{figure*}

\begin{figure*}
  \centering
  \includegraphics[width=0.9\textwidth]{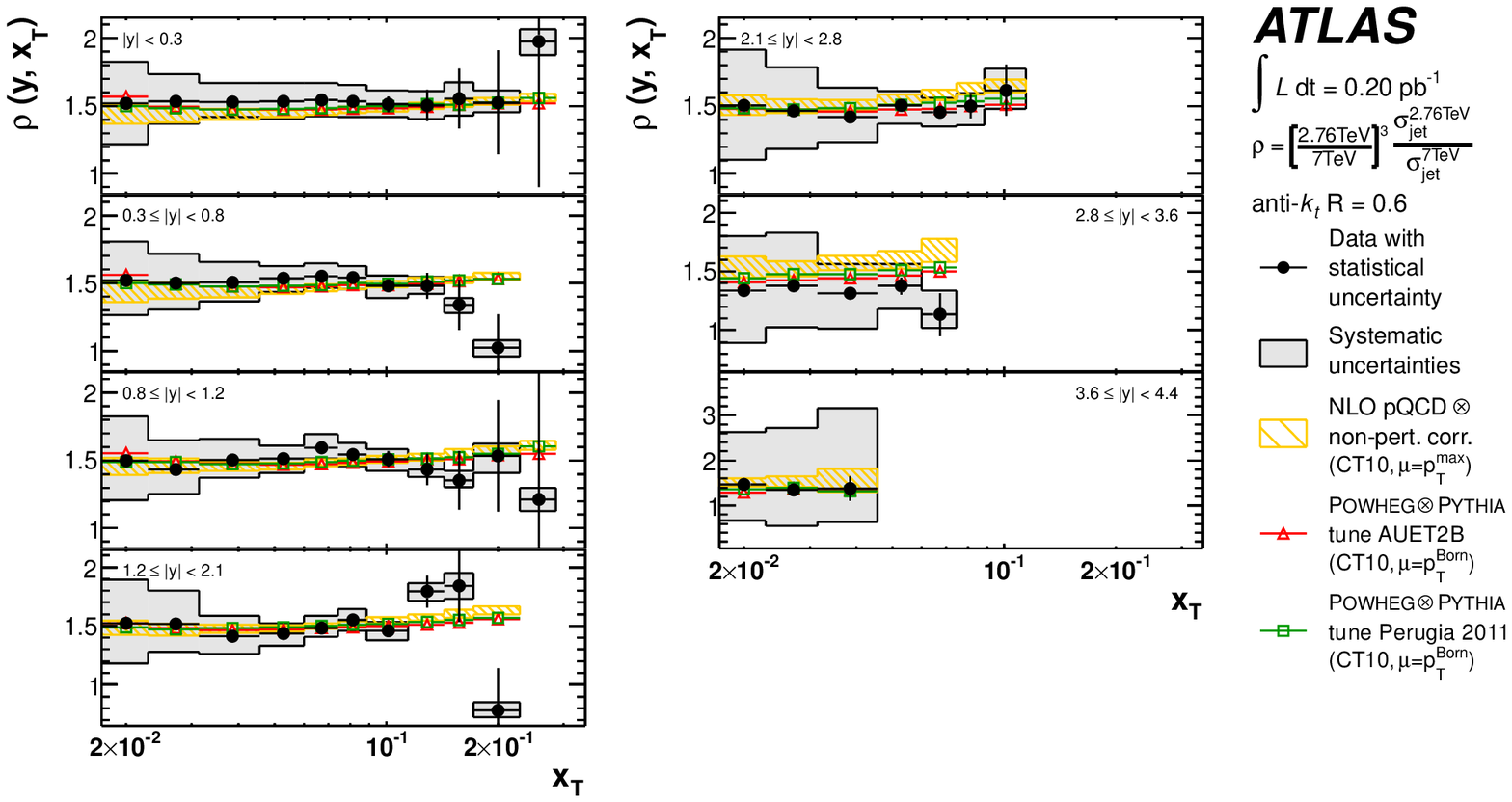}
  \caption{Ratio of the measured inclusive jet double-differential cross-section at $\sqrt{s}=2.76$ TeV to the one at $\sqrt{s}=7$ TeV as a function of the jet \xt{} in bins of jet rapidity, for \antikt{} jet with $R=0.6$.
The theoretical prediction from \nlojetpp{} is calculated using the CT10 PDF set with corrections for non-perturbative effects applied.
Also shown are \powheg{} predictions using \pythia{} for the simulation of the parton shower and hadronisation with the AUET2B tune and the \Perugia{}~2011 tune. Only the statistical uncertainty is shown on the \powheg{} predictions. Statistically insignificant data points at large \xt{} are omitted. The 4.3\% uncertainty from the luminosity measurements is not shown.}
  \label{fig:ratioxt_pwg_06}
\end{figure*}

Figures~\ref{fig:xsrbands_pt_r04} and \ref{fig:xsrbands_pt_r06} show  
the cross-section ratio as a function of the jet \pt{}, plotted as the double ratio with respect to 
the NLO pQCD prediction using the CT10 PDF set with non-perturbative corrections applied, for \antikt{} jets with $R=0.4$ and $R=0.6$.\footnote{As written in Sect.~\ref{sec:unfold}, the measurement at $\sqrt{s}=2.76$ \TeV{} uses a quality selection for jets with low \pt{} in Monte Carlo simulation at $|\eta|\sim1$, which is a different treatment than was done for the published measurement at $\sqrt{s}=7$ \TeV{}~\cite{Aad:2011fc}. The ratio is extracted using the coherent treatment in the two measurements at the different beam energies, shifting the measured cross-section at $\sqrt{s}=7$ \TeV{} from the published result within its uncertainty. 
The shifts are sizable in the bin $0.8\leq|y|<1.2$ only. For jets with $R=0.4$ ($R=0.6$), they are $13 \%$ ($10 \%$) in the  $20\leq\pt<30\GeV{}$ bin, and $1.5 \%$ ($2.6 \%$) in the $30\leq\pt<45$ \GeV{} bin.
In the rapidity range $1.2\leq|y|<2.1$, the shift is $1.8 \%$ ($1.9 \%$) at $20\leq\pt<30$ \GeV{}. 
These bins in the $\sqrt{s}=7$ \TeV{} measurement only enter in the extraction of $\rho(y,\pt)$ and not in that of $\rho(y,\xt)$.} 

The systematic uncertainty on the measurement is significantly reduced and is generally smaller than the theory uncertainties. 
The measurement is also compared to the predictions using different PDF sets, namely MSTW2008, \linebreak NNPDF~2.1, HERAPDF~1.5 and ABM~11.
In general, the measured points are slightly higher than the predictions in the central rapidity regions and are lower in the forward rapidity regions. 
The deviation is more pronounced for the prediction using the ABM~11 PDF set in the barrel region, which yields a different shape with respect to the other PDF sets.

The very small systematic uncertainty in the $\rho(y,\pt)$ measurement suggests that the measured jet cross-section at $\sqrt{s}=2.76$ TeV may contribute to constrain the PDF uncertainties in a global PDF fit in the pQCD framework by correctly taking the correlation of systematic uncertainties to the previous $\sqrt{s}=7$ TeV measurement into account. 
Such an NLO pQCD analysis is described in Sect.~\ref{sec:pdffit}.
\begin{figure*}
  \centering
  \includegraphics[width=0.9\textwidth]{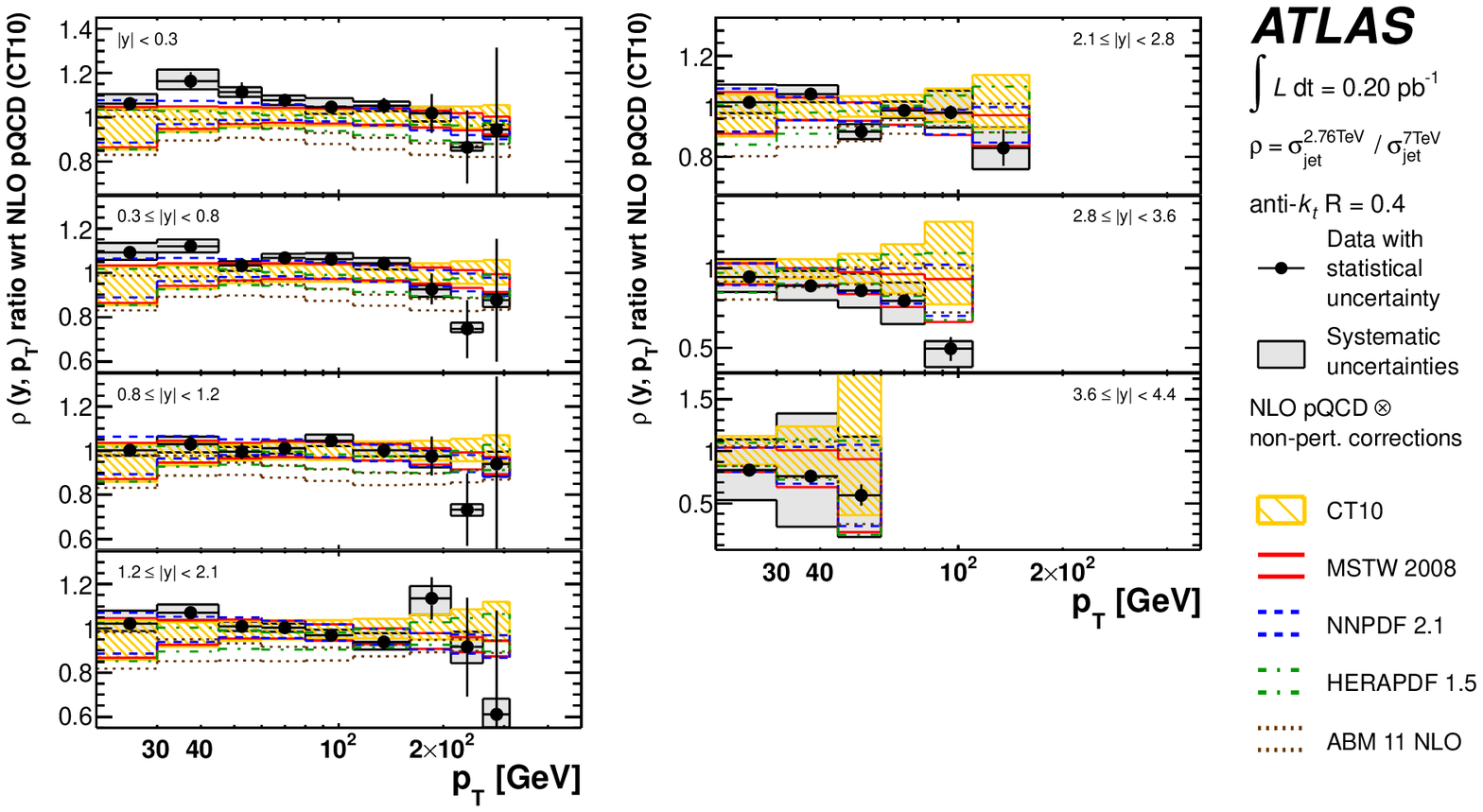}  
  \caption{Ratio of the inclusive jet cross-section at $\sqrt{s}=2.76$ \TeV{} to the one at $\sqrt{s}=7$ \TeV, 
shown as a double ratio to the theoretical prediction calculated with the CT10 PDFs as a function of the jet \pt{} 
in bins of jet rapidity, for \antikt{} jets with $R=0.4$. Statistically insignificant data points at large \pt{} are omitted. 
The 4.3\% uncertainty from the luminosity measurements is not shown.}
  \label{fig:xsrbands_pt_r04}
\end{figure*}

\begin{figure*}
  \centering
  \includegraphics[width=0.9\textwidth]{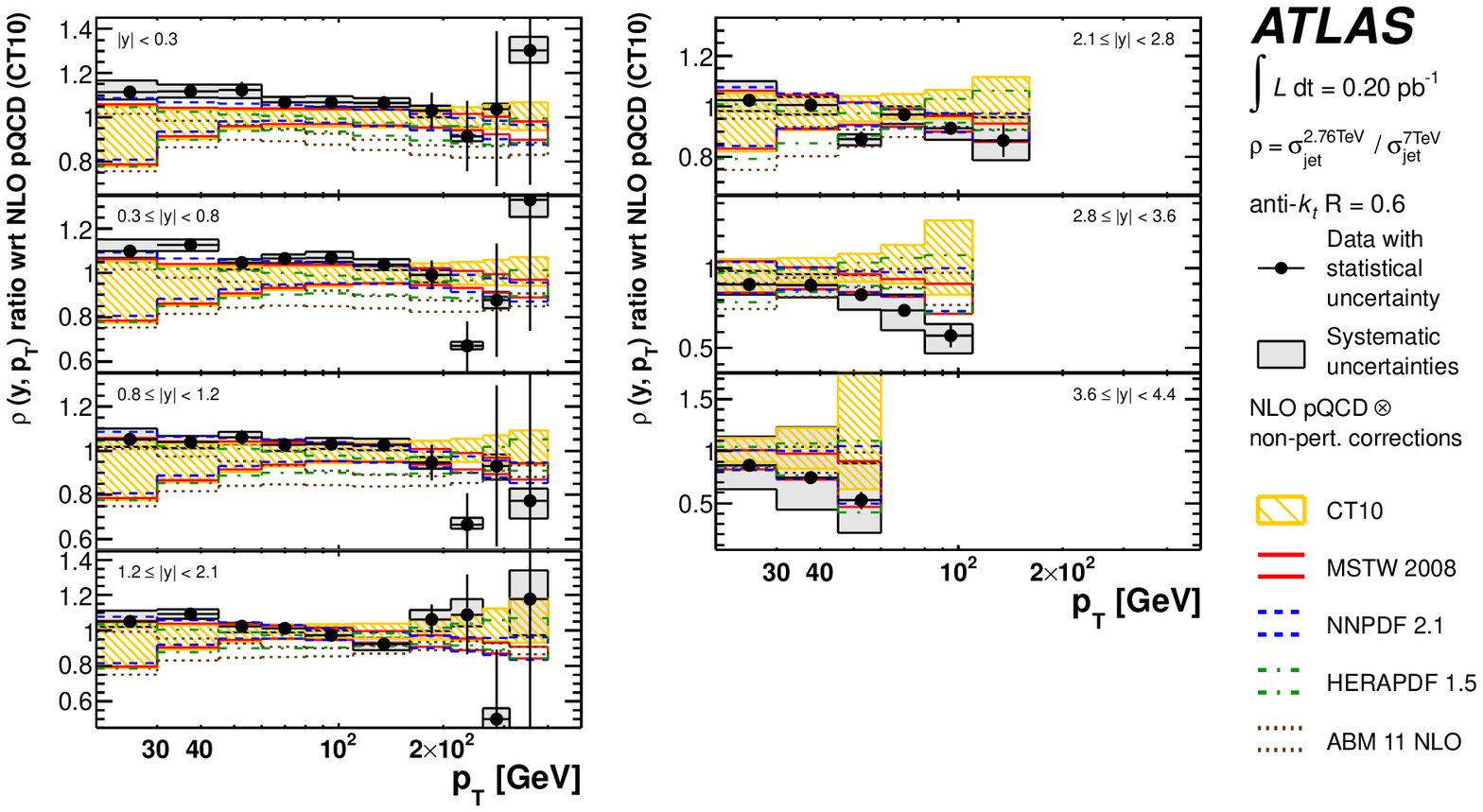}  
  \caption{Ratio of the inclusive jet cross-section at $\sqrt{s}=2.76$ \TeV{} to the one at $\sqrt{s}=7$ \TeV{}, 
shown as a double ratio to the theoretical prediction calculated with the CT10 PDFs as a function of the jet \pt{} 
in bins of jet rapidity, for \antikt{} jets with $R=0.6$. Statistically insignificant data points at large \pt{} are omitted. 
The 4.3\% uncertainty from the luminosity measurements is not shown.}
  \label{fig:xsrbands_pt_r06}
\end{figure*}

A comparison of the jet cross-section ratio as a function of \pt{} to the \powheg{} prediction is made in Figs.~\ref{fig:ratiopt_pwg_bands04} and~\ref{fig:ratiopt_pwg_bands06}. Differences between the tunes used in \pythia{} for the parton shower are very small, and deviations are seen only in the forward region for large \pT. Like the NLO pQCD prediction with non-perturbative corrections, the \powheg{} prediction has a different trend in the central rapidity region with respect to data, deviating by more than $10\%$. However, it follows the data very well in the forward region.

\begin{figure*}
  \centering
  \includegraphics[width=0.9\textwidth]{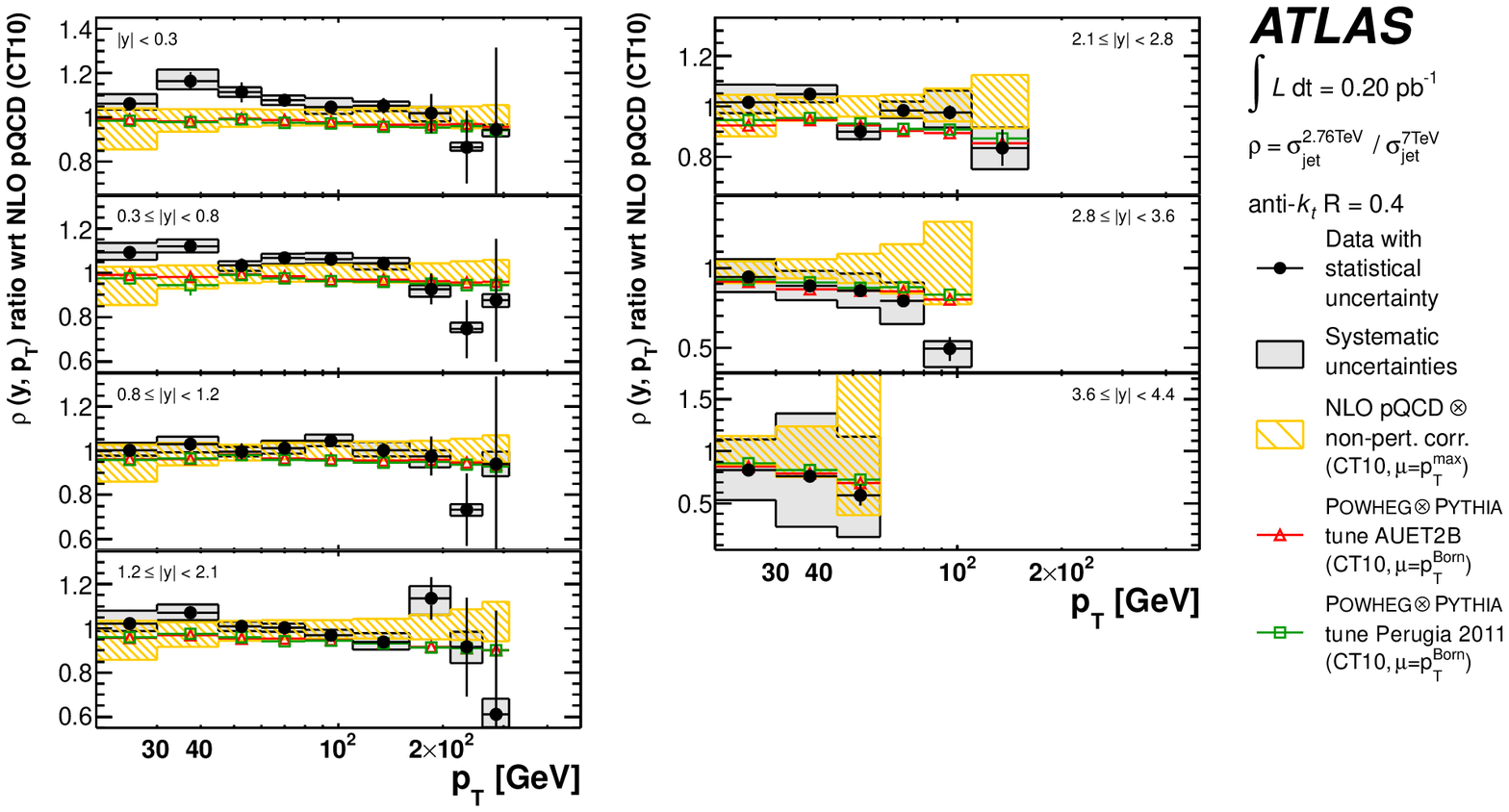}
  \caption{Ratio of the inclusive jet cross-section at $\sqrt{s}=2.76$ TeV to the one at $\sqrt{s}=7$ TeV, shown as a double ratio to the theoretical prediction  calculated with the CT10 PDFs as a function of $\pT$ in bins of jet rapidity, for anti-$k_t$ jets with $R=0.4$. 
Also shown are \powheg{} predictions using \pythia{} for the simulation of the parton shower and hadronisation with the AUET2B tune and the \Perugia{}~2011 tune. Only the statistical uncertainty is shown on the \powheg{} predictions. Statistically insignificant data points at large \pt{} are omitted. The 4.3\% uncertainty from the luminosity measurements is not shown.}
  \label{fig:ratiopt_pwg_bands04}
\end{figure*}

\begin{figure*}
  \centering
  \includegraphics[width=0.9\textwidth]{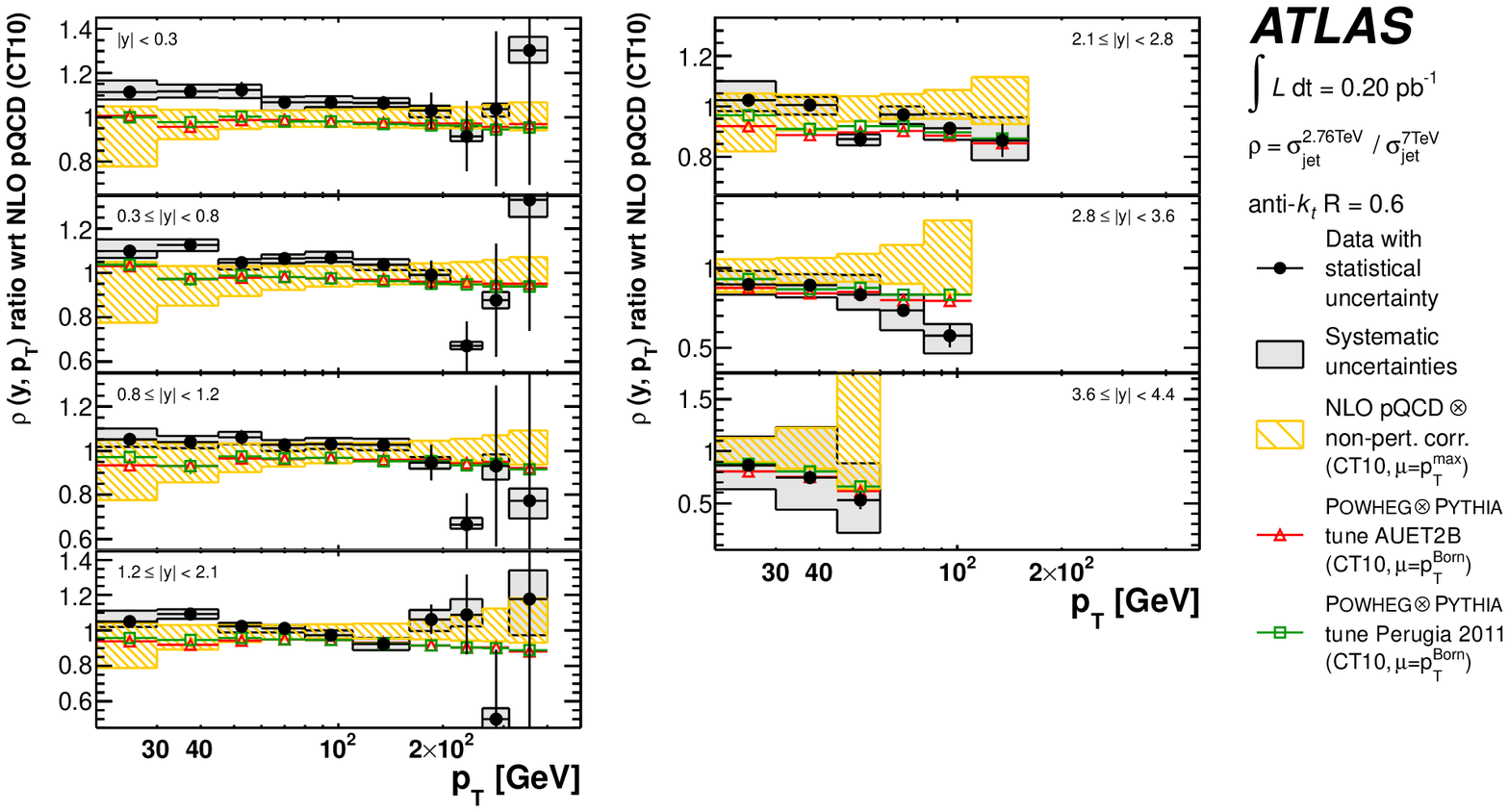}
  \caption{Ratio of the inclusive jet cross-section at $\sqrt{s}=2.76$ TeV to the one at $\sqrt{s}=7$ TeV, shown as a double ratio to the theoretical prediction  calculated with the CT10 PDFs as a function of $\pT$ in bins of jet rapidity, for anti-$k_t$ jets with $R=0.6$. 
Also shown are \powheg{} predictions using \pythia{} for the simulation of the parton shower and hadronisation with the AUET2B tune and the \Perugia{}~2011 tune. Only the statistical uncertainty is shown on the \powheg{} predictions. Statistically insignificant data points at large \pt{} are omitted. The 4.3\% uncertainty from the luminosity measurements is not shown.}
  \label{fig:ratiopt_pwg_bands06}
\end{figure*}

\section{NLO pQCD analysis of \heraI and ATLAS jet data}
\label{sec:pdffit}
Knowledge of the PDFs of the proton comes mainly from deep-inelastic lepton-proton scattering experiments covering a broad range of momentum-transfer squared $Q^2$ and of Bjorken $x$.
 The PDFs are  determined from data using pQCD in the DGLAP formalism~\cite{Gribov:1972ri,Gribov:1972rt,Lipatov:1974qm,Altarelli:1977zs,Dokshitzer:1977sg}. The quark distributions in the region~$x\lesssim 0.01$  are in general well constrained by the precise measurement of the proton structure function $F_2(x,Q^2)$ at HERA~\cite{HERA:2009wt}. However, the gluon momentum distribution~$xg(x,Q^2)$ at~$x$ values above~$0.01$ has not been as precisely determined in deep-inelastic scattering.
The inclusive jet \pt{} spectrum at low and moderate \pt{} is sensitive to the gluon distribution function. 

The systematic uncertainty on the jet cross-section at $\sqrt{s}=2.76$~\TeV{} is strongly correlated with the ATLAS jet cross-section measured at $\sqrt{s}=7$~\TeV{}, as described in Sect.~\ref{sec:jets}. 
Therefore, increased sensitivity to the PDFs is expected when these two jet cross-section datasets are analysed together, with proper treatment of correlation between the measurements. 

A combined NLO pQCD analysis of the inclusive jet cross-section in \pp{} collisions at \cme2.76~\TeV{} together with the ATLAS inclusive jet cross-section in \pp{} collisions at \cme7~\TeV~\cite{Aad:2011fc} and \heraI I data \cite{HERA:2009wt} is presented here.
The analysis is performed using the HERAFitter package~\cite{HERAFitter,HERA:2009wt,Aaron:2009kv}, which uses the light-quark coefficient functions calculated to NLO as implemented in QCDNUM~\cite{Botje:2010ay} and the heavy-quark coefficient functions from the variable-flavour number scheme (VFNS)~\cite{Thorne:1997ga,Thorne:2006qt} for the PDF evolution, as well as MINUIT~\cite{minuit} for minimisation of $\chi^2$. The data are compared to the theory using the $\chi^2$ function defined in \linebreak Refs.~\cite{H1:2009bp,Adloff:2003uh,Chekanov:2002pv}. 
The heavy quark masses are chosen to be $m_c=1.4~$GeV and $m_b=4.75~$GeV~\cite{Martin:2009iq}. The strong coupling constant is fixed to  $\alphas(M_Z) =  0.1176$, as in Ref.~\cite{HERA:2009wt}. 
A minimum $Q^2$ cut of $Q^2_{\mathrm{min}} = 3.5$~GeV$^2$ is imposed on the \heraI data to avoid the non-perturbative region. 
The prediction for the ATLAS jet data is obtained from the NLO pQCD calculation to which the non-perturbative correction is applied as described in Sect.~\ref{sec:theory}.
Due to the large values of the non-per\-tur\-ba\-tive corrections and their large uncertainties at low \pt{} of the jet, all the bins with $\pt<45$~\GeV\ are excluded from the analysis. 

The DGLAP evolution equations yield the PDFs at any value of $Q^2$, given that they are parameterised as functions of $x$ at an initial scale $Q^2_0$.  In the present analysis, this scale is  chosen to be $Q^2_0 = 1.9~$GeV$^2$ such that it is below  $m_c^2$.  PDFs are parameterised at the evolution starting scale $Q^{2}_{0}$ using a HERAPDF-inspired ansatz as in Ref.~\cite{Aad:2012sb}:
\begin{eqnarray}
  x u_v(x) &=&  A_{u_v} x^{B_{u_v}} (1-x)^{C_{u_v}} ( 1 + E_{u_v} x^2), \nonumber\\
  x d_v(x) &=&  A_{d_v} x^{B_{d_v}} (1-x)^{C_{d_v}}, \nonumber\\
  x \bar{U} (x) &=& A_{\bar{U}} x^{B_{\bar{U}}} (1-x)^{C_{\bar{U}}}, \nonumber\\  
  x \bar{D} (x) &=& A_{\bar{D}} x^{B_{\bar{D}}} (1-x)^{C_{\bar{D}}}, \nonumber\\
  x g(x)   &=& A_g x^{B_g} (1-x)^{C_g} - A'_gx^{B'_g}(1-x)^{C'_g} \,.\label{eqn:pdfparametrisation}
\end{eqnarray} 
Here $\bar{U}=\bar{u}$ whereas $\bar{D}=\bar{d} +  \bar{s}$.
The parameters $A_{u_v}$ and $A_{d_v}$ are fixed using  the quark counting rule and $A_g$ using the momentum sum rule. The normalisation and slope parameters, $A$ and $B$, of $\bar{u}$ and $\bar{d}$  are set equal such that  $x\bar{u} = x\bar{d}$ at  $x\to 0$.
An extra term for the valence distribution ($E_{u_v}$) is observed to improve the fit quality significantly.
The strange-quark distribution is constrained to a certain fraction of $\bar{D}$ as $x\bar{s}=f_sx\bar{D}$, where $f_s=0.31$ is chosen in this analysis.
The gluon distribution uses the so-called flexible form, suggested by MSTW analyses, with  $C'_g=25$~\cite{Martin:2009iq}. This value of the $C'_g$ parameter
ensures that the additional term contributes at low $x$ only. With all these additional constraints applied, the fit has $13$ free
parameters to describe the parton densities. 

To see the impact of the ATLAS jet data on the PDFs, a fit only to the \heraI dataset
is performed first. Then, the fit parameters are fixed and the $\chi^2$ value between jet data and the fit prediction is calculated using experimental uncertainties only. The data are included taking into account bin-to-bin correlations. 
Finally, fits to \heraI + ATLAS~jet data are performed for $R=0.4$ and $R=0.6$ jet sizes independently, since correlations of uncertainties between measurements based on two different jet radius parameters have not been determined.
The correlations of systematic uncertainties between the \cme$7$~\TeV\ and \cme$2.76$~\TeV\ datasets are treated as described in Sect.~\ref{sec:ratio}. The PDF uncertainties are determined using the Hessian method~\cite{Pumplin:2002vw,Alekhin:2002fv}.

The consistency of the PDF fit with different datasets in terms of the $\chi^2$ values is given in~\ref{sec:chi2}. Very good fit quality is found for both radius parameters. The $\chi^2$ values also show the pull of ATLAS jet data for both jet radius parameters, while the description of the \heraI data is almost unaffected. 

The fits determine shifts for the correlated systematic uncertainties in the data, which are applied to the theory predictions. 
Typically these shifts are smaller than half a standard deviation and comparable in size for the fits to the two different jet radius parameters. 
Larger differences are found for the normalisation parameters in the fit using the $\sqrt{s}=2.76\TeV$ jet data, being $0.0\%$ for $R=0.4$ and $-2.4\%$ for $R=0.6$, respectively, in spite of the fact that the integrated luminosity is the same in the two cases and thus $100\%$ correlated. 
Since this correlation is not implemented in the fitting method, the differences between the data and theory prediction
for jets with $R=0.4$ and $R=0.6$~(see Sect.~\ref{sec:xsec}) are compensated using shifts of the nuisance parameters. Interestingly, the gluon PDFs obtained from the two fits are very similar. 
Additional studies where the normalisation is fixed in the fit show that the impact of the difference in normalisation on the parton distributions is small.

In the fits using the HERA data and both of the ATLAS jet datasets at the different centre-of-mass energies, the shifts of jet-related systematic uncertainties modelled by $88$ nuisance parameters contribute 19~(12) units in total to the correlated components of the $\chi^2$ for the fit using $R=0.4$~($R=0.6$). A few shifts of jet systematic uncertainties are found to be different between the $R=0.4$ and $R=0.6$ fits, e.g. the jet energy resolution in forward rapidity bins differs by $\sim 0.5 \sigma$. In order to evaluate the impact of the larger shifts on the fit parameters, a special fit is performed in which several systematic uncertainties with the largest shifts are treated as uncorrelated. In these special fits, the PDF parameters in Eq.~\ref{eqn:pdfparametrisation} are found to be compatible with the results of the default fits.

In the following, the results for the fit using jet data with $R=0.6$ are presented.
The results for $R=0.4$ are compatible.
The results of the fits to \heraI data and to the combined data from \heraI and ATLAS jet measurements are presented in Fig.~\ref{fig:fitfit}, which shows the momentum distribution of the gluon $xg$ and sea quarks $xS=2(x\bar{u}+x\bar{d}+x\bar{s})$ at the scale $Q^2=1.9 \GeV^2$. The gluon momentum distribution tends to be harder after the inclusion of the jet data than that obtained from HERA data alone. Furthermore, the uncertainty in $xg$ is reduced if the ATLAS jet data are included in the fit. 
Being smaller in the $\text{high-}x$ region, the sea quark momentum distribution tends to be softer with the ATLAS jet data used in the fit.  
This reduction of the central value results in a larger relative uncertainty on $xS$. 

The fit is also performed for \heraI data in combination with the ATLAS jet data at $\sqrt{s}=2.76\TeV$ and $\sqrt{s}=7\TeV$ separately (see Fig.~\ref{fig:fitfit}). It is found that the impact on the gluon momentum distribution is largely reduced. Hence, the full potential of the ATLAS jet data for PDF fits can be exploited only when both datasets and the information about the correlations are used.

\begin{figure*}
\centering
\subfigure[$xg$]{\includegraphics[width=0.67\linewidth]{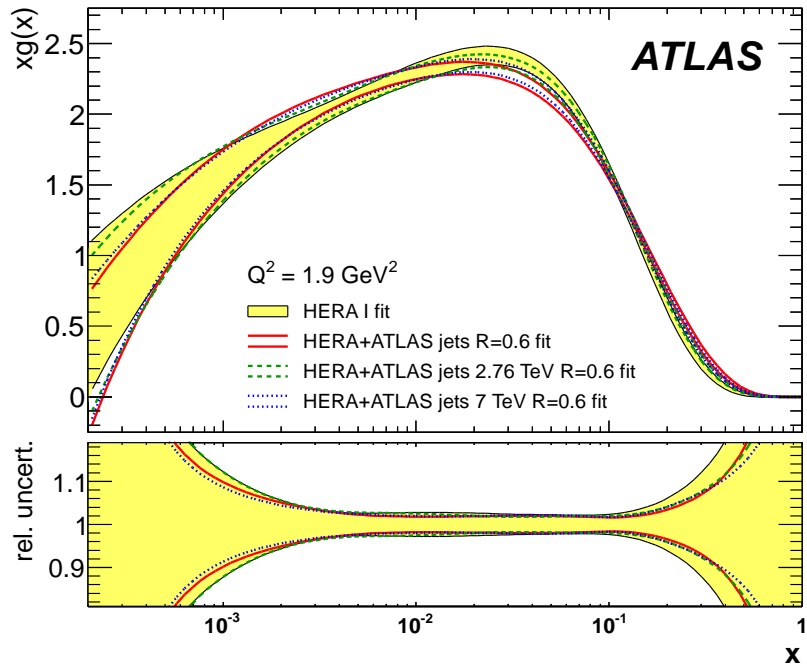}}\\
\subfigure[$xS$]{\includegraphics[width=0.67\linewidth]{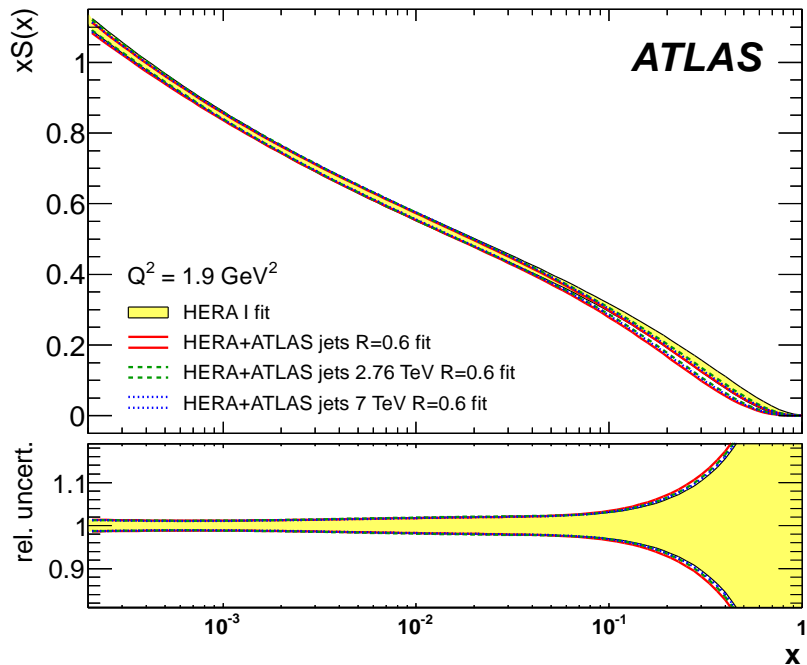}}
\caption{\label{fig:fitfit}Momentum distributions of the (a) gluon $xg(x)$ and (b) sea quarks $xS(x)$ together with their relative experimental uncertainty as a function of~$x$ for $Q^2=1.9$~GeV$^2$. The filled area indicates a fit to \heraI data only. 
The bands show fits to \heraI data in combination with both ATLAS jet datasets, and with the individual ATLAS jet datasets separately, each for jets with $R=0.6$.
For each fit the uncertainty in the PDF is centred on unity. 
}
\end{figure*} 

The measured jet cross-section and the cross-section ratio, $\rho(y,\pt)$, are compared to the predictions based on fitted PDF sets in Figs.~\ref{fig:fitdatatotheory6}~and~\ref{fig:fitdatatoratio}, respectively. 
The data are well described by the prediction based on the refit PDFs after the addition of the jet data. 
The description is particularly improved in the forward region.
\begin{figure*}
\centering
{\includegraphics[width=0.9\textwidth]{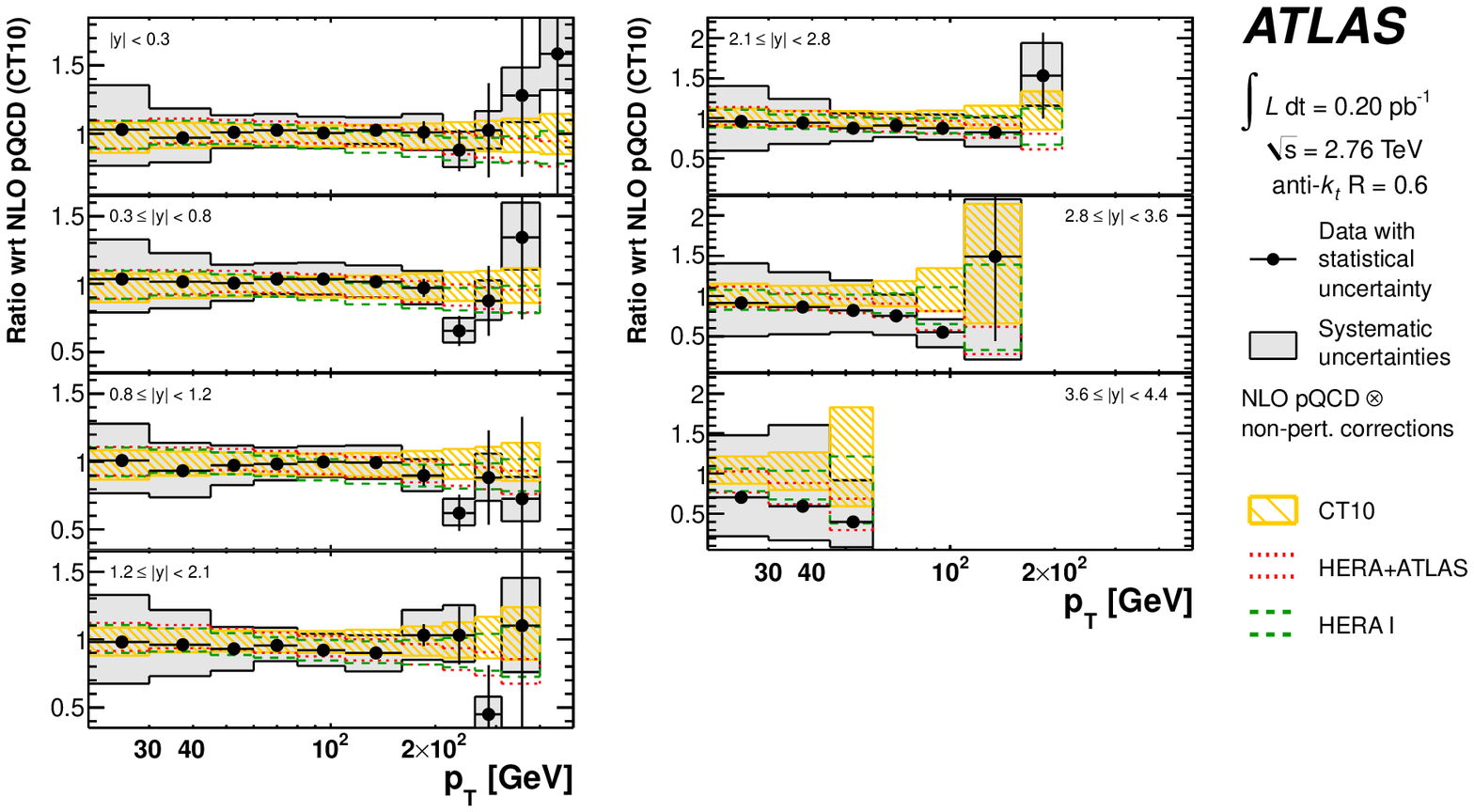}}
\caption{\label{fig:fitdatatotheory6}  
Comparison of NLO pQCD predictions of the jet cross-section at $\sqrt{s}=2.76$~\TeV{} calculated with the CT10 PDF set, the fitted PDF set using the HERA data only and the one using HERA data and the ATLAS jet data with $R=0.6$.
The predictions are normalised to the one using the CT10 PDF set. Also shown is the measured jet cross-section. 
The 2.7\% uncertainty from the luminosity measurement is not shown.
}
\end{figure*} 

\begin{figure*}
\centering
{\includegraphics[width=0.9\textwidth]{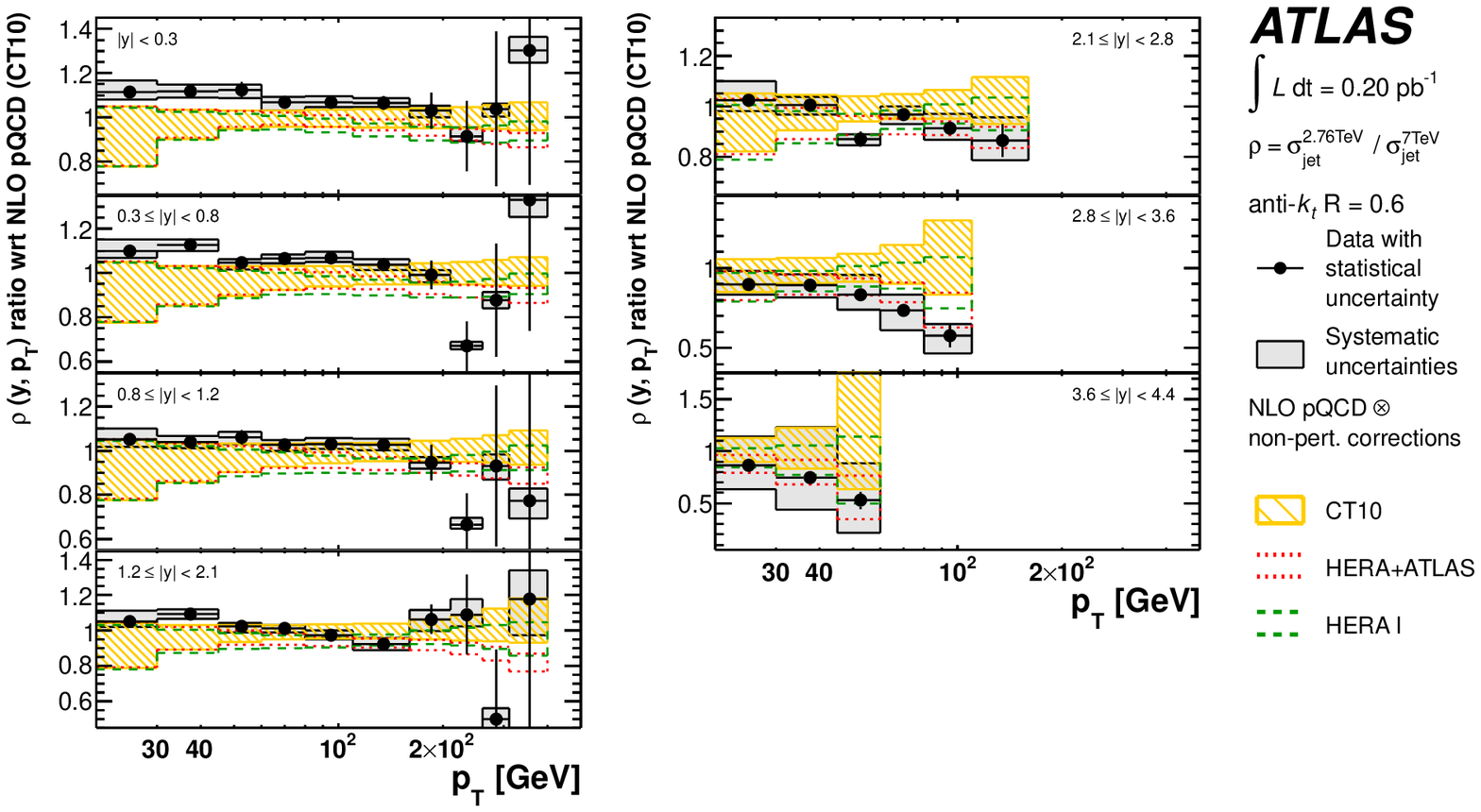}}
\caption{\label{fig:fitdatatoratio}
Comparison of NLO pQCD predictions of the jet cross-section ratio of $\sqrt{s}=2.76$~\TeV{} to $\sqrt{s}=7$~\TeV{} calculated with the CT10 PDF set, the fitted PDF set using the HERA data only and the one using HERA data and the ATLAS jet data with $R=0.6$.
The predictions are normalised to the one using the CT10 PDF set. Also shown is the measured jet cross-section ratio. 
The 4.3\% uncertainty from the luminosity measurements is not shown.
}
\end{figure*} 

\section{Conclusion}\label{sec:conclusion}
The inclusive jet cross-section in $pp$ collisions at $\sqrt{s}=\linebreak2.76\TeV$ has been measured for jets reconstructed with the \antikt{} algorithm with two radius parameters of $R=0.4$ and $R=0.6$, based on the data collected using the ATLAS detector at the beginning of 2011 LHC operation, corresponding to an integrated luminosity of $0.20~\ipb$. 
The measurement is performed as a function of the jet transverse momentum, in bins of jet rapidity.

The ratio of the inclusive jet cross-sections at $\sqrt{s}=2.76\TeV$ and $\sqrt{s}=7\TeV$ is shown in this paper.
The correlation of the sources of uncertainty common to the two measurements is fully taken into account, resulting in a reduction of systematic uncertainties in the ratio measurement. 

The measurements are compared to fixed-order NLO perturbative QCD calculations,
to which corrections for non-perturbative effects are applied. The comparison is
performed with five different PDF sets. The predictions are in good agreement
with the data in general, in both the jet cross-section and the cross-section 
ratio. This confirms that perturbative QCD can describe jet production at high 
jet transverse momentum.  Due to the reduced systematic uncertainties, the ratio
measurement starts to show preferences for certain PDF sets.
The measurement is also compared to predictions from NLO matrix elements with
matched parton-shower Monte Carlo simulation. In particular in the forward
region, the central value of the prediction describes the data well.

An NLO pQCD analysis in the DGLAP formalism has been performed using 
the ATLAS inclusive jet cross-section data at $\sqrt{s}=2.76\TeV$ and $\sqrt{s}=7\TeV$, together with \heraI I data. 
By including the ATLAS jet data, a harder gluon distribution and a softer sea-quark distribution in the high Bjorken-$x$ region are obtained with respect to the fit of \heraI data only. Furthermore, it is shown that the full potential of the ATLAS jet data for PDF fits can be exploited further when the information about the correlations between the measurements at $\sqrt{s}=2.76\TeV$ and $\sqrt{s}=7\TeV$ is used.

\section*{Acknowledgements}
We thank CERN for the very successful operation of the LHC, as well as the
support staff from our institutions without whom ATLAS could not be
operated efficiently.

We acknowledge the support of ANPCyT, Argentina; YerPhI, Armenia; ARC,
Australia; BMWF and FWF, Austria; ANAS, Azerbaijan; SSTC, Belarus; CNPq and FAPESP,
Brazil; NSERC, NRC and CFI, Canada; CERN; CONICYT, Chile; CAS, MOST and NSFC,
China; COLCIENCIAS, Co\-lom\-bia; MSMT CR, MPO CR and VSC CR, Czech Republic;
DNRF, DNSRC and Lundbeck Foundation, Denmark; EPLANET, ERC and NSRF, European Union;
IN2P3-CNRS, CEA-DSM/IRFU, France; GNSF, Georgia; BMBF, DFG, HGF, MPG and AvH
Foundation, Germany; GSRT and NSRF, Greece; ISF, MINERVA, GIF, DIP and Benoziyo Center,
Israel; INFN, Italy; MEXT and JSPS, Japan; CNRST, Morocco; FOM and NWO,
Netherlands; BRF and RCN, Norway; MNiSW, Poland; GRICES and FCT, Portugal; MERYS
(MECTS), Romania; MES of Russia and ROSATOM, Russian Federation; JINR; MSTD,
Serbia; MSSR, Slovakia; \linebreak ARRS and MIZ\v{S}, Slovenia; DST/NRF, South Africa;
MICINN, Spain; SRC and Wallenberg Foundation, Sweden; SER, SNSF and Cantons of
Bern and Geneva, Switzerland; NSC, Taiwan; TAEK, Turkey; STFC, the Royal
Society and Leverhulme Trust, United Kingdom; DOE and NSF, United States of
America.

The crucial computing support from all WLCG partners is acknowledged
gratefully, in particular from CERN and the ATLAS Tier-1 facilities at
TRIUMF (Canada), NDGF (Denmark, Norway, Sweden), CC-IN2P3 (France),
KIT\linebreak/GridKA (Germany), INFN-CNAF (Italy), NL-T1 (Netherlands), PIC (Spain),
ASGC (Taiwan), RAL (UK) and BNL (USA) and in the Tier-2 facilities
worldwide.
\appendix
\clearpage
\section{Bin boundaries of the measured cross-section}
\label{sec:bins}
\begin{table}[hbt]
    \caption{Bin boundaries in the variable \xt{} used in the extraction of $\rho(y,x_T)$, 
the cross-section ratio as a function of \xt{} at different centre-of-mass energies. 
Also shown are the corresponding jet \pt{} values at each centre-of-mass energy.}
    \label{tab:bins}
    \centering
    \begin{tabular}{ccc}
      \hline \hline
      $\xt$ & \pt{} [\GeV]         & \pt{} [\GeV] \\
            & at $\sqrt{s}=7$ \TeV & at $\sqrt{s}=2.76$ \TeV\\
      \hline      
            0.0171   &              60 &      23.65   \\
            0.0229   &              80 &      31.54   \\
            0.0314   &             110 &      43.37   \\
            0.0457   &             160 &      63.08   \\
            0.0600   &             210 &      82.80   \\
            0.0743   &             260 &      102.5   \\
            0.0886   &             310 &      122.2   \\
            0.1143   &             400 &      157.7   \\
            0.1429   &             500 &      197.1   \\
            0.1714   &             600 &      236.5   \\
            0.2286   &             800 &      315.4   \\
            0.2857   &            1000 &      394.2   \\
            0.3429   &            1200 &      473.1   \\
	    \hline\hline
    \end{tabular}
\end{table}
The bin boundaries of the cross-section measurement are given in Table~\ref{tab:bins} for the \xt{} binning, and for the \pt{} binning at both centre-of-mass energies, $\sqrt{s}=7\TeV$ and $\sqrt{s}=2.76\TeV$.
\clearpage

\section[Tables of chi2 values for the PDF fit]{Tables of $\chi^2$ values for the PDF fit}
\label{sec:chi2}
\begin{table*}[hbt]
\caption{\label{tab:chi2}
Agreement of the PDF fit with different datasets in terms of the $\chi^2$ values, using various combinations of input datasets for the fit. \heraI is used as baseline and in combination with ATLAS jet data from the single-jet double-differential cross-section measurements at $\sqrt{s} = 2.76\TeV$ and $\sqrt{s} = 7\TeV$ for anti-$k_t$ jets with two radius parameters, $R=0.4$ and $R=0.6$. When both ATLAS jet datasets at $\sqrt{s} = 2.76\TeV$ and $\sqrt{s} = 7\TeV$ are used, all correlations between the measurements are taken into account.
The $\chi^2$ value of the fit with respect to the individual datasets tested is given separately for the uncorrelated and the correlated components as $\chi^2_\mathrm{uncor}$ and $\chi^2_\mathrm{cor}$, where the shifts of the systematic uncertainties are summed in quadrature for each category.
In general, a very good fit quality is found. Comparison of $\chi^2$ values of the fit using \heraI data only with fits including both \heraI and ATLAS jet data shows the pull of ATLAS jet data for both jet radius parameters, while the description of the \heraI\ data is almost unaffected. For example, $\chi^2_\mathrm{uncor}$ of the \heraI dataset worsens only slightly by 8 units from 556 to 564 when the ATLAS jet data for $R=0.6$ is included in the fit, while $\chi^2_\mathrm{uncor}$ improved from 33~(50) to 29~(40) for the $2.76\TeV$~($7\TeV$) jet data. There is also an improvement from 22 to 12 in $\chi^2_\mathrm{cor}$ for jets.
}
\centering
\begin{tabular}{l|lccc}
 \hline\hline
 input datasets                & test dataset                        & $\chi^2_\mathrm{uncor}$ & $\chi^2_\mathrm{cor}$        & $N_\mathrm{points}$ \\
 \hline 
                                & \heraI                             &  556               & 3.0                   & 592 \\
                                & ATLAS jets $2.76\TeV$, $R=0.4$     &  29                &\multirow{2}{*}{21}    & 40 \\
 \heraI                         & ATLAS jets $7\TeV$,    $R=0.4$     &  44                &                       & 76 \\
                                & ATLAS jets $2.76\TeV$, $R=0.6$     &  33                &\multirow{2}{*}{22}    & 40 \\
                                & ATLAS jets $7\TeV$,    $R=0.6$     &  50                &                       & 76 \\
 \hline\hline
                                & \heraI                              &  562              & 3.6                   & 592 \\
 \heraI                         & ATLAS jets $2.76\TeV$, $R=0.4$      &  27               &\multirow{2}{*}{19}    & 40 \\
 ATLAS jets $2.76\TeV$, $R=0.4$ & ATLAS jets $7\TeV$,    $R=0.4$      &  33               &                       & 76 \\
 ATLAS jets $7\TeV$, $R=0.4$    & ATLAS jets $2.76\TeV$, $R=0.6$      &  29               &\multirow{2}{*}{13}    & 40 \\
                                & ATLAS jets $7\TeV$,    $R=0.6$      &  41               &                       & 76 \\

 \hline
 \heraI                         & \heraI                              & 557               & 3.1                   & 592 \\
 ATLAS jets $2.76\TeV$, $R=0.4$ & ATLAS jets $2.76\TeV$, $R=0.4$      & 20                & 7.4                   & 40 \\
 \hline                                                                               
 \heraI                         & \heraI                              & 559               & 3.4                   & 592 \\
 ATLAS jets $7\TeV$, $R=0.4$    & ATLAS jets $7\TeV$, $R=0.4$         & 28                & 14                    & 76 \\
 \hline\hline
                                & \heraI                              & 564               & 4.0                   & 592 \\
 \heraI                         & ATLAS jets $2.76\TeV$, $R=0.6$ jets & 29                & \multirow{2}{*}{12}   & 40 \\
 ATLAS jets $2.76\TeV$, $R=0.6$ & ATLAS jets $7\TeV$, $R=0.6$         & 40                &                       & 76 \\
 ATLAS jets $7\TeV$, $R=0.6$    & ATLAS jets $2.76\TeV$, $R=0.4$      & 26                & \multirow{2}{*}{18}   & 40 \\
                                & ATLAS jets $7\TeV$, $R=0.4$         & 32                &                       & 76 \\
\hline
\heraI                          & \heraI                              & 558               & 3.2                   & 592 \\
ATLAS jets $2.76\TeV$, $R=0.6$  & ATLAS jets $2.76\TeV$, $R=0.6$      & 21                & 4.9                   & 40 \\
\hline                                                                                
\heraI                          & \heraI                              & 560               & 3.6                   & 592 \\
ATLAS jets $7\TeV$, $R=0.6$     & ATLAS jets $7\TeV$, $R=0.6$         & 34                & 9.4                   & 76 \\
\hline\hline                 
\end{tabular}
\end{table*}
Table~\ref{tab:chi2} shows the $\chi^2$ values for the PDF fits with various combinations of \heraI and ATLAS jet data for different radius parameters $R$ of the \antikt{} algorithm.

\clearpage
\section{Tables of the measured jet cross-sections and cross-section ratios}
The measured inclusive single-jet cross-sections are shown in Tables~\ref{tab:xs_r04_y0}--\ref{tab:xs_r04_y6} and Tables~\ref{tab:xs_r06_y0}--\ref{tab:xs_r06_y6} for jets with $R=0.4$ and $R=0.6$, respectively. Tables~\ref{tab:rxt_r04_y0}--\ref{tab:rxt_r04_y6} and \ref{tab:rxt_r06_y0}--\ref{tab:rxt_r06_y6} show the measured cross-section ratio, $\rho(y,\xt{})$, for $R=0.4$ and $R=0.6$, and Tables~\ref{tab:rpt_r04_y0}--\ref{tab:rpt_r04_y6} and Tables~\ref{tab:rpt_r06_y0}--\ref{tab:rpt_r06_y6} show the measured cross-section ratio, $\rho(y,\pt{})$, for $R=0.4$ and $R=0.6$, respectively.

\clearpage
\begin{table*}
\caption{Measured jet cross section for anti-$k_t$ jets with $R=0.4$ in the rapidity bin $|y|<0.3$. NPCorr stands for multiplicative non-perturbative corrections, $\sigma$ is the cross section and $\delta_\mathrm{stat.}$ is the statistical uncertainty. $\gamma_i$ and $u_j$ correspond to the correlated and uncorrelated systematic uncertainties given in \%. They are described in Table~\ref{tab:uncert}, where $i$ in $\gamma_i$ denotes a nuisance parameter. For each table entry, the outcome of shifting the corresponding nuisance parameter by one standard deviation, up or down, is shown as superscript or subscript, respectively. In some bins these shifts may lead to cross-section variations in the same direction. The 2.7\% uncertainty from the luminosity measurement is not in the table.}\label{tab:xs_r04_y0}
\tiny
\centering
\begin{tabular}{@{}c@{}@{}c@{}@{}c@{}@{}c@{}@{}r@{}@{}r@{}@{}r@{}@{}r@{}@{}r@{}@{}r@{}@{}r@{}@{}r@{}@{}r@{}@{}r@{}@{}r@{}@{}r@{}@{}r@{}@{}r@{}@{}c@{}@{}c@{}@{}c@{}@{}c@{}@{}c@{}@{}c@{}@{}c@{}}
\hline\hline
 \pt &  NPCorr &  $\sigma$ & $\delta_\mathrm{stat.}$ & $\gamma_{1}$ & $\gamma_{7}$ & $\gamma_{13}$ & $\gamma_{19}$ & $\gamma_{25}$ & $\gamma_{88}$ & $\gamma_{32}$ & $\gamma_{38}$ & $\gamma_{44}$ & $\gamma_{50}$ & $\gamma_{56}$ & $\gamma_{62}$ & $\gamma_{68}$ & $\gamma_{89}$ & $\gamma_{76}$ & $\gamma_{82}$ & $\gamma_{74}$ & $\gamma_{75}$ & $\gamma_{83}$ & $u_1$ & $u_2$ \\
(GeV) & & (nb/GeV)& \% &  \%&  \%&  \%&  \%&  \%&  \%&  \%&  \%&  \%&  \%&  \%&  \%&  \%&  \%&  \%&  \%&  \%&  \%&  \%&  \%&  \%\\
\hline
\ \ 20--30\ \ &\ \  0.88\!\!$\begin{array}{r} + 0.07\\- 0.11\end{array}$\ \ &\ \ 1.308e+03\ \ & \ \ 1.7\ \ &\ \ $^{+13.}_{-9.6}$\ \ & \ \ $^{ +12}_{ -10}$\ \ & \ \ $^{+ 7.8}_{-7.4}$\ \ & \ \ $^{+ 7.2}_{-9.2}$\ \ & \ \ $^{+ 4.9}_{-4.0}$\ \ & \ \ 0.0\ \ &\ \ $^{+ 1.8}_{-2.1}$\ \ & \ \ $^{+ 5.3}_{-5.0}$\ \ & \ \ $^{+ 4.5}_{-4.7}$\ \ & \ \ $^{-0.0}_{+ 0.0}$\ \ & \ \ $^{+ 0.0}_{-0.0}$\ \ & \ \ $^{+ 1.3}_{-2.1}$\ \ & \ \ $^{-0.0}_{+ 0.0}$\ \ & \ \ 0.0\ \ &\ \ $\pm4.6$\ \ & \ \ $\mp 0.1$\ \ & \ \ $\mp 0.1$\ \ & \ \ $\pm 0.3$\ \ & \ \ $\pm 2.0$\ \ & \ \ $\pm 1.0$\ \ & \ \ $\pm 0.5$\ \ \\
\ \ 30--45\ \ &\ \  0.91\!\!$\begin{array}{r} + 0.06\\- 0.09\end{array}$\ \ &\ \ 1.945e+02\ \ & \ \ 2.9\ \ &\ \ $^{+ 4.6}_{-3.5}$\ \ & \ \ $^{+ 9.6}_{-8.0}$\ \ & \ \ $^{+ 6.7}_{-5.4}$\ \ & \ \ $^{+12.}_{-9.4}$\ \ & \ \ $^{+ 1.8}_{-1.2}$\ \ & \ \ 0.0\ \ &\ \ $^{+ 3.3}_{-2.5}$\ \ & \ \ $^{+ 4.4}_{-3.5}$\ \ & \ \ $^{+ 5.1}_{-4.3}$\ \ & \ \ $^{-0.0}_{+ 0.0}$\ \ & \ \ $^{+ 0.0}_{-0.0}$\ \ & \ \ $^{+ 4.4}_{-3.7}$\ \ & \ \ $^{-0.0}_{+ 0.0}$\ \ & \ \ 0.0\ \ &\ \ $\pm3.5$\ \ & \ \ $\mp 0.1$\ \ & \ \ $\pm 0.0$\ \ & \ \ $\pm 0.1$\ \ & \ \ $\pm 1.0$\ \ & \ \ $\pm 1.0$\ \ & \ \ $\pm 0.5$\ \ \\
\hline\hline\end{tabular}
\end{table*}
\begin{table*}
\caption{Measured jet cross section for anti-$k_t$ jets with $R=0.4$ in the rapidity bin $0.3\leq|y|<0.8$. See caption of Table~\ref{tab:xs_r04_y0} for details.}
\label{tab:xs_r04_y1}
\tiny
\centering
\begin{tabular}{@{}c@{}@{}c@{}@{}c@{}@{}c@{}@{}r@{}@{}r@{}@{}r@{}@{}r@{}@{}r@{}@{}r@{}@{}r@{}@{}r@{}@{}r@{}@{}r@{}@{}r@{}@{}r@{}@{}r@{}@{}r@{}@{}c@{}@{}c@{}@{}c@{}@{}c@{}@{}c@{}@{}c@{}@{}c@{}}
\hline\hline
 \pt &  NPCorr &  $\sigma$ & $\delta_\mathrm{stat.}$ & $\gamma_{1}$ & $\gamma_{7}$ & $\gamma_{13}$ & $\gamma_{19}$ & $\gamma_{25}$ & $\gamma_{88}$ & $\gamma_{32}$ & $\gamma_{38}$ & $\gamma_{44}$ & $\gamma_{50}$ & $\gamma_{56}$ & $\gamma_{62}$ & $\gamma_{68}$ & $\gamma_{89}$ & $\gamma_{76}$ & $\gamma_{82}$ & $\gamma_{74}$ & $\gamma_{75}$ & $\gamma_{83}$ & $u_1$ & $u_2$ \\
(GeV) & & (nb/GeV)& \% &  \%&  \%&  \%&  \%&  \%&  \%&  \%&  \%&  \%&  \%&  \%&  \%&  \%&  \%&  \%&  \%&  \%&  \%&  \%&  \%&  \%\\
\hline
\ \ 20--30\ \ &\ \  0.88\!\!$\begin{array}{r} + 0.07\\- 0.12\end{array}$\ \ &\ \ 1.304e+03\ \ & \ \ 1.4\ \ &\ \ $^{+13.}_{-9.1}$\ \ & \ \ $^{+ 6.6}_{-8.4}$\ \ & \ \ $^{+12.}_{-9.3}$\ \ & \ \ $^{+ 6.2}_{-7.5}$\ \ & \ \ $^{+ 4.3}_{-4.0}$\ \ & \ \ $^{+ 0.1}_{-0.1}$\ \ & \ \ $^{+ 2.0}_{-2.3}$\ \ & \ \ $^{+ 4.5}_{-4.3}$\ \ & \ \ $^{+ 4.6}_{-4.7}$\ \ & \ \ $^{-0.0}_{+ 0.0}$\ \ & \ \ $^{+ 0.0}_{-0.0}$\ \ & \ \ $^{+ 2.7}_{-3.0}$\ \ & \ \ $^{-0.0}_{+ 0.0}$\ \ & \ \ $^{+ 0.1}_{-0.0}$\ \ & \ \ $\pm5.7$\ \ & \ \ $\pm 0.0$\ \ & \ \ $\mp 0.1$\ \ & \ \ $\pm 0.2$\ \ & \ \ $\pm 2.0$\ \ & \ \ $\pm 1.0$\ \ & \ \ $\pm 0.5$\ \ \\
\ \ 30--45\ \ &\ \  0.91\!\!$\begin{array}{r} + 0.06\\- 0.09\end{array}$\ \ &\ \ 1.887e+02\ \ & \ \ 2.3\ \ &\ \ $^{+ 4.2}_{-3.4}$\ \ & \ \ $^{ +12}_{ -10}$\ \ & \ \ $^{+ 6.8}_{-5.7}$\ \ & \ \ $^{+ 9.6}_{-7.9}$\ \ & \ \ $^{+ 2.3}_{-2.0}$\ \ & \ \ $^{+ 0.1}_{-0.1}$\ \ & \ \ $^{+ 2.8}_{-2.7}$\ \ & \ \ $^{+ 3.4}_{-3.0}$\ \ & \ \ $^{+ 5.0}_{-4.5}$\ \ & \ \ $^{-0.0}_{-0.0}$\ \ & \ \ $^{+ 0.0}_{-0.1}$\ \ & \ \ $^{+ 3.6}_{-3.3}$\ \ & \ \ $^{-0.2}_{+ 0.2}$\ \ & \ \ $^{+ 0.1}_{-0.1}$\ \ & \ \ $\pm2.6$\ \ & \ \ $\mp 0.1$\ \ & \ \ $\pm 0.0$\ \ & \ \ $\pm 0.1$\ \ & \ \ $\pm 1.0$\ \ & \ \ $\pm 1.0$\ \ & \ \ $\pm 0.5$\ \ \\
\hline\hline\end{tabular}
\end{table*}
\begin{table*}
\caption{Measured jet cross section for anti-$k_t$ jets with $R=0.4$ in the rapidity bin $0.8\leq|y|<1.2$. See caption of Table~\ref{tab:xs_r04_y0} for details.}
\label{tab:xs_r04_y2}
\tiny
\centering
\begin{tabular}{@{}c@{}@{}c@{}@{}c@{}@{}c@{}@{}r@{}@{}r@{}@{}r@{}@{}r@{}@{}r@{}@{}r@{}@{}r@{}@{}r@{}@{}r@{}@{}r@{}@{}r@{}@{}r@{}@{}r@{}@{}r@{}@{}c@{}@{}c@{}@{}c@{}@{}c@{}@{}c@{}@{}c@{}@{}c@{}}
\hline\hline
 \pt &  NPCorr &  $\sigma$ & $\delta_\mathrm{stat.}$ & $\gamma_{2}$ & $\gamma_{8}$ & $\gamma_{14}$ & $\gamma_{20}$ & $\gamma_{26}$ & $\gamma_{88}$ & $\gamma_{33}$ & $\gamma_{39}$ & $\gamma_{45}$ & $\gamma_{51}$ & $\gamma_{57}$ & $\gamma_{63}$ & $\gamma_{69}$ & $\gamma_{89}$ & $\gamma_{77}$ & $\gamma_{82}$ & $\gamma_{74}$ & $\gamma_{75}$ & $\gamma_{83}$ & $u_1$ & $u_2$ \\
(GeV) & & (nb/GeV)& \% &  \%&  \%&  \%&  \%&  \%&  \%&  \%&  \%&  \%&  \%&  \%&  \%&  \%&  \%&  \%&  \%&  \%&  \%&  \%&  \%&  \%\\
\hline
\ \ 20--30\ \ &\ \  0.89\!\!$\begin{array}{r} + 0.07\\- 0.11\end{array}$\ \ &\ \ 1.144e+03\ \ & \ \ 1.6\ \ &\ \ $^{+11.}_{-8.7}$\ \ & \ \ $^{+ 5.7}_{-8.1}$\ \ & \ \ $^{+ 9.8}_{-8.9}$\ \ & \ \ $^{ +14}_{ -12}$\ \ & \ \ $^{+ 3.9}_{-3.7}$\ \ & \ \ $^{+ 6.2}_{-6.8}$\ \ & \ \ $^{+ 1.9}_{-2.0}$\ \ & \ \ $^{+ 4.1}_{-4.0}$\ \ & \ \ $^{+ 4.2}_{-4.3}$\ \ & \ \ $^{-0.0}_{+ 0.0}$\ \ & \ \ $^{+ 0.0}_{-0.0}$\ \ & \ \ $^{+ 2.5}_{-2.6}$\ \ & \ \ $^{-0.0}_{+ 0.0}$\ \ & \ \ $^{+ 3.6}_{-3.8}$\ \ & \ \ $\pm9.5$\ \ & \ \ $\pm 1.2$\ \ & \ \ $\mp 0.3$\ \ & \ \ $\pm 0.1$\ \ & \ \ $\pm 2.0$\ \ & \ \ $\pm 1.0$\ \ & \ \ $\pm 0.5$\ \ \\
\ \ 30--45\ \ &\ \  0.92\!\!$\begin{array}{r} + 0.06\\- 0.09\end{array}$\ \ &\ \ 1.599e+02\ \ & \ \ 2.7\ \ &\ \ $^{+ 3.6}_{-3.6}$\ \ & \ \ $^{ +12}_{ -10}$\ \ & \ \ $^{+ 6.3}_{-6.2}$\ \ & \ \ $^{+ 3.5}_{-3.3}$\ \ & \ \ $^{+ 1.9}_{-2.0}$\ \ & \ \ $^{+ 7.4}_{-7.1}$\ \ & \ \ $^{+ 2.5}_{-2.8}$\ \ & \ \ $^{+ 3.0}_{-3.2}$\ \ & \ \ $^{+ 4.5}_{-4.7}$\ \ & \ \ $^{-0.0}_{+ 0.0}$\ \ & \ \ $^{+ 0.0}_{-0.0}$\ \ & \ \ $^{+ 3.2}_{-3.4}$\ \ & \ \ $^{-0.2}_{+ 0.2}$\ \ & \ \ $^{+ 4.1}_{-4.1}$\ \ & \ \ $\pm5.3$\ \ & \ \ $\pm 0.1$\ \ & \ \ $\mp 0.0$\ \ & \ \ $\pm 0.7$\ \ & \ \ $\pm 1.0$\ \ & \ \ $\pm 1.0$\ \ & \ \ $\pm 0.5$\ \ \\
\hline\hline\end{tabular}
\end{table*}
\begin{table*}
\caption{Measured jet cross section for anti-$k_t$ jets with $R=0.4$ in the rapidity bin $1.2\leq|y|<2.1$. See caption of Table~\ref{tab:xs_r04_y0} for details.}
\label{tab:xs_r04_y3}
\tiny
\centering
\begin{tabular}{@{}c@{}@{}c@{}@{}c@{}@{}c@{}@{}r@{}@{}r@{}@{}r@{}@{}r@{}@{}r@{}@{}r@{}@{}r@{}@{}r@{}@{}r@{}@{}r@{}@{}r@{}@{}r@{}@{}r@{}@{}r@{}@{}c@{}@{}c@{}@{}c@{}@{}c@{}@{}c@{}@{}c@{}@{}c@{}}
\hline\hline
 \pt &  NPCorr &  $\sigma$ & $\delta_\mathrm{stat.}$ & $\gamma_{3}$ & $\gamma_{9}$ & $\gamma_{15}$ & $\gamma_{21}$ & $\gamma_{27}$ & $\gamma_{88}$ & $\gamma_{34}$ & $\gamma_{40}$ & $\gamma_{46}$ & $\gamma_{52}$ & $\gamma_{58}$ & $\gamma_{64}$ & $\gamma_{70}$ & $\gamma_{89}$ & $\gamma_{78}$ & $\gamma_{82}$ & $\gamma_{74}$ & $\gamma_{75}$ & $\gamma_{83}$ & $u_1$ & $u_2$ \\
(GeV) & & (nb/GeV)& \% &  \%&  \%&  \%&  \%&  \%&  \%&  \%&  \%&  \%&  \%&  \%&  \%&  \%&  \%&  \%&  \%&  \%&  \%&  \%&  \%&  \%\\
\hline
\ \ 20--30\ \ &\ \  0.89\!\!$\begin{array}{r} + 0.08\\- 0.11\end{array}$\ \ &\ \ 9.462e+02\ \ & \ \ 1.2\ \ &\ \ $^{+13.}_{-9.1}$\ \ & \ \ $^{+ 6.5}_{-8.3}$\ \ & \ \ $^{+12.}_{-9.3}$\ \ & \ \ $^{+13.}_{-8.5}$\ \ & \ \ $^{+ 4.4}_{-3.9}$\ \ & \ \ $^{ +15}_{ -14}$\ \ & \ \ $^{+ 2.1}_{-2.2}$\ \ & \ \ $^{+ 4.5}_{-4.1}$\ \ & \ \ $^{+ 4.5}_{-4.5}$\ \ & \ \ $^{-0.0}_{+ 0.0}$\ \ & \ \ $^{+ 0.0}_{-0.0}$\ \ & \ \ $^{+ 2.6}_{-2.9}$\ \ & \ \ $^{-0.0}_{+ 0.0}$\ \ & \ \ $^{+ 5.0}_{-5.2}$\ \ & \ \ $\pm12.$\ \ & \ \ $\mp 0.6$\ \ & \ \ $\mp 0.1$\ \ & \ \ $\pm 0.6$\ \ & \ \ $\pm 2.0$\ \ & \ \ $\pm 1.0$\ \ & \ \ $\pm 0.5$\ \ \\
\ \ 30--45\ \ &\ \  0.92\!\!$\begin{array}{r} + 0.06\\- 0.08\end{array}$\ \ &\ \ 1.270e+02\ \ & \ \ 2.1\ \ &\ \ $^{+ 4.3}_{-3.8}$\ \ & \ \ $^{ +13}_{ -11}$\ \ & \ \ $^{+ 7.0}_{-6.5}$\ \ & \ \ $^{+ 1.0}_{-0.6}$\ \ & \ \ $^{+ 2.3}_{-2.0}$\ \ & \ \ $^{ +17}_{ -14}$\ \ & \ \ $^{+ 2.9}_{-2.8}$\ \ & \ \ $^{+ 3.4}_{-3.3}$\ \ & \ \ $^{+ 5.1}_{-4.9}$\ \ & \ \ $^{-0.0}_{+ 0.0}$\ \ & \ \ $^{+ 0.0}_{-0.0}$\ \ & \ \ $^{+ 3.6}_{-3.6}$\ \ & \ \ $^{-0.2}_{+ 0.1}$\ \ & \ \ $^{+ 6.0}_{-5.7}$\ \ & \ \ $\pm5.5$\ \ & \ \ $\mp 0.4$\ \ & \ \ $\pm 0.1$\ \ & \ \ $\pm 0.1$\ \ & \ \ $\pm 1.0$\ \ & \ \ $\pm 1.0$\ \ & \ \ $\pm 0.5$\ \ \\
\hline\hline\end{tabular}
\end{table*}
\begin{table*}
\caption{Measured jet cross section for anti-$k_t$ jets with $R=0.4$ in the rapidity bin $2.1\leq|y|<2.8$. See caption of Table~\ref{tab:xs_r04_y0} for details.}
\label{tab:xs_r04_y4}
\tiny
\centering
\begin{tabular}{@{}c@{}@{}c@{}@{}c@{}@{}c@{}@{}r@{}@{}r@{}@{}r@{}@{}r@{}@{}r@{}@{}r@{}@{}r@{}@{}r@{}@{}r@{}@{}r@{}@{}r@{}@{}r@{}@{}r@{}@{}r@{}@{}c@{}@{}c@{}@{}c@{}@{}c@{}@{}c@{}@{}c@{}@{}c@{}}
\hline\hline
 \pt &  NPCorr &  $\sigma$ & $\delta_\mathrm{stat.}$ & $\gamma_{4}$ & $\gamma_{10}$ & $\gamma_{16}$ & $\gamma_{22}$ & $\gamma_{28}$ & $\gamma_{88}$ & $\gamma_{35}$ & $\gamma_{41}$ & $\gamma_{47}$ & $\gamma_{53}$ & $\gamma_{59}$ & $\gamma_{65}$ & $\gamma_{71}$ & $\gamma_{89}$ & $\gamma_{79}$ & $\gamma_{82}$ & $\gamma_{74}$ & $\gamma_{75}$ & $\gamma_{84}$ & $u_1$ & $u_2$ \\
(GeV) & & (nb/GeV)& \% &  \%&  \%&  \%&  \%&  \%&  \%&  \%&  \%&  \%&  \%&  \%&  \%&  \%&  \%&  \%&  \%&  \%&  \%&  \%&  \%&  \%\\
\hline
\ \ 20--30\ \ &\ \  0.88\!\!$\begin{array}{r} + 0.09\\- 0.11\end{array}$\ \ &\ \ 6.251e+02\ \ & \ \ 1.9\ \ &\ \ $^{+14.}_{-9.3}$\ \ & \ \ $^{+ 6.7}_{-8.1}$\ \ & \ \ $^{+13.}_{-9.3}$\ \ & \ \ $^{+ 8.1}_{-6.3}$\ \ & \ \ $^{+ 4.9}_{-3.9}$\ \ & \ \ $^{ +29}_{ -24}$\ \ & \ \ $^{+ 2.1}_{-2.1}$\ \ & \ \ $^{+ 5.0}_{-4.1}$\ \ & \ \ $^{+ 4.9}_{-4.4}$\ \ & \ \ $^{-0.0}_{+ 0.0}$\ \ & \ \ $^{+ 0.0}_{-0.0}$\ \ & \ \ $^{+ 2.8}_{-2.7}$\ \ & \ \ $^{-0.0}_{+ 0.0}$\ \ & \ \ $^{+ 4.2}_{-3.8}$\ \ & \ \ $\pm18.$\ \ & \ \ $\mp 0.4$\ \ & \ \ $\mp 0.0$\ \ & \ \ $\pm 0.0$\ \ & \ \ $\pm 2.0$\ \ & \ \ $\pm 1.0$\ \ & \ \ $\pm 0.5$\ \ \\
\ \ 30--45\ \ &\ \  0.92\!\!$\begin{array}{r} + 0.07\\- 0.08\end{array}$\ \ &\ \ 7.790e+01\ \ & \ \ 1.5\ \ &\ \ $^{+ 4.2}_{-3.5}$\ \ & \ \ $^{ +13}_{ -10}$\ \ & \ \ $^{+ 7.0}_{-6.2}$\ \ & \ \ $^{+ 4.6}_{-3.9}$\ \ & \ \ $^{+ 2.2}_{-1.8}$\ \ & \ \ $^{ +30}_{ -22}$\ \ & \ \ $^{+ 3.0}_{-2.5}$\ \ & \ \ $^{+ 3.6}_{-2.9}$\ \ & \ \ $^{+ 5.0}_{-4.3}$\ \ & \ \ $^{-0.0}_{+ 0.0}$\ \ & \ \ $^{+ 0.0}_{-0.0}$\ \ & \ \ $^{+ 3.7}_{-3.2}$\ \ & \ \ $^{-0.2}_{+ 0.1}$\ \ & \ \ $^{+ 4.5}_{-3.9}$\ \ & \ \ $\pm5.1$\ \ & \ \ $\mp 0.8$\ \ & \ \ $\pm 0.3$\ \ & \ \ $\pm 0.1$\ \ & \ \ $\pm 1.0$\ \ & \ \ $\pm 1.0$\ \ & \ \ $\pm 0.5$\ \ \\
\hline\hline\end{tabular}
\end{table*}
\begin{table*}
\caption{Measured jet cross section for anti-$k_t$ jets with $R=0.4$ in the rapidity bin $2.8\leq|y|<3.6$. See caption of Table~\ref{tab:xs_r04_y0} for details.}
\label{tab:xs_r04_y5}
\tiny
\centering
\begin{tabular}{@{}c@{}@{}c@{}@{}c@{}@{}c@{}@{}r@{}@{}r@{}@{}r@{}@{}r@{}@{}r@{}@{}r@{}@{}r@{}@{}r@{}@{}r@{}@{}r@{}@{}r@{}@{}r@{}@{}r@{}@{}r@{}@{}c@{}@{}c@{}@{}c@{}@{}c@{}@{}c@{}@{}c@{}@{}c@{}}
\hline\hline
 \pt &  NPCorr &  $\sigma$ & $\delta_\mathrm{stat.}$ & $\gamma_{5}$ & $\gamma_{11}$ & $\gamma_{17}$ & $\gamma_{23}$ & $\gamma_{29}$ & $\gamma_{88}$ & $\gamma_{36}$ & $\gamma_{42}$ & $\gamma_{48}$ & $\gamma_{54}$ & $\gamma_{60}$ & $\gamma_{66}$ & $\gamma_{72}$ & $\gamma_{89}$ & $\gamma_{80}$ & $\gamma_{82}$ & $\gamma_{74}$ & $\gamma_{75}$ & $\gamma_{85}$ & $u_1$ & $u_2$ \\
(GeV) & & (nb/GeV)& \% &  \%&  \%&  \%&  \%&  \%&  \%&  \%&  \%&  \%&  \%&  \%&  \%&  \%&  \%&  \%&  \%&  \%&  \%&  \%&  \%&  \%\\
\hline
\ \ 20--30\ \ &\ \  0.86\!\!$\begin{array}{r} + 0.11\\- 0.09\end{array}$\ \ &\ \ 3.133e+02\ \ & \ \ 2.3\ \ &\ \ $^{+15.}_{-10.}$\ \ & \ \ $^{+ 7.4}_{-9.4}$\ \ & \ \ $^{ +13}_{ -10}$\ \ & \ \ $^{+11.}_{-8.3}$\ \ & \ \ $^{+ 5.0}_{-4.4}$\ \ & \ \ $^{ +43}_{ -33}$\ \ & \ \ $^{+ 2.6}_{-2.5}$\ \ & \ \ $^{+ 5.2}_{-4.6}$\ \ & \ \ $^{+ 5.3}_{-5.1}$\ \ & \ \ $^{-0.0}_{+ 0.0}$\ \ & \ \ $^{+ 0.0}_{-0.0}$\ \ & \ \ $^{+ 3.2}_{-3.2}$\ \ & \ \ $^{-0.0}_{+ 0.0}$\ \ & \ \ $^{+ 4.9}_{-4.9}$\ \ & \ \ $\pm19.$\ \ & \ \ $\pm 0.1$\ \ & \ \ $\mp 0.1$\ \ & \ \ $\pm 0.4$\ \ & \ \ $\pm 2.0$\ \ & \ \ $\pm 2.0$\ \ & \ \ $\pm 0.5$\ \ \\
\ \ 30--45\ \ &\ \  0.89\!\!$\begin{array}{r} + 0.09\\- 0.06\end{array}$\ \ &\ \ 2.821e+01\ \ & \ \ 2.3\ \ &\ \ $^{+ 4.2}_{-4.0}$\ \ & \ \ $^{ +15}_{ -13}$\ \ & \ \ $^{+ 8.0}_{-7.9}$\ \ & \ \ $^{+ 6.2}_{-5.8}$\ \ & \ \ $^{+ 2.4}_{-2.4}$\ \ & \ \ $^{ +51}_{ -34}$\ \ & \ \ $^{+ 3.5}_{-3.3}$\ \ & \ \ $^{+ 3.9}_{-3.7}$\ \ & \ \ $^{+ 6.4}_{-5.6}$\ \ & \ \ $^{+ 0.1}_{+ 0.0}$\ \ & \ \ $^{+ 0.0}_{+ 0.0}$\ \ & \ \ $^{+ 4.2}_{-4.0}$\ \ & \ \ $^{-0.1}_{+ 0.1}$\ \ & \ \ $^{+ 5.9}_{-5.5}$\ \ & \ \ $\pm5.8$\ \ & \ \ $\pm 0.3$\ \ & \ \ $\mp 0.5$\ \ & \ \ $\pm 0.5$\ \ & \ \ $\pm 1.0$\ \ & \ \ $\pm 2.0$\ \ & \ \ $\pm 0.5$\ \ \\
\hline\hline\end{tabular}
\end{table*}
\begin{table*}
\caption{Measured jet cross section for anti-$k_t$ jets with $R=0.4$ in the rapidity bin $3.6\leq|y|<4.4$. See caption of Table~\ref{tab:xs_r04_y0} for details.}
\label{tab:xs_r04_y6}
\tiny
\centering
\begin{tabular}{@{}c@{}@{}c@{}@{}c@{}@{}c@{}@{}r@{}@{}r@{}@{}r@{}@{}r@{}@{}r@{}@{}r@{}@{}r@{}@{}r@{}@{}r@{}@{}r@{}@{}r@{}@{}r@{}@{}r@{}@{}r@{}@{}c@{}@{}c@{}@{}c@{}@{}c@{}@{}c@{}@{}c@{}@{}c@{}}
\hline\hline
 \pt &  NPCorr &  $\sigma$ & $\delta_\mathrm{stat.}$ & $\gamma_{6}$ & $\gamma_{12}$ & $\gamma_{18}$ & $\gamma_{24}$ & $\gamma_{30}$ & $\gamma_{88}$ & $\gamma_{37}$ & $\gamma_{43}$ & $\gamma_{49}$ & $\gamma_{55}$ & $\gamma_{61}$ & $\gamma_{67}$ & $\gamma_{73}$ & $\gamma_{89}$ & $\gamma_{81}$ & $\gamma_{82}$ & $\gamma_{74}$ & $\gamma_{75}$ & $\gamma_{86}$ & $u_1$ & $u_2$ \\
(GeV) & & (nb/GeV)& \% &  \%&  \%&  \%&  \%&  \%&  \%&  \%&  \%&  \%&  \%&  \%&  \%&  \%&  \%&  \%&  \%&  \%&  \%&  \%&  \%&  \%\\
\hline
\ \ 20--30\ \ &\ \  0.78\!\!$\begin{array}{r} + 0.14\\- 0.08\end{array}$\ \ &\ \ 4.651e+01\ \ & \ \ 5.8\ \ &\ \ $^{ +23}_{ -14}$\ \ & \ \ $^{ +12}_{ -14}$\ \ & \ \ $^{ +21}_{ -15}$\ \ & \ \ $^{+ 5.4}_{-5.0}$\ \ & \ \ $^{+ 7.4}_{-6.6}$\ \ & \ \ $^{+116}_{ -62}$\ \ & \ \ $^{+ 3.5}_{-4.1}$\ \ & \ \ $^{+ 7.6}_{-7.0}$\ \ & \ \ $^{+ 7.9}_{-8.2}$\ \ & \ \ $^{ 0.0}_{+ 0.0}$\ \ & \ \ $^{+ 0.0}_{-0.0}$\ \ & \ \ $^{+ 4.6}_{-5.1}$\ \ & \ \ $^{-0.0}_{+ 0.0}$\ \ & \ \ $^{ +21}_{ -18}$\ \ & \ \ $\pm31.$\ \ & \ \ $\pm 0.0$\ \ & \ \ $\mp 0.3$\ \ & \ \ $\pm 1.5$\ \ & \ \ $\pm 2.0$\ \ & \ \ $\pm 2.0$\ \ & \ \ $\pm 0.5$\ \ \\
\ \ 30--45\ \ &\ \  0.78\!\!$\begin{array}{r} + 0.14\\- 0.09\end{array}$\ \ &\ \ 1.330e+00\ \ & \ \ 6.4\ \ &\ \ $^{+7.2}_{-13.}$\ \ & \ \ $^{ +28}_{ -28}$\ \ & \ \ $^{ +16}_{ -19}$\ \ & \ \ $^{+ 3.7}_{-4.2}$\ \ & \ \ $^{+ 5.5}_{-5.6}$\ \ & \ \ $^{+218}_{ -68}$\ \ & \ \ $^{+ 7.1}_{-8.0}$\ \ & \ \ $^{+6.9}_{-13.}$\ \ & \ \ $^{ +11}_{ -16}$\ \ & \ \ $^{-0.0}_{-0.0}$\ \ & \ \ $^{-0.0}_{-0.0}$\ \ & \ \ $^{+7.6}_{-13.}$\ \ & \ \ $^{+ 0.0}_{-0.0}$\ \ & \ \ $^{ +32}_{ -30}$\ \ & \ \ $\pm17.$\ \ & \ \ $\pm 1.9$\ \ & \ \ $\pm 3.0$\ \ & \ \ $\pm 1.9$\ \ & \ \ $\pm 1.0$\ \ & \ \ $\pm 1.0$\ \ & \ \ $\pm 0.5$\ \ \\
\hline\hline\end{tabular}
\end{table*}

\begin{table*}
\caption{Measured jet cross section for anti-$k_t$ jets with $R=0.6$ in the rapidity bin $|y|<0.3$. See caption of Table~\ref{tab:xs_r04_y0} for details.}
\label{tab:xs_r06_y0}
\tiny
\centering
\begin{tabular}{@{}c@{}@{}c@{}@{}c@{}@{}c@{}@{}r@{}@{}r@{}@{}r@{}@{}r@{}@{}r@{}@{}r@{}@{}r@{}@{}r@{}@{}r@{}@{}r@{}@{}r@{}@{}r@{}@{}r@{}@{}r@{}@{}c@{}@{}c@{}@{}c@{}@{}c@{}@{}c@{}@{}c@{}@{}c@{}}
\hline\hline
 \pt &  NPCorr &  $\sigma$ & $\delta_\mathrm{stat.}$ & $\gamma_{1}$ & $\gamma_{7}$ & $\gamma_{13}$ & $\gamma_{19}$ & $\gamma_{25}$ & $\gamma_{88}$ & $\gamma_{32}$ & $\gamma_{38}$ & $\gamma_{44}$ & $\gamma_{50}$ & $\gamma_{56}$ & $\gamma_{62}$ & $\gamma_{68}$ & $\gamma_{89}$ & $\gamma_{76}$ & $\gamma_{82}$ & $\gamma_{74}$ & $\gamma_{75}$ & $\gamma_{83}$ & $u_1$ & $u_2$ \\
(GeV) & & (nb/GeV)& \% &  \%&  \%&  \%&  \%&  \%&  \%&  \%&  \%&  \%&  \%&  \%&  \%&  \%&  \%&  \%&  \%&  \%&  \%&  \%&  \%&  \%\\
\hline
\ \ 20--30\ \ &\ \  1.24\!\!$\begin{array}{r} + 0.06\\- 0.13\end{array}$\ \ &\ \ 2.542e+03\ \ & \ \ 1.4\ \ &\ \ $^{ +14}_{ -12}$\ \ & \ \ $^{ +11}_{ -11}$\ \ & \ \ $^{ +20}_{ -15}$\ \ & \ \ $^{+10.}_{-9.3}$\ \ & \ \ $^{+ 7.3}_{-5.3}$\ \ & \ \ 0.0\ \ &\ \ $^{+ 2.1}_{-2.3}$\ \ & \ \ $^{+ 6.1}_{-5.3}$\ \ & \ \ $^{+ 5.1}_{-4.8}$\ \ & \ \ $^{-0.0}_{+ 0.0}$\ \ & \ \ $^{+ 0.0}_{-0.0}$\ \ & \ \ $^{+ 1.7}_{-2.3}$\ \ & \ \ $^{-0.0}_{+ 0.0}$\ \ & \ \ 0.0\ \ &\ \ $\pm5.3$\ \ & \ \ $\mp 0.3$\ \ & \ \ $\mp 0.1$\ \ & \ \ $\pm 0.5$\ \ & \ \ $\pm 2.0$\ \ & \ \ $\pm 1.0$\ \ & \ \ $\pm 0.5$\ \ \\
\ \ 30--45\ \ &\ \  1.12\!\!$\begin{array}{r} + 0.07\\- 0.07\end{array}$\ \ &\ \ 3.004e+02\ \ & \ \ 2.4\ \ &\ \ $^{+ 8.4}_{-7.2}$\ \ & \ \ $^{+12.}_{-9.9}$\ \ & \ \ $^{+11.}_{-8.6}$\ \ & \ \ $^{+ 8.7}_{-7.5}$\ \ & \ \ $^{+ 1.7}_{-1.9}$\ \ & \ \ 0.0\ \ &\ \ $^{+ 2.6}_{-2.8}$\ \ & \ \ $^{+ 3.9}_{-3.9}$\ \ & \ \ $^{+ 5.0}_{-4.7}$\ \ & \ \ $^{-0.0}_{-0.0}$\ \ & \ \ $^{-0.0}_{-0.1}$\ \ & \ \ $^{+ 4.0}_{-4.1}$\ \ & \ \ $^{-0.0}_{+ 0.0}$\ \ & \ \ 0.0\ \ &\ \ $\pm3.6$\ \ & \ \ $\mp 0.1$\ \ & \ \ $\pm 0.0$\ \ & \ \ $\pm 0.1$\ \ & \ \ $\pm 1.0$\ \ & \ \ $\pm 1.0$\ \ & \ \ $\pm 0.5$\ \ \\
\hline\hline\end{tabular}
\end{table*}
\begin{table*}
\caption{Measured jet cross section for anti-$k_t$ jets with $R=0.6$ in the rapidity bin $0.3\leq|y|<0.8$. See caption of Table~\ref{tab:xs_r04_y0} for details.}
\label{tab:xs_r06_y1}
\tiny
\centering
\begin{tabular}{@{}c@{}@{}c@{}@{}c@{}@{}c@{}@{}r@{}@{}r@{}@{}r@{}@{}r@{}@{}r@{}@{}r@{}@{}r@{}@{}r@{}@{}r@{}@{}r@{}@{}r@{}@{}r@{}@{}r@{}@{}r@{}@{}c@{}@{}c@{}@{}c@{}@{}c@{}@{}c@{}@{}c@{}@{}c@{}}
\hline\hline
 \pt &  NPCorr &  $\sigma$ & $\delta_\mathrm{stat.}$ & $\gamma_{1}$ & $\gamma_{7}$ & $\gamma_{13}$ & $\gamma_{19}$ & $\gamma_{25}$ & $\gamma_{88}$ & $\gamma_{32}$ & $\gamma_{38}$ & $\gamma_{44}$ & $\gamma_{50}$ & $\gamma_{56}$ & $\gamma_{62}$ & $\gamma_{68}$ & $\gamma_{89}$ & $\gamma_{76}$ & $\gamma_{82}$ & $\gamma_{74}$ & $\gamma_{75}$ & $\gamma_{83}$ & $u_1$ & $u_2$ \\
(GeV) & & (nb/GeV)& \% &  \%&  \%&  \%&  \%&  \%&  \%&  \%&  \%&  \%&  \%&  \%&  \%&  \%&  \%&  \%&  \%&  \%&  \%&  \%&  \%&  \%\\
\hline
\ \ 20--30\ \ &\ \  1.24\!\!$\begin{array}{r} + 0.07\\- 0.13\end{array}$\ \ &\ \ 2.446e+03\ \ & \ \ 1.1\ \ &\ \ $^{ +14}_{ -11}$\ \ & \ \ $^{+11.}_{-9.6}$\ \ & \ \ $^{ +17}_{ -13}$\ \ & \ \ $^{+ 8.8}_{-8.4}$\ \ & \ \ $^{+ 5.3}_{-4.7}$\ \ & \ \ $^{+ 0.1}_{-0.1}$\ \ & \ \ $^{+ 2.3}_{-2.3}$\ \ & \ \ $^{+ 4.8}_{-4.4}$\ \ & \ \ $^{+ 5.0}_{-4.9}$\ \ & \ \ $^{-0.0}_{+ 0.0}$\ \ & \ \ $^{+ 0.0}_{-0.0}$\ \ & \ \ $^{+ 3.0}_{-3.0}$\ \ & \ \ $^{-0.0}_{+ 0.0}$\ \ & \ \ $^{+ 0.1}_{-0.1}$\ \ & \ \ $\pm4.6$\ \ & \ \ $\mp 0.1$\ \ & \ \ $\mp 0.2$\ \ & \ \ $\pm 0.6$\ \ & \ \ $\pm 2.0$\ \ & \ \ $\pm 1.0$\ \ & \ \ $\pm 0.5$\ \ \\
\ \ 30--45\ \ &\ \  1.13\!\!$\begin{array}{r} + 0.07\\- 0.08\end{array}$\ \ &\ \ 3.025e+02\ \ & \ \ 1.9\ \ &\ \ $^{+ 8.1}_{-7.2}$\ \ & \ \ $^{+11.}_{-10.}$\ \ & \ \ $^{+10.0}_{-9.3}$\ \ & \ \ $^{+ 8.0}_{-7.3}$\ \ & \ \ $^{+ 2.7}_{-2.8}$\ \ & \ \ $^{+ 0.1}_{-0.1}$\ \ & \ \ $^{+ 2.8}_{-3.1}$\ \ & \ \ $^{+ 3.4}_{-3.4}$\ \ & \ \ $^{+ 5.2}_{-5.0}$\ \ & \ \ $^{-0.0}_{+ 0.0}$\ \ & \ \ $^{+ 0.0}_{-0.0}$\ \ & \ \ $^{+ 3.6}_{-3.7}$\ \ & \ \ $^{-0.2}_{+ 0.1}$\ \ & \ \ $^{+ 0.1}_{-0.1}$\ \ & \ \ $\pm3.3$\ \ & \ \ $\mp 0.7$\ \ & \ \ $\pm 0.1$\ \ & \ \ $\pm 0.3$\ \ & \ \ $\pm 1.0$\ \ & \ \ $\pm 1.0$\ \ & \ \ $\pm 0.5$\ \ \\
\hline\hline\end{tabular}
\end{table*}
\begin{table*}
\caption{Measured jet cross section for anti-$k_t$ jets with $R=0.6$ in the rapidity bin $0.8\leq|y|<1.2$. See caption of Table~\ref{tab:xs_r04_y0} for details.}
\label{tab:xs_r06_y2}
\tiny
\centering
\begin{tabular}{@{}c@{}@{}c@{}@{}c@{}@{}c@{}@{}r@{}@{}r@{}@{}r@{}@{}r@{}@{}r@{}@{}r@{}@{}r@{}@{}r@{}@{}r@{}@{}r@{}@{}r@{}@{}r@{}@{}r@{}@{}r@{}@{}c@{}@{}c@{}@{}c@{}@{}c@{}@{}c@{}@{}c@{}@{}c@{}}
\hline\hline
 \pt &  NPCorr &  $\sigma$ & $\delta_\mathrm{stat.}$ & $\gamma_{2}$ & $\gamma_{8}$ & $\gamma_{14}$ & $\gamma_{20}$ & $\gamma_{26}$ & $\gamma_{88}$ & $\gamma_{33}$ & $\gamma_{39}$ & $\gamma_{45}$ & $\gamma_{51}$ & $\gamma_{57}$ & $\gamma_{63}$ & $\gamma_{69}$ & $\gamma_{89}$ & $\gamma_{77}$ & $\gamma_{82}$ & $\gamma_{74}$ & $\gamma_{75}$ & $\gamma_{83}$ & $u_1$ & $u_2$ \\
(GeV) & & (nb/GeV)& \% &  \%&  \%&  \%&  \%&  \%&  \%&  \%&  \%&  \%&  \%&  \%&  \%&  \%&  \%&  \%&  \%&  \%&  \%&  \%&  \%&  \%\\
\hline
\ \ 20--30\ \ &\ \  1.25\!\!$\begin{array}{r} + 0.08\\- 0.12\end{array}$\ \ &\ \ 2.275e+03\ \ & \ \ 1.3\ \ &\ \ $^{ +12}_{ -10}$\ \ & \ \ $^{+ 8.6}_{-8.7}$\ \ & \ \ $^{ +14}_{ -12}$\ \ & \ \ $^{+10.}_{-8.2}$\ \ & \ \ $^{+ 4.7}_{-4.0}$\ \ & \ \ $^{+ 3.8}_{-3.9}$\ \ & \ \ $^{+ 1.9}_{-2.0}$\ \ & \ \ $^{+ 4.3}_{-3.8}$\ \ & \ \ $^{+ 4.4}_{-4.2}$\ \ & \ \ $^{-0.0}_{+ 0.0}$\ \ & \ \ $^{+ 0.0}_{-0.0}$\ \ & \ \ $^{+ 2.5}_{-2.7}$\ \ & \ \ $^{-0.0}_{+ 0.0}$\ \ & \ \ $^{+ 3.2}_{-3.3}$\ \ & \ \ $\pm10.$\ \ & \ \ $\pm 1.3$\ \ & \ \ $\mp 0.3$\ \ & \ \ $\pm 0.4$\ \ & \ \ $\pm 2.0$\ \ & \ \ $\pm 1.0$\ \ & \ \ $\pm 0.5$\ \ \\
\ \ 30--45\ \ &\ \  1.13\!\!$\begin{array}{r} + 0.07\\- 0.07\end{array}$\ \ &\ \ 2.557e+02\ \ & \ \ 2.2\ \ &\ \ $^{+ 8.2}_{-7.6}$\ \ & \ \ $^{+11.}_{-10.}$\ \ & \ \ $^{+10.}_{-9.3}$\ \ & \ \ $^{+ 5.0}_{-4.8}$\ \ & \ \ $^{+ 3.0}_{-2.9}$\ \ & \ \ $^{+ 4.8}_{-4.6}$\ \ & \ \ $^{+ 3.0}_{-3.2}$\ \ & \ \ $^{+ 3.7}_{-3.6}$\ \ & \ \ $^{+ 5.5}_{-5.3}$\ \ & \ \ $^{-0.0}_{-0.0}$\ \ & \ \ $^{+ 0.0}_{-0.1}$\ \ & \ \ $^{+ 3.9}_{-3.9}$\ \ & \ \ $^{-0.1}_{+ 0.2}$\ \ & \ \ $^{+ 4.3}_{-4.2}$\ \ & \ \ $\pm7.6$\ \ & \ \ $\pm 0.0$\ \ & \ \ $\pm 0.1$\ \ & \ \ $\pm 0.2$\ \ & \ \ $\pm 1.0$\ \ & \ \ $\pm 1.0$\ \ & \ \ $\pm 0.5$\ \ \\
\hline\hline\end{tabular}
\end{table*}
\begin{table*}
\caption{Measured jet cross section for anti-$k_t$ jets with $R=0.6$ in the rapidity bin $1.2\leq|y|<2.1$. See caption of Table~\ref{tab:xs_r04_y0} for details.}
\label{tab:xs_r06_y3}
\tiny
\centering
\begin{tabular}{@{}c@{}@{}c@{}@{}c@{}@{}c@{}@{}r@{}@{}r@{}@{}r@{}@{}r@{}@{}r@{}@{}r@{}@{}r@{}@{}r@{}@{}r@{}@{}r@{}@{}r@{}@{}r@{}@{}r@{}@{}r@{}@{}c@{}@{}c@{}@{}c@{}@{}c@{}@{}c@{}@{}c@{}@{}c@{}}
\hline\hline
 \pt &  NPCorr &  $\sigma$ & $\delta_\mathrm{stat.}$ & $\gamma_{3}$ & $\gamma_{9}$ & $\gamma_{15}$ & $\gamma_{21}$ & $\gamma_{27}$ & $\gamma_{88}$ & $\gamma_{34}$ & $\gamma_{40}$ & $\gamma_{46}$ & $\gamma_{52}$ & $\gamma_{58}$ & $\gamma_{64}$ & $\gamma_{70}$ & $\gamma_{89}$ & $\gamma_{78}$ & $\gamma_{82}$ & $\gamma_{74}$ & $\gamma_{75}$ & $\gamma_{83}$ & $u_1$ & $u_2$ \\
(GeV) & & (nb/GeV)& \% &  \%&  \%&  \%&  \%&  \%&  \%&  \%&  \%&  \%&  \%&  \%&  \%&  \%&  \%&  \%&  \%&  \%&  \%&  \%&  \%&  \%\\
\hline
\ \ 20--30\ \ &\ \  1.24\!\!$\begin{array}{r} + 0.09\\- 0.11\end{array}$\ \ &\ \ 1.775e+03\ \ & \ \  1.0\ \ &\ \ $^{ +14}_{ -11}$\ \ & \ \ $^{+10.}_{-9.8}$\ \ & \ \ $^{ +17}_{ -13}$\ \ & \ \ $^{ +14}_{ -10}$\ \ & \ \ $^{+ 5.2}_{-4.7}$\ \ & \ \ $^{ +14}_{ -14}$\ \ & \ \ $^{+ 2.3}_{-2.4}$\ \ & \ \ $^{+ 4.8}_{-4.3}$\ \ & \ \ $^{+ 4.9}_{-4.8}$\ \ & \ \ $^{-0.0}_{+ 0.0}$\ \ & \ \ $^{+ 0.0}_{-0.0}$\ \ & \ \ $^{+ 3.0}_{-3.1}$\ \ & \ \ $^{-0.0}_{+ 0.0}$\ \ & \ \ $^{+ 5.7}_{-5.7}$\ \ & \ \ $\pm13.$\ \ & \ \ $\mp 0.4$\ \ & \ \ $\mp 0.1$\ \ & \ \ $\pm 0.4$\ \ & \ \ $\pm 2.0$\ \ & \ \ $\pm 1.0$\ \ & \ \ $\pm 0.5$\ \ \\
\ \ 30--45\ \ &\ \  1.12\!\!$\begin{array}{r} + 0.08\\- 0.06\end{array}$\ \ &\ \ 2.067e+02\ \ & \ \ 1.8\ \ &\ \ $^{+ 8.4}_{-7.5}$\ \ & \ \ $^{+11.}_{-9.9}$\ \ & \ \ $^{+10.}_{-9.2}$\ \ & \ \ $^{+ 4.9}_{-4.6}$\ \ & \ \ $^{+ 3.2}_{-2.7}$\ \ & \ \ $^{ +15}_{ -13}$\ \ & \ \ $^{+ 3.3}_{-2.9}$\ \ & \ \ $^{+ 3.8}_{-3.5}$\ \ & \ \ $^{+ 5.3}_{-5.1}$\ \ & \ \ $^{+ 0.0}_{+ 0.0}$\ \ & \ \ $^{+ 0.0}_{-0.0}$\ \ & \ \ $^{+ 4.0}_{-3.8}$\ \ & \ \ $^{-0.1}_{+ 0.1}$\ \ & \ \ $^{+ 6.6}_{-6.3}$\ \ & \ \ $\pm6.8$\ \ & \ \ $\mp 0.2$\ \ & \ \ $\pm 0.1$\ \ & \ \ $\pm 0.2$\ \ & \ \ $\pm 1.0$\ \ & \ \ $\pm 1.0$\ \ & \ \ $\pm 0.5$\ \ \\
\hline\hline\end{tabular}
\end{table*}
\begin{table*}
\caption{Measured jet cross section for anti-$k_t$ jets with $R=0.6$ in the rapidity bin $2.1\leq|y|<2.8$. See caption of Table~\ref{tab:xs_r04_y0} for details.}
\label{tab:xs_r06_y4}
\tiny
\centering
\begin{tabular}{@{}c@{}@{}c@{}@{}c@{}@{}c@{}@{}r@{}@{}r@{}@{}r@{}@{}r@{}@{}r@{}@{}r@{}@{}r@{}@{}r@{}@{}r@{}@{}r@{}@{}r@{}@{}r@{}@{}r@{}@{}r@{}@{}c@{}@{}c@{}@{}c@{}@{}c@{}@{}c@{}@{}c@{}@{}c@{}}
\hline\hline
 \pt &  NPCorr &  $\sigma$ & $\delta_\mathrm{stat.}$ & $\gamma_{4}$ & $\gamma_{10}$ & $\gamma_{16}$ & $\gamma_{22}$ & $\gamma_{28}$ & $\gamma_{88}$ & $\gamma_{35}$ & $\gamma_{41}$ & $\gamma_{47}$ & $\gamma_{53}$ & $\gamma_{59}$ & $\gamma_{65}$ & $\gamma_{71}$ & $\gamma_{89}$ & $\gamma_{79}$ & $\gamma_{82}$ & $\gamma_{74}$ & $\gamma_{75}$ & $\gamma_{84}$ & $u_1$ & $u_2$ \\
(GeV) & & (nb/GeV)& \% &  \%&  \%&  \%&  \%&  \%&  \%&  \%&  \%&  \%&  \%&  \%&  \%&  \%&  \%&  \%&  \%&  \%&  \%&  \%&  \%&  \%\\
\hline
\ \ 20--30\ \ &\ \  1.19\!\!$\begin{array}{r} + 0.13\\- 0.07\end{array}$\ \ &\ \ 1.154e+03\ \ & \ \ 1.5\ \ &\ \ $^{ +16}_{ -12}$\ \ & \ \ $^{ +12}_{ -10}$\ \ & \ \ $^{ +20}_{ -14}$\ \ & \ \ $^{+ 6.4}_{-6.0}$\ \ & \ \ $^{+ 5.9}_{-5.0}$\ \ & \ \ $^{ +26}_{ -21}$\ \ & \ \ $^{+ 2.2}_{-2.6}$\ \ & \ \ $^{+ 5.4}_{-4.7}$\ \ & \ \ $^{+ 5.4}_{-5.1}$\ \ & \ \ $^{-0.0}_{+ 0.0}$\ \ & \ \ $^{+ 0.0}_{-0.0}$\ \ & \ \ $^{+ 3.0}_{-3.2}$\ \ & \ \ $^{-0.0}_{+ 0.0}$\ \ & \ \ $^{+ 4.5}_{-4.5}$\ \ & \ \ $\pm20.$\ \ & \ \ $\mp 0.7$\ \ & \ \ $\pm 0.0$\ \ & \ \ $\pm 0.3$\ \ & \ \ $\pm 2.0$\ \ & \ \ $\pm 1.0$\ \ & \ \ $\pm 0.5$\ \ \\
\ \ 30--45\ \ &\ \  1.09\!\!$\begin{array}{r} + 0.09\\- 0.03\end{array}$\ \ &\ \ 1.225e+02\ \ & \ \ 1.3\ \ &\ \ $^{+ 8.4}_{-7.4}$\ \ & \ \ $^{ +11}_{ -10}$\ \ & \ \ $^{+10.}_{-9.6}$\ \ & \ \ $^{+ 6.0}_{-5.8}$\ \ & \ \ $^{+ 3.0}_{-2.9}$\ \ & \ \ $^{ +23}_{ -19}$\ \ & \ \ $^{+ 3.1}_{-3.1}$\ \ & \ \ $^{+ 3.6}_{-3.6}$\ \ & \ \ $^{+ 5.4}_{-5.4}$\ \ & \ \ $^{+ 0.0}_{+ 0.0}$\ \ & \ \ $^{+ 0.0}_{-0.0}$\ \ & \ \ $^{+ 3.8}_{-3.9}$\ \ & \ \ $^{-0.1}_{+ 0.1}$\ \ & \ \ $^{+ 4.6}_{-4.8}$\ \ & \ \ $\pm5.7$\ \ & \ \ $\mp 1.1$\ \ & \ \ $\pm 0.3$\ \ & \ \ $\pm 0.2$\ \ & \ \ $\pm 1.0$\ \ & \ \ $\pm 1.0$\ \ & \ \ $\pm 0.5$\ \ \\
\hline\hline\end{tabular}
\end{table*}
\begin{table*}
\caption{Measured jet cross section for anti-$k_t$ jets with $R=0.6$ in the rapidity bin $2.8\leq|y|<3.6$. See caption of Table~\ref{tab:xs_r04_y0} for details.}
\label{tab:xs_r06_y5}
\tiny
\centering
\begin{tabular}{@{}c@{}@{}c@{}@{}c@{}@{}c@{}@{}r@{}@{}r@{}@{}r@{}@{}r@{}@{}r@{}@{}r@{}@{}r@{}@{}r@{}@{}r@{}@{}r@{}@{}r@{}@{}r@{}@{}r@{}@{}r@{}@{}c@{}@{}c@{}@{}c@{}@{}c@{}@{}c@{}@{}c@{}@{}c@{}}
\hline\hline
 \pt &  NPCorr &  $\sigma$ & $\delta_\mathrm{stat.}$ & $\gamma_{5}$ & $\gamma_{11}$ & $\gamma_{17}$ & $\gamma_{23}$ & $\gamma_{29}$ & $\gamma_{88}$ & $\gamma_{36}$ & $\gamma_{42}$ & $\gamma_{48}$ & $\gamma_{54}$ & $\gamma_{60}$ & $\gamma_{66}$ & $\gamma_{72}$ & $\gamma_{89}$ & $\gamma_{80}$ & $\gamma_{82}$ & $\gamma_{74}$ & $\gamma_{75}$ & $\gamma_{85}$ & $u_1$ & $u_2$ \\
(GeV) & & (nb/GeV)& \% &  \%&  \%&  \%&  \%&  \%&  \%&  \%&  \%&  \%&  \%&  \%&  \%&  \%&  \%&  \%&  \%&  \%&  \%&  \%&  \%&  \%\\
\hline
\ \ 20--30\ \ &\ \  1.12\!\!$\begin{array}{r} + 0.16\\- 0.06\end{array}$\ \ &\ \ 5.583e+02\ \ & \ \ 1.8\ \ &\ \ $^{ +17}_{ -13}$\ \ & \ \ $^{ +12}_{ -11}$\ \ & \ \ $^{ +21}_{ -15}$\ \ & \ \ $^{+ 7.2}_{-6.4}$\ \ & \ \ $^{+ 6.2}_{-5.2}$\ \ & \ \ $^{ +38}_{ -30}$\ \ & \ \ $^{+ 2.7}_{-2.7}$\ \ & \ \ $^{+ 5.7}_{-4.9}$\ \ & \ \ $^{+ 5.8}_{-5.3}$\ \ & \ \ $^{-0.0}_{+ 0.0}$\ \ & \ \ $^{+ 0.0}_{-0.0}$\ \ & \ \ $^{+ 3.4}_{-3.4}$\ \ & \ \ $^{-0.0}_{+ 0.0}$\ \ & \ \ $^{+ 7.3}_{-6.7}$\ \ & \ \ $\pm21.$\ \ & \ \ $\mp 0.4$\ \ & \ \ $\pm 0.1$\ \ & \ \ $\pm 0.4$\ \ & \ \ $\pm 2.0$\ \ & \ \ $\pm 2.0$\ \ & \ \ $\pm 0.5$\ \ \\
\ \ 30--45\ \ &\ \  1.05\!\!$\begin{array}{r} + 0.12\\- 0.04\end{array}$\ \ &\ \ 4.400e+01\ \ & \ \ 2.0\ \ &\ \ $^{+ 8.9}_{-8.6}$\ \ & \ \ $^{ +13}_{ -12}$\ \ & \ \ $^{ +12}_{ -11}$\ \ & \ \ $^{+ 6.4}_{-6.1}$\ \ & \ \ $^{+ 3.5}_{-3.4}$\ \ & \ \ $^{ +43}_{ -30}$\ \ & \ \ $^{+ 3.8}_{-3.7}$\ \ & \ \ $^{+ 4.2}_{-4.0}$\ \ & \ \ $^{+ 6.0}_{-6.2}$\ \ & \ \ $^{+ 0.0}_{-0.0}$\ \ & \ \ $^{+ 0.0}_{-0.0}$\ \ & \ \ $^{+ 4.5}_{-4.4}$\ \ & \ \ $^{-0.1}_{+ 0.1}$\ \ & \ \ $^{+ 7.7}_{-7.8}$\ \ & \ \ $\pm10.$\ \ & \ \ $\pm 0.0$\ \ & \ \ $\pm 0.5$\ \ & \ \ $\pm 0.3$\ \ & \ \ $\pm 1.0$\ \ & \ \ $\pm 2.0$\ \ & \ \ $\pm 0.5$\ \ \\
\hline\hline\end{tabular}
\end{table*}
\begin{table*}
\caption{Measured jet cross section for anti-$k_t$ jets with $R=0.6$ in the rapidity bin $3.6\leq|y|<4.4$. See caption of Table~\ref{tab:xs_r04_y0} for details.}
\label{tab:xs_r06_y6}
\tiny
\centering
\begin{tabular}{@{}c@{}@{}c@{}@{}c@{}@{}c@{}@{}r@{}@{}r@{}@{}r@{}@{}r@{}@{}r@{}@{}r@{}@{}r@{}@{}r@{}@{}r@{}@{}r@{}@{}r@{}@{}r@{}@{}r@{}@{}r@{}@{}c@{}@{}c@{}@{}c@{}@{}c@{}@{}c@{}@{}c@{}@{}c@{}}
\hline\hline
 \pt &  NPCorr &  $\sigma$ & $\delta_\mathrm{stat.}$ & $\gamma_{6}$ & $\gamma_{12}$ & $\gamma_{18}$ & $\gamma_{24}$ & $\gamma_{30}$ & $\gamma_{88}$ & $\gamma_{37}$ & $\gamma_{43}$ & $\gamma_{49}$ & $\gamma_{55}$ & $\gamma_{61}$ & $\gamma_{67}$ & $\gamma_{73}$ & $\gamma_{89}$ & $\gamma_{81}$ & $\gamma_{82}$ & $\gamma_{74}$ & $\gamma_{75}$ & $\gamma_{86}$ & $u_1$ & $u_2$ \\
(GeV) & & (nb/GeV)& \% &  \%&  \%&  \%&  \%&  \%&  \%&  \%&  \%&  \%&  \%&  \%&  \%&  \%&  \%&  \%&  \%&  \%&  \%&  \%&  \%&  \%\\
\hline
\ \ 20--30\ \ &\ \  0.99\!\!$\begin{array}{r} + 0.19\\- 0.06\end{array}$\ \ &\ \ 8.729e+01\ \ & \ \ 4.6\ \ &\ \ $^{ +26}_{ -16}$\ \ & \ \ $^{ +19}_{ -15}$\ \ & \ \ $^{ +31}_{ -19}$\ \ & \ \ $^{+ 5.7}_{-3.9}$\ \ & \ \ $^{+ 9.2}_{-6.2}$\ \ & \ \ $^{ +93}_{ -54}$\ \ & \ \ $^{+ 4.9}_{-3.4}$\ \ & \ \ $^{+ 8.5}_{-5.9}$\ \ & \ \ $^{+ 8.9}_{-6.7}$\ \ & \ \ $^{-0.0}_{+ 0.0}$\ \ & \ \ $^{-0.0}_{-0.0}$\ \ & \ \ $^{+ 5.9}_{-4.3}$\ \ & \ \ $^{-0.0}_{+ 0.0}$\ \ & \ \ $^{ +15}_{ -12}$\ \ & \ \ $\pm25.$\ \ & \ \ $\mp 2.5$\ \ & \ \ $\pm 1.4$\ \ & \ \ $\pm 0.6$\ \ & \ \ $\pm 2.0$\ \ & \ \ $\pm 2.0$\ \ & \ \ $\pm 0.5$\ \ \\
\ \ 30--45\ \ &\ \  0.95\!\!$\begin{array}{r} + 0.13\\- 0.08\end{array}$\ \ &\ \ 2.220e+00\ \ & \ \ 5.0\ \ &\ \ $^{ +21}_{ -14}$\ \ & \ \ $^{ +29}_{ -20}$\ \ & \ \ $^{ +26}_{ -19}$\ \ & \ \ $^{+ 8.2}_{-5.3}$\ \ & \ \ $^{+ 8.8}_{-6.1}$\ \ & \ \ $^{+159}_{ -56}$\ \ & \ \ $^{+10.0}_{-6.8}$\ \ & \ \ $^{+ 9.8}_{-7.6}$\ \ & \ \ $^{ +14}_{ -11}$\ \ & \ \ $^{+ 0.0}_{-0.0}$\ \ & \ \ $^{-0.0}_{+ 0.0}$\ \ & \ \ $^{+11.}_{-8.2}$\ \ & \ \ $^{+ 0.0}_{-0.0}$\ \ & \ \ $^{ +28}_{ -18}$\ \ & \ \ $\pm14.$\ \ & \ \ $\mp 2.4$\ \ & \ \ $\pm 5.4$\ \ & \ \ $\pm 1.5$\ \ & \ \ $\pm 1.0$\ \ & \ \ $\pm 1.0$\ \ & \ \ $\pm 0.5$\ \ \\
\hline\hline\end{tabular}
\end{table*}

\begin{table*}
\caption{Measured jet cross section ratio in bins of \xt{} for anti-$k_t$ jets with $R=0.4$ in the rapidity bin $|y|<0.3$. NPCorr stands for multiplicative non-perturbative corrections, $\rho$ is the measured cross section ratio and $\delta_\mathrm{stat.}$ is the statistical uncertainty. $\gamma_i$ and $u_j$ correspond to the correlated and uncorrelated systematic uncertainties given in \%. They are described in Table~\ref{tab:uncert}, where $i$ in $\gamma_i$ denotes a nuisance parameter. $u_3$ is the uncertainty due to pile-up effects in the cross section measurement at $\sqrt{s}=7$ \TeV{}. For each table entry, the outcome of shifting the corresponding nuisance parameter by one standard deviation, up or down, is shown as superscript or subscript, respectively. In some bins these shifts may lead to cross-section variations in the same direction. The 4.3\% uncertainty from the luminosity measurements is not in this table.}
\label{tab:rxt_r04_y0}
\tiny
\centering
\begin{tabular}{@{}c@{}@{}c@{}@{}c@{}@{}c@{}@{}r@{}@{}r@{}@{}r@{}@{}r@{}@{}r@{}@{}r@{}@{}r@{}@{}r@{}@{}r@{}@{}r@{}@{}r@{}@{}r@{}@{}r@{}@{}r@{}@{}c@{}@{}c@{}@{}c@{}@{}c@{}@{}c@{}@{}c@{}@{}c@{}@{}c@{}}
\hline\hline
 \xt &  NPCorr &  $\rho$ & $\delta_\mathrm{stat.}$ & $\gamma_{1}$ & $\gamma_{7}$ & $\gamma_{13}$ & $\gamma_{19}$ & $\gamma_{25}$ & $\gamma_{31}$ & $\gamma_{32}$ & $\gamma_{38}$ & $\gamma_{44}$ & $\gamma_{50}$ & $\gamma_{56}$ & $\gamma_{62}$ & $\gamma_{68}$ & $\gamma_{76}$ & $\gamma_{82}$ & $\gamma_{74}$ & $\gamma_{75}$ & $\gamma_{83}$ & $u_1$ & $u_2$ & $u_3$ \\
 & & & \% &  \%&  \%&  \%&  \%&  \%&  \%&  \%&  \%&  \%&  \%&  \%&  \%&  \%&  \%&  \%&  \%&  \%&  \%&  \%&  \%&  \%\\
\hline
\ \ (1.71--2.29) e-02\ \ &\ \  0.90\!\!$\begin{array}{r} + 0.03\\- 0.06\end{array}$\ \ &\ \ 1.210\ \ & \ \ 2.5\ \ &\ \ $^{+6.5}_{-5.7}$\ \ & \ \ $^{+2.6}_{-2.5}$\ \ & \ \ $^{+3.5}_{-3.5}$\ \ & \ \ $^{+7.6}_{-8.2}$\ \ & \ \ $^{+1.2}_{-1.0}$\ \ & \ \ 0.0\ \ &\ \ $^{-1.1}_{+ 1.0}$\ \ & \ \ $^{+2.9}_{-3.1}$\ \ & \ \ $^{+ 0.3}_{-0.7}$\ \ & \ \ $^{-0.4}_{+ 0.3}$\ \ & \ \ $^{-0.0}_{-0.1}$\ \ & \ \ $^{-2.1}_{+1.4}$\ \ & \ \ $^{+ 0.1}_{-0.1}$\ \ & \ \ $\mp 5.0$\ \ & \ \ $\pm 0.2$\ \ & \ \ $\pm 0.1$\ \ & \ \ $\pm 0.4$\ \ & \ \ $\pm 0.9$\ \ & \ \ $\pm 1.4$\ \ & \ \ $^{+ 0.6}_{-0.5}$\ \ & \ \ $\pm 0.3$\ \ \\
\ \ (2.29--3.14) e-02\ \ &\ \  0.92\!\!$\begin{array}{r} + 0.03\\- 0.05\end{array}$\ \ &\ \ 1.289\ \ & \ \ 2.9\ \ &\ \ $^{+4.1}_{-3.9}$\ \ & \ \ $^{+2.1}_{-2.4}$\ \ & \ \ $^{+2.2}_{-1.9}$\ \ & \ \ $^{+7.0}_{-6.9}$\ \ & \ \ $^{-0.5}_{+ 1.1}$\ \ & \ \ 0.0\ \ &\ \ $^{-0.0}_{+ 0.5}$\ \ & \ \ $^{+2.7}_{-2.0}$\ \ & \ \ $^{+ 0.3}_{+ 0.6}$\ \ & \ \ $^{-0.5}_{+ 1.0}$\ \ & \ \ $^{+ 0.0}_{+ 0.3}$\ \ & \ \ $^{+1.6}_{-1.5}$\ \ & \ \ $^{+ 0.6}_{-0.2}$\ \ & \ \ $\mp 0.3$\ \ & \ \ $\pm 0.6$\ \ & \ \ $\pm 0.0$\ \ & \ \ $\pm 2.0$\ \ & \ \ 0.0 \ \ & \ \ $\pm 1.4$\ \ & \ \ $^{+ 0.5}_{-0.5}$\ \ & \ \ $\pm 0.4$\ \ \\
\hline\hline\end{tabular}
\end{table*}
\begin{table*}
\caption{Measured jet cross section ratio in \xt{} bin for anti-$k_t$ jets with $R=0.4$ in the rapidity bin $0.3\leq|y|<0.8$. See caption of Table~\ref{tab:rxt_r04_y0} for details.}
\label{tab:rxt_r04_y1}
\tiny
\centering
\begin{tabular}{@{}c@{}@{}c@{}@{}c@{}@{}c@{}@{}r@{}@{}r@{}@{}r@{}@{}r@{}@{}r@{}@{}r@{}@{}r@{}@{}r@{}@{}r@{}@{}r@{}@{}r@{}@{}r@{}@{}r@{}@{}r@{}@{}c@{}@{}c@{}@{}c@{}@{}c@{}@{}c@{}@{}c@{}@{}c@{}@{}c@{}}
\hline\hline
 \xt &  NPCorr &  $\rho$ & $\delta_\mathrm{stat.}$ & $\gamma_{1}$ & $\gamma_{7}$ & $\gamma_{13}$ & $\gamma_{19}$ & $\gamma_{25}$ & $\gamma_{31}$ & $\gamma_{32}$ & $\gamma_{38}$ & $\gamma_{44}$ & $\gamma_{50}$ & $\gamma_{56}$ & $\gamma_{62}$ & $\gamma_{68}$ & $\gamma_{76}$ & $\gamma_{82}$ & $\gamma_{74}$ & $\gamma_{75}$ & $\gamma_{83}$ & $u_1$ & $u_2$ & $u_3$ \\
 & & & \% &  \%&  \%&  \%&  \%&  \%&  \%&  \%&  \%&  \%&  \%&  \%&  \%&  \%&  \%&  \%&  \%&  \%&  \%&  \%&  \%&  \%\\
\hline
\ \ (1.71--2.29) e-02\ \ &\ \  0.90\!\!$\begin{array}{r} + 0.04\\- 0.06\end{array}$\ \ &\ \ 1.278\ \ & \ \ 1.9\ \ &\ \ $^{+7.1}_{-5.7}$\ \ & \ \ $^{+4.5}_{-6.5}$\ \ & \ \ $^{+2.8}_{-2.4}$\ \ & \ \ $^{+5.7}_{-6.5}$\ \ & \ \ $^{+ 0.9}_{-1.1}$\ \ & \ \ $^{+ 0.1}_{-0.1}$\ \ & \ \ $^{-0.7}_{+ 0.5}$\ \ & \ \ $^{+2.4}_{-2.5}$\ \ & \ \ $^{+ 0.1}_{-0.5}$\ \ & \ \ $^{-0.5}_{+ 0.2}$\ \ & \ \ $^{-0.0}_{-0.1}$\ \ & \ \ $^{+2.2}_{-2.4}$\ \ & \ \ $^{-1.6}_{+1.4}$\ \ & \ \ $\mp 5.1$\ \ & \ \ $\pm 0.2$\ \ & \ \ $\pm 0.1$\ \ & \ \ $\pm 0.1$\ \ & \ \ $\pm 0.9$\ \ & \ \ $\pm 1.4$\ \ & \ \ $^{+ 0.5}_{-0.5}$\ \ & \ \ $\pm 0.4$\ \ \\
\ \ (2.29--3.14) e-02\ \ &\ \  0.92\!\!$\begin{array}{r} + 0.04\\- 0.05\end{array}$\ \ &\ \ 1.309\ \ & \ \ 2.3\ \ &\ \ $^{+4.2}_{-3.9}$\ \ & \ \ $^{+5.4}_{-4.8}$\ \ & \ \ $^{+2.9}_{-2.4}$\ \ & \ \ $^{+6.3}_{-5.1}$\ \ & \ \ $^{+ 1.4}_{-0.8}$\ \ & \ \ $^{+ 0.1}_{-0.1}$\ \ & \ \ $^{+ 0.2}_{+ 0.1}$\ \ & \ \ $^{+2.3}_{-1.8}$\ \ & \ \ $^{+ 0.1}_{+ 0.3}$\ \ & \ \ $^{-0.7}_{+ 1.2}$\ \ & \ \ $^{+ 0.0}_{+ 0.3}$\ \ & \ \ $^{+3.7}_{-3.7}$\ \ & \ \ $^{-3.6}_{+4.0}$\ \ & \ \ $\mp 1.6$\ \ & \ \ $\pm 0.1$\ \ & \ \ $\pm 0.1$\ \ & \ \ $\pm 0.3$\ \ & \ \ 0.0 \ \ & \ \ $\pm 1.4$\ \ & \ \ $^{+ 0.5}_{-0.5}$\ \ & \ \ $\pm 0.4$\ \ \\
\hline\hline\end{tabular}
\end{table*}
\begin{table*}
\caption{Measured jet cross section ratio in \xt{} bin for anti-$k_t$ jets with $R=0.4$ in the rapidity bin $0.8\leq|y|<1.2$. See caption of Table~\ref{tab:rxt_r04_y0} for details.}
\label{tab:rxt_r04_y2}
\tiny
\centering
\begin{tabular}{@{}c@{}@{}c@{}@{}c@{}@{}c@{}@{}r@{}@{}r@{}@{}r@{}@{}r@{}@{}r@{}@{}r@{}@{}r@{}@{}r@{}@{}r@{}@{}r@{}@{}r@{}@{}r@{}@{}r@{}@{}r@{}@{}c@{}@{}c@{}@{}c@{}@{}c@{}@{}c@{}@{}c@{}@{}c@{}@{}c@{}}
\hline\hline
 \xt &  NPCorr &  $\rho$ & $\delta_\mathrm{stat.}$ & $\gamma_{2}$ & $\gamma_{8}$ & $\gamma_{14}$ & $\gamma_{20}$ & $\gamma_{26}$ & $\gamma_{31}$ & $\gamma_{33}$ & $\gamma_{39}$ & $\gamma_{45}$ & $\gamma_{51}$ & $\gamma_{57}$ & $\gamma_{63}$ & $\gamma_{69}$ & $\gamma_{77}$ & $\gamma_{82}$ & $\gamma_{74}$ & $\gamma_{75}$ & $\gamma_{83}$ & $u_1$ & $u_2$ & $u_3$ \\
 & & & \% &  \%&  \%&  \%&  \%&  \%&  \%&  \%&  \%&  \%&  \%&  \%&  \%&  \%&  \%&  \%&  \%&  \%&  \%&  \%&  \%&  \%\\
\hline
\ \ (1.71--2.29) e-02\ \ &\ \  0.90\!\!$\begin{array}{r} + 0.04\\- 0.06\end{array}$\ \ &\ \ 1.158\ \ & \ \ 2.3\ \ &\ \ $^{+6.9}_{-6.0}$\ \ & \ \ $^{+4.8}_{-6.2}$\ \ & \ \ $^{+2.7}_{-2.7}$\ \ & \ \ $^{+6.3}_{-5.1}$\ \ & \ \ $^{+ 0.9}_{-0.9}$\ \ & \ \ $^{+3.1}_{-3.7}$\ \ & \ \ $^{-0.4}_{+ 0.6}$\ \ & \ \ $^{+2.2}_{-2.4}$\ \ & \ \ $^{+ 0.1}_{-0.6}$\ \ & \ \ $^{-0.4}_{+ 0.2}$\ \ & \ \ $^{-0.0}_{-0.1}$\ \ & \ \ $^{+2.2}_{-2.2}$\ \ & \ \ $^{-1.7}_{+1.6}$\ \ & \ \ $\mp  11$\ \ & \ \ $\pm 0.5$\ \ & \ \ $\pm 0.2$\ \ & \ \ $\pm 0.2$\ \ & \ \ $\pm 0.9$\ \ & \ \ $\pm 1.4$\ \ & \ \ $^{+ 0.6}_{-0.5}$\ \ & \ \ $\pm 0.5$\ \ \\
\ \ (2.29--3.14) e-02\ \ &\ \  0.93\!\!$\begin{array}{r} + 0.03\\- 0.04\end{array}$\ \ &\ \ 1.209\ \ & \ \ 2.7\ \ &\ \ $^{+4.3}_{-3.8}$\ \ & \ \ $^{+5.7}_{-4.6}$\ \ & \ \ $^{+2.5}_{-2.1}$\ \ & \ \ $^{+ 0.9}_{+ 0.1}$\ \ & \ \ $^{+ 1.3}_{-0.6}$\ \ & \ \ $^{+4.4}_{-4.2}$\ \ & \ \ $^{+ 0.0}_{+ 0.4}$\ \ & \ \ $^{+2.2}_{-1.8}$\ \ & \ \ $^{+ 0.2}_{+ 0.2}$\ \ & \ \ $^{-1.0}_{+ 1.3}$\ \ & \ \ $^{+ 0.1}_{+ 0.3}$\ \ & \ \ $^{+3.9}_{-3.7}$\ \ & \ \ $^{-3.9}_{+4.2}$\ \ & \ \ $\mp 3.5$\ \ & \ \ $\mp 0.1$\ \ & \ \ $\pm 0.2$\ \ & \ \ $\pm 0.5$\ \ & \ \ 0.0 \ \ & \ \ $\pm 1.4$\ \ & \ \ $^{+ 0.5}_{-0.5}$\ \ & \ \ $\pm 0.8$\ \ \\
\hline\hline\end{tabular}
\end{table*}
\begin{table*}
\caption{Measured jet cross section ratio in \xt{} bin for anti-$k_t$ jets with $R=0.4$ in the rapidity bin $1.2\leq|y|<2.1$. See caption of Table~\ref{tab:rxt_r04_y0} for details.}
\label{tab:rxt_r04_y3}
\tiny
\centering
\begin{tabular}{@{}c@{}@{}c@{}@{}c@{}@{}c@{}@{}r@{}@{}r@{}@{}r@{}@{}r@{}@{}r@{}@{}r@{}@{}r@{}@{}r@{}@{}r@{}@{}r@{}@{}r@{}@{}r@{}@{}r@{}@{}r@{}@{}c@{}@{}c@{}@{}c@{}@{}c@{}@{}c@{}@{}c@{}@{}c@{}@{}c@{}}
\hline\hline
 \xt &  NPCorr &  $\rho$ & $\delta_\mathrm{stat.}$ & $\gamma_{3}$ & $\gamma_{9}$ & $\gamma_{15}$ & $\gamma_{21}$ & $\gamma_{27}$ & $\gamma_{31}$ & $\gamma_{34}$ & $\gamma_{40}$ & $\gamma_{46}$ & $\gamma_{52}$ & $\gamma_{58}$ & $\gamma_{64}$ & $\gamma_{70}$ & $\gamma_{78}$ & $\gamma_{82}$ & $\gamma_{74}$ & $\gamma_{75}$ & $\gamma_{83}$ & $u_1$ & $u_2$ & $u_3$ \\
 & & & \% &  \%&  \%&  \%&  \%&  \%&  \%&  \%&  \%&  \%&  \%&  \%&  \%&  \%&  \%&  \%&  \%&  \%&  \%&  \%&  \%&  \%\\
\hline
\ \ (1.71--2.29) e-02\ \ &\ \  0.91\!\!$\begin{array}{r} + 0.03\\- 0.06\end{array}$\ \ &\ \ 1.236\ \ & \ \ 1.8\ \ &\ \ $^{+7.1}_{-5.9}$\ \ & \ \ $^{+4.2}_{-5.3}$\ \ & \ \ $^{+2.5}_{-1.7}$\ \ & \ \ $^{+4.2}_{-3.4}$\ \ & \ \ $^{+ 1.0}_{-0.9}$\ \ & \ \ $^{+8.9}_{-9.4}$\ \ & \ \ $^{-0.6}_{+ 0.7}$\ \ & \ \ $^{+2.5}_{-2.3}$\ \ & \ \ $^{-0.0}_{+ 0.3}$\ \ & \ \ $^{-0.3}_{+ 0.2}$\ \ & \ \ $^{+ 0.0}_{-0.1}$\ \ & \ \ $^{+2.2}_{-2.3}$\ \ & \ \ $^{-1.7}_{+1.8}$\ \ & \ \ $\mp  11$\ \ & \ \ $\pm 0.5$\ \ & \ \ $\pm 0.1$\ \ & \ \ $\pm 0.4$\ \ & \ \ $\pm 0.9$\ \ & \ \ $\pm 1.4$\ \ & \ \ $^{+ 0.5}_{-0.5}$\ \ & \ \ $\pm 0.5$\ \ \\
\ \ (2.29--3.14) e-02\ \ &\ \  0.93\!\!$\begin{array}{r} + 0.03\\- 0.04\end{array}$\ \ &\ \ 1.275\ \ & \ \ 2.4\ \ &\ \ $^{+4.3}_{-3.9}$\ \ & \ \ $^{+5.4}_{-4.9}$\ \ & \ \ $^{+2.4}_{-2.2}$\ \ & \ \ $^{-0.4}_{+ 0.8}$\ \ & \ \ $^{+ 1.0}_{-0.6}$\ \ & \ \ $^{+11.}_{-9.6}$\ \ & \ \ $^{-0.3}_{+ 0.5}$\ \ & \ \ $^{+2.0}_{-1.6}$\ \ & \ \ $^{-0.3}_{+ 0.4}$\ \ & \ \ $^{-1.1}_{+1.3}$\ \ & \ \ $^{+ 0.1}_{+ 0.3}$\ \ & \ \ $^{+3.9}_{-3.6}$\ \ & \ \ $^{-4.2}_{+4.3}$\ \ & \ \ $\mp 2.2$\ \ & \ \ $\pm 0.5$\ \ & \ \ $\pm 0.1$\ \ & \ \ $\pm 0.3$\ \ & \ \ 0.0 \ \ & \ \ $\pm 1.4$\ \ & \ \ $^{+ 0.5}_{-0.5}$\ \ & \ \ $\pm 0.5$\ \ \\
\hline\hline\end{tabular}
\end{table*}
\begin{table*}
\caption{Measured jet cross section ratio in \xt{} bin for anti-$k_t$ jets with $R=0.4$ in the rapidity bin $2.1\leq|y|<2.8$. See caption of Table~\ref{tab:rxt_r04_y0} for details.}
\label{tab:rxt_r04_y4}
\tiny
\centering
\begin{tabular}{@{}c@{}@{}c@{}@{}c@{}@{}c@{}@{}r@{}@{}r@{}@{}r@{}@{}r@{}@{}r@{}@{}r@{}@{}r@{}@{}r@{}@{}r@{}@{}r@{}@{}r@{}@{}r@{}@{}r@{}@{}r@{}@{}c@{}@{}c@{}@{}c@{}@{}c@{}@{}c@{}@{}c@{}@{}c@{}@{}c@{}}
\hline\hline
 \xt &  NPCorr &  $\rho$ & $\delta_\mathrm{stat.}$ & $\gamma_{4}$ & $\gamma_{10}$ & $\gamma_{16}$ & $\gamma_{22}$ & $\gamma_{28}$ & $\gamma_{31}$ & $\gamma_{35}$ & $\gamma_{41}$ & $\gamma_{47}$ & $\gamma_{53}$ & $\gamma_{59}$ & $\gamma_{65}$ & $\gamma_{71}$ & $\gamma_{79}$ & $\gamma_{82}$ & $\gamma_{74}$ & $\gamma_{75}$ & $\gamma_{84}$ & $u_1$ & $u_2$ & $u_3$ \\
 & & & \% &  \%&  \%&  \%&  \%&  \%&  \%&  \%&  \%&  \%&  \%&  \%&  \%&  \%&  \%&  \%&  \%&  \%&  \%&  \%&  \%&  \%\\
\hline
\ \ (1.71--2.29) e-02\ \ &\ \  0.91\!\!$\begin{array}{r} + 0.04\\- 0.05\end{array}$\ \ &\ \ 1.242\ \ & \ \ 2.6\ \ &\ \ $^{+6.7}_{-6.4}$\ \ & \ \ $^{+3.4}_{-6.0}$\ \ & \ \ $^{+2.0}_{-2.2}$\ \ & \ \ $^{+4.4}_{-4.7}$\ \ & \ \ $^{+ 1.1}_{-0.9}$\ \ & \ \ $^{ +18}_{ -20}$\ \ & \ \ $^{-0.7}_{+ 0.9}$\ \ & \ \ $^{+2.2}_{-2.3}$\ \ & \ \ $^{-0.5}_{+ 0.2}$\ \ & \ \ $^{-0.3}_{+ 0.4}$\ \ & \ \ $^{-0.0}_{-0.1}$\ \ & \ \ $^{+2.2}_{-2.4}$\ \ & \ \ $^{-1.7}_{+2.0}$\ \ & \ \ $\mp  10$\ \ & \ \ $\pm 0.2$\ \ & \ \ $\pm 0.2$\ \ & \ \ $\pm 0.3$\ \ & \ \ $\pm 0.9$\ \ & \ \ $\pm 1.4$\ \ & \ \ $^{+ 0.5}_{-0.5}$\ \ & \ \ $\pm 0.6$\ \ \\
\ \ (2.29--3.14) e-02\ \ &\ \  0.94\!\!$\begin{array}{r} + 0.03\\- 0.04\end{array}$\ \ &\ \ 1.296\ \ & \ \ 1.7\ \ &\ \ $^{+4.4}_{-3.8}$\ \ & \ \ $^{+6.1}_{-5.0}$\ \ & \ \ $^{+3.0}_{-1.9}$\ \ & \ \ $^{+3.8}_{-3.1}$\ \ & \ \ $^{+ 1.4}_{-0.5}$\ \ & \ \ $^{ +20}_{ -16}$\ \ & \ \ $^{+ 0.1}_{+ 0.9}$\ \ & \ \ $^{+2.2}_{-1.4}$\ \ & \ \ $^{+ 0.2}_{+ 0.9}$\ \ & \ \ $^{-1.0}_{+ 1.5}$\ \ & \ \ $^{+ 0.1}_{+ 0.4}$\ \ & \ \ $^{+3.9}_{-3.6}$\ \ & \ \ $^{-3.9}_{+4.8}$\ \ & \ \ $\mp 1.6$\ \ & \ \ $\pm 0.2$\ \ & \ \ $\pm 0.3$\ \ & \ \ $\pm 0.1$\ \ & \ \ 0.0 \ \ & \ \ $\pm 1.4$\ \ & \ \ $^{+ 0.5}_{-0.5}$\ \ & \ \ $\pm 0.9$\ \ \\
\hline\hline\end{tabular}
\end{table*}
\begin{table*}
\caption{Measured jet cross section ratio in \xt{} bin for anti-$k_t$ jets with $R=0.4$ in the rapidity bin $2.8\leq|y|<3.6$. See caption of Table~\ref{tab:rxt_r04_y0} for details.}
\label{tab:rxt_r04_y5}
\tiny
\centering
\begin{tabular}{@{}c@{}@{}c@{}@{}c@{}@{}c@{}@{}r@{}@{}r@{}@{}r@{}@{}r@{}@{}r@{}@{}r@{}@{}r@{}@{}r@{}@{}r@{}@{}r@{}@{}r@{}@{}r@{}@{}r@{}@{}r@{}@{}c@{}@{}c@{}@{}c@{}@{}c@{}@{}c@{}@{}c@{}@{}c@{}@{}c@{}}
\hline\hline
 \xt &  NPCorr &  $\rho$ & $\delta_\mathrm{stat.}$ & $\gamma_{5}$ & $\gamma_{11}$ & $\gamma_{17}$ & $\gamma_{23}$ & $\gamma_{29}$ & $\gamma_{31}$ & $\gamma_{36}$ & $\gamma_{42}$ & $\gamma_{48}$ & $\gamma_{54}$ & $\gamma_{60}$ & $\gamma_{66}$ & $\gamma_{72}$ & $\gamma_{80}$ & $\gamma_{82}$ & $\gamma_{74}$ & $\gamma_{75}$ & $\gamma_{85}$ & $u_1$ & $u_2$ & $u_3$ \\
 & & & \% &  \%&  \%&  \%&  \%&  \%&  \%&  \%&  \%&  \%&  \%&  \%&  \%&  \%&  \%&  \%&  \%&  \%&  \%&  \%&  \%&  \%\\
\hline
\ \ (1.71--2.29) e-02\ \ &\ \  0.91\!\!$\begin{array}{r} + 0.05\\- 0.04\end{array}$\ \ &\ \ 1.147\ \ & \ \ 3.2\ \ &\ \ $^{+8.2}_{-6.5}$\ \ & \ \ $^{+7.0}_{-8.3}$\ \ & \ \ $^{+4.0}_{-3.4}$\ \ & \ \ $^{+ 1.1}_{-0.3}$\ \ & \ \ $^{+1.5}_{-1.2}$\ \ & \ \ $^{ +29}_{ -26}$\ \ & \ \ $^{-0.5}_{+ 0.4}$\ \ & \ \ $^{+3.0}_{-2.8}$\ \ & \ \ $^{+ 0.5}_{-0.7}$\ \ & \ \ $^{-0.4}_{+ 0.4}$\ \ & \ \ $^{-0.0}_{-0.1}$\ \ & \ \ $^{+2.7}_{-2.9}$\ \ & \ \ $^{-2.1}_{+2.1}$\ \ & \ \ $\mp  14$\ \ & \ \ $\pm 0.6$\ \ & \ \ $\pm 0.3$\ \ & \ \ $\pm 0.3$\ \ & \ \ $\pm 0.9$\ \ & \ \ $\pm 2.2$\ \ & \ \ $^{+ 0.5}_{-0.5}$\ \ & \ \ $\pm 3.9$\ \ \\
\ \ (2.29--3.14) e-02\ \ &\ \  0.92\!\!$\begin{array}{r} + 0.05\\- 0.03\end{array}$\ \ &\ \ 1.196\ \ & \ \ 4.2\ \ &\ \ $^{+4.8}_{-5.2}$\ \ & \ \ $^{+7.4}_{-5.2}$\ \ & \ \ $^{+4.1}_{-3.0}$\ \ & \ \ $^{-1.7}_{+2.3}$\ \ & \ \ $^{+1.4}_{-1.9}$\ \ & \ \ $^{ +33}_{ -24}$\ \ & \ \ $^{+ 0.2}_{-0.4}$\ \ & \ \ $^{+2.3}_{-2.8}$\ \ & \ \ $^{-0.2}_{+ 1.0}$\ \ & \ \ $^{-1.3}_{+1.7}$\ \ & \ \ $^{+ 0.1}_{+ 0.3}$\ \ & \ \ $^{+4.4}_{-5.2}$\ \ & \ \ $^{-5.0}_{+6.2}$\ \ & \ \ $\mp 1.6$\ \ & \ \ $\pm 0.2$\ \ & \ \ $\pm 0.8$\ \ & \ \ $\pm 0.7$\ \ & \ \ 0.0 \ \ & \ \ $\pm 2.2$\ \ & \ \ $^{+ 0.5}_{-0.5}$\ \ & \ \ $\pm 4.8$\ \ \\
\hline\hline\end{tabular}
\end{table*}
\begin{table*}
\caption{Measured jet cross section ratio in \xt{} bin for anti-$k_t$ jets with $R=0.4$ in the rapidity bin $3.6\leq|y|<4.4$. See caption of Table~\ref{tab:rxt_r04_y0} for details.}
\label{tab:rxt_r04_y6}
\tiny
\centering
\begin{tabular}{@{}c@{}@{}c@{}@{}c@{}@{}c@{}@{}r@{}@{}r@{}@{}r@{}@{}r@{}@{}r@{}@{}r@{}@{}r@{}@{}r@{}@{}r@{}@{}r@{}@{}r@{}@{}r@{}@{}r@{}@{}r@{}@{}c@{}@{}c@{}@{}c@{}@{}c@{}@{}c@{}@{}c@{}@{}c@{}@{}c@{}}
\hline\hline
 \xt &  NPCorr &  $\rho$ & $\delta_\mathrm{stat.}$ & $\gamma_{6}$ & $\gamma_{12}$ & $\gamma_{18}$ & $\gamma_{24}$ & $\gamma_{30}$ & $\gamma_{31}$ & $\gamma_{37}$ & $\gamma_{43}$ & $\gamma_{49}$ & $\gamma_{55}$ & $\gamma_{61}$ & $\gamma_{67}$ & $\gamma_{73}$ & $\gamma_{81}$ & $\gamma_{82}$ & $\gamma_{74}$ & $\gamma_{75}$ & $\gamma_{86}$ & $u_1$ & $u_2$ & $u_3$ \\
 & & & \% &  \%&  \%&  \%&  \%&  \%&  \%&  \%&  \%&  \%&  \%&  \%&  \%&  \%&  \%&  \%&  \%&  \%&  \%&  \%&  \%&  \%\\
\hline
\ \ (1.71--2.29) e-02\ \ &\ \  0.85\!\!$\begin{array}{r} + 0.08\\- 0.04\end{array}$\ \ &\ \ 1.144\ \ & \ \ 8.0\ \ &\ \ $^{+11.}_{-8.8}$\ \ & \ \ $^{+9.3}_{-9.6}$\ \ & \ \ $^{+4.3}_{-2.8}$\ \ & \ \ $^{+1.6}_{-2.9}$\ \ & \ \ $^{+ 0.9}_{-1.9}$\ \ & \ \ $^{ +78}_{ -57}$\ \ & \ \ $^{-1.3}_{+ 1.0}$\ \ & \ \ $^{+3.0}_{-4.1}$\ \ & \ \ $^{+ 0.4}_{-0.3}$\ \ & \ \ $^{-1.1}_{+ 0.7}$\ \ & \ \ $^{+ 0.1}_{-0.1}$\ \ & \ \ $^{+4.8}_{-4.7}$\ \ & \ \ $^{-4.7}_{+3.7}$\ \ & \ \ $\mp  19$\ \ & \ \ $\pm 0.7$\ \ & \ \ $\pm 0.2$\ \ & \ \ $\pm 2.2$\ \ & \ \ $\pm 0.9$\ \ & \ \ $\pm 2.2$\ \ & \ \ $^{+ 0.5}_{-0.5}$\ \ & \ \ $\pm 6.4$\ \ \\
\ \ (2.29--3.14) e-02\ \ &\ \  0.85\!\!$\begin{array}{r} + 0.09\\- 0.06\end{array}$\ \ &\ \ 1.245\ \ & \ \ 7.1\ \ &\ \ $^{+8.7}_{-9.2}$\ \ & \ \ $^{+11.}_{-8.3}$\ \ & \ \ $^{+6.2}_{-5.1}$\ \ & \ \ $^{+ 1.9}_{-0.7}$\ \ & \ \ $^{+ 1.1}_{-0.6}$\ \ & \ \ $^{+144}_{ -54}$\ \ & \ \ $^{-0.4}_{+ 0.9}$\ \ & \ \ $^{+3.5}_{-3.1}$\ \ & \ \ $^{+ 0.5}_{+ 0.4}$\ \ & \ \ $^{-2.5}_{+3.7}$\ \ & \ \ $^{+ 0.0}_{+ 1.3}$\ \ & \ \ $^{+7.4}_{-8.5}$\ \ & \ \ $^{-8.3}_{+12.}$\ \ & \ \ $\mp 9.0$\ \ & \ \ $\pm 5.1$\ \ & \ \ $\pm 2.5$\ \ & \ \ $\pm 4.9$\ \ & \ \ 0.0 \ \ & \ \ $\pm 1.4$\ \ & \ \ $^{+ 0.5}_{-0.5}$\ \ & \ \ $\pm 7.7$\ \ \\
\hline\hline\end{tabular}
\end{table*}

\begin{table*}
\caption{Measured jet cross section ratio in \xt{} bin for anti-$k_t$ jets with $R=0.6$ in the rapidity bin $|y|<0.3$. See caption of Table~\ref{tab:rxt_r04_y0} for details.}
\label{tab:rxt_r06_y0}
\tiny
\centering
\begin{tabular}{@{}c@{}@{}c@{}@{}c@{}@{}c@{}@{}r@{}@{}r@{}@{}r@{}@{}r@{}@{}r@{}@{}r@{}@{}r@{}@{}r@{}@{}r@{}@{}r@{}@{}r@{}@{}r@{}@{}r@{}@{}r@{}@{}c@{}@{}c@{}@{}c@{}@{}c@{}@{}c@{}@{}c@{}@{}c@{}@{}c@{}}
\hline\hline
 \xt &  NPCorr &  $\rho$ & $\delta_\mathrm{stat.}$ & $\gamma_{1}$ & $\gamma_{7}$ & $\gamma_{13}$ & $\gamma_{19}$ & $\gamma_{25}$ & $\gamma_{31}$ & $\gamma_{32}$ & $\gamma_{38}$ & $\gamma_{44}$ & $\gamma_{50}$ & $\gamma_{56}$ & $\gamma_{62}$ & $\gamma_{68}$ & $\gamma_{76}$ & $\gamma_{82}$ & $\gamma_{74}$ & $\gamma_{75}$ & $\gamma_{83}$ & $u_1$ & $u_2$ & $u_3$ \\
 & & & \% &  \%&  \%&  \%&  \%&  \%&  \%&  \%&  \%&  \%&  \%&  \%&  \%&  \%&  \%&  \%&  \%&  \%&  \%&  \%&  \%&  \%\\
\hline
\ \ (1.71--2.29) e-02\ \ &\ \  1.07\!\!$\begin{array}{r} + 0.02\\- 0.07\end{array}$\ \ &\ \ 1.519\ \ & \ \ 2.0\ \ &\ \ $^{ +12}_{ -11}$\ \ & \ \ $^{+4.0}_{-5.6}$\ \ & \ \ $^{ +12}_{ -11}$\ \ & \ \ $^{+6.1}_{-7.1}$\ \ & \ \ $^{+2.6}_{-2.6}$\ \ & \ \ 0.0\ \ &\ \ $^{-0.8}_{+ 0.3}$\ \ & \ \ $^{+3.0}_{-3.6}$\ \ & \ \ $^{+ 0.4}_{-1.4}$\ \ & \ \ $^{-0.4}_{+ 0.3}$\ \ & \ \ $^{-0.0}_{-0.1}$\ \ & \ \ $^{-1.5}_{+ 0.7}$\ \ & \ \ $^{+ 0.1}_{-0.2}$\ \ & \ \ $\mp 5.0$\ \ & \ \ $\pm 1.0$\ \ & \ \ $\pm 0.1$\ \ & \ \ $\pm 0.8$\ \ & \ \ $\pm 0.9$\ \ & \ \ $\pm 1.4$\ \ & \ \ $^{+ 0.6}_{-0.5}$\ \ & \ \ $\pm 0.3$\ \ \\
\ \ (2.29--3.14) e-02\ \ &\ \  1.04\!\!$\begin{array}{r} + 0.03\\- 0.05\end{array}$\ \ &\ \ 1.536\ \ & \ \ 2.6\ \ &\ \ $^{+8.0}_{-6.9}$\ \ & \ \ $^{+3.6}_{-3.4}$\ \ & \ \ $^{+7.8}_{-6.4}$\ \ & \ \ $^{+3.7}_{-2.5}$\ \ & \ \ $^{+ 0.8}_{-0.2}$\ \ & \ \ 0.0\ \ &\ \ $^{-0.2}_{+ 0.6}$\ \ & \ \ $^{+2.5}_{-1.7}$\ \ & \ \ $^{-0.1}_{+ 0.7}$\ \ & \ \ $^{-0.7}_{+ 1.1}$\ \ & \ \ $^{+ 0.1}_{+ 0.4}$\ \ & \ \ $^{+1.5}_{-1.3}$\ \ & \ \ $^{+ 0.5}_{-0.0}$\ \ & \ \ $\mp 1.7$\ \ & \ \ $\pm 0.4$\ \ & \ \ $\pm 0.0$\ \ & \ \ $\pm 0.7$\ \ & \ \ 0.0 \ \ & \ \ $\pm 1.4$\ \ & \ \ $^{+ 0.5}_{-0.5}$\ \ & \ \ $\pm 0.6$\ \ \\
\hline\hline\end{tabular}
\end{table*}
\begin{table*}
\caption{Measured jet cross section ratio in \xt{} bin for anti-$k_t$ jets with $R=0.6$ in the rapidity bin $0.3\leq|y|<0.8$. See caption of Table~\ref{tab:rxt_r04_y0} for details.}
\label{tab:rxt_r06_y1}
\tiny
\centering
\begin{tabular}{@{}c@{}@{}c@{}@{}c@{}@{}c@{}@{}r@{}@{}r@{}@{}r@{}@{}r@{}@{}r@{}@{}r@{}@{}r@{}@{}r@{}@{}r@{}@{}r@{}@{}r@{}@{}r@{}@{}r@{}@{}r@{}@{}c@{}@{}c@{}@{}c@{}@{}c@{}@{}c@{}@{}c@{}@{}c@{}@{}c@{}}
\hline\hline
 \xt &  NPCorr &  $\rho$ & $\delta_\mathrm{stat.}$ & $\gamma_{1}$ & $\gamma_{7}$ & $\gamma_{13}$ & $\gamma_{19}$ & $\gamma_{25}$ & $\gamma_{31}$ & $\gamma_{32}$ & $\gamma_{38}$ & $\gamma_{44}$ & $\gamma_{50}$ & $\gamma_{56}$ & $\gamma_{62}$ & $\gamma_{68}$ & $\gamma_{76}$ & $\gamma_{82}$ & $\gamma_{74}$ & $\gamma_{75}$ & $\gamma_{83}$ & $u_1$ & $u_2$ & $u_3$ \\
 & & & \% &  \%&  \%&  \%&  \%&  \%&  \%&  \%&  \%&  \%&  \%&  \%&  \%&  \%&  \%&  \%&  \%&  \%&  \%&  \%&  \%&  \%\\
\hline
\ \ (1.71--2.29) e-02\ \ &\ \  1.08\!\!$\begin{array}{r} + 0.02\\- 0.08\end{array}$\ \ &\ \ 1.523\ \ & \ \ 1.6\ \ &\ \ $^{ +13}_{ -11}$\ \ & \ \ $^{+4.2}_{-4.6}$\ \ & \ \ $^{+8.3}_{-7.1}$\ \ & \ \ $^{+7.0}_{-6.7}$\ \ & \ \ $^{+1.2}_{-1.0}$\ \ & \ \ $^{+ 0.3}_{-0.2}$\ \ & \ \ $^{-0.2}_{+ 0.1}$\ \ & \ \ $^{+2.9}_{-2.9}$\ \ & \ \ $^{+ 0.6}_{-1.0}$\ \ & \ \ $^{-0.5}_{+ 0.2}$\ \ & \ \ $^{-0.0}_{-0.1}$\ \ & \ \ $^{+2.6}_{-2.8}$\ \ & \ \ $^{-1.7}_{+1.5}$\ \ & \ \ $\mp 5.0$\ \ & \ \ $\pm 0.2$\ \ & \ \ $\pm 0.1$\ \ & \ \ $\pm 0.4$\ \ & \ \ $\pm 0.9$\ \ & \ \ $\pm 1.4$\ \ & \ \ $^{+ 0.5}_{-0.5}$\ \ & \ \ $\pm 0.3$\ \ \\
\ \ (2.29--3.14) e-02\ \ &\ \  1.05\!\!$\begin{array}{r} + 0.02\\- 0.05\end{array}$\ \ &\ \ 1.500\ \ & \ \ 1.9\ \ &\ \ $^{+8.0}_{-7.4}$\ \ & \ \ $^{+4.8}_{-4.0}$\ \ & \ \ $^{+6.8}_{-5.9}$\ \ & \ \ $^{+5.0}_{-4.2}$\ \ & \ \ $^{+ 0.3}_{+ 0.2}$\ \ & \ \ $^{+ 0.2}_{-0.1}$\ \ & \ \ $^{+ 0.2}_{+ 0.2}$\ \ & \ \ $^{+2.2}_{-1.8}$\ \ & \ \ $^{+ 0.6}_{+ 0.3}$\ \ & \ \ $^{-0.9}_{+ 1.5}$\ \ & \ \ $^{+ 0.1}_{+ 0.4}$\ \ & \ \ $^{+3.9}_{-3.9}$\ \ & \ \ $^{-3.6}_{+4.5}$\ \ & \ \ $\mp 2.8$\ \ & \ \ $\pm 0.1$\ \ & \ \ $\pm 0.1$\ \ & \ \ $\pm 0.3$\ \ & \ \ 0.0 \ \ & \ \ $\pm 1.4$\ \ & \ \ $^{+ 0.5}_{-0.5}$\ \ & \ \ $\pm 0.5$\ \ \\
\hline\hline\end{tabular}%
\end{table*}

\begin{table*}
\caption{Measured jet cross section ratio in bins of \pt{} for anti-$k_t$ jets with $R=0.4$ in the rapidity bin $|y|<0.3$. NPCorr stands for multiplicative non-perturbative corrections, $\rho$ is the measured cross section and $\delta_\mathrm{stat.}$ is the statistical uncertainty. $\gamma_i$ and $u_j$ correspond to the correlated and uncorrelated systematic uncertainties given in \%. They are described in Table~\ref{tab:uncert}, where $i$ in $\gamma_i$ denotes a nuisance parameter. $u_3$ is the uncertainty due to pile-up effects in the cross section measurement at $\sqrt{s}=7$ \TeV{}. For each table entry, the outcome of shifting the corresponding nuisance parameter by one standard deviation, up or down, is shown as superscript or subscript, respectively. In some bins these shifts may lead to cross-section variations in the same direction. The 4.3\% uncertainty from the luminosity measurements is not in this table.}
\label{tab:rpt_r04_y0}
\tiny
\centering

\end{table*}
\begin{table*}
\caption{Measured jet cross section ratio in \pt{} bin for anti-$k_t$ jets with $R=0.4$ in the rapidity bin $0.3\leq|y|<0.8$. See caption of Table~\ref{tab:rpt_r04_y0} for details.}
\label{tab:rpt_r04_y1}
\tiny
\centering
%
\end{table*}
\begin{table*}
\caption{Measured jet cross section ratio in \pt{} bin for anti-$k_t$ jets with $R=0.4$ in the rapidity bin $0.8\leq|y|<1.2$. See caption of Table~\ref{tab:rpt_r04_y0} for details.}
\label{tab:rpt_r04_y2}
\tiny
\centering
%
\end{table*}
\begin{table*}
\caption{Measured jet cross section ratio in \pt{} bin for anti-$k_t$ jets with $R=0.4$ in the rapidity bin $1.2\leq|y|<2.1$. See caption of Table~\ref{tab:rpt_r04_y0} for details.}
\label{tab:rpt_r04_y3}
\tiny
\centering
%
\end{table*}

\begin{table*}
\caption{Measured jet cross section ratio in \pt{} bin for anti-$k_t$ jets with $R=0.6$ in the rapidity bin $|y|<0.3$. See caption of Table~\ref{tab:rpt_r04_y0} for details.}
\label{tab:rpt_r06_y0}
\tiny
\centering
%
\end{table*}

\bibliographystyle{atlasnote}
\bibliography{2p76paper}

\onecolumn
\clearpage
\begin{flushleft}
{\Large The ATLAS Collaboration}

\bigskip

G.~Aad$^{\rm 48}$,
T.~Abajyan$^{\rm 21}$,
B.~Abbott$^{\rm 112}$,
J.~Abdallah$^{\rm 12}$,
S.~Abdel~Khalek$^{\rm 116}$,
A.A.~Abdelalim$^{\rm 49}$,
O.~Abdinov$^{\rm 11}$,
R.~Aben$^{\rm 106}$,
B.~Abi$^{\rm 113}$,
M.~Abolins$^{\rm 89}$
% ...
\bigskip
\\
% ...
$^{11}$ Institute of Physics, Azerbaijan Academy of Sciences, Baku, Azerbaijan\\
$^{12}$ Institut de F{\'\i}sica d'Altes Energies and Departament de F{\'\i}sica de la Universitat Aut{\`o}noma de Barcelona, Barcelona, Spain\\
% ...
$^{21}$ Physikalisches Institut, University of Bonn, Bonn, Germany\\
% ...
$^{48}$ Fakult{\"a}t f{\"u}r Mathematik und Physik, Albert-Ludwigs-Universit{\"a}t, Freiburg, Germany\\
$^{49}$ Section de Physique, Universit{\'e} de Gen{\`e}ve, Geneva, Switzerland\\
% ...
$^{89}$ Department of Physics and Astronomy, Michigan State University, East Lansing MI, United States of America\\
% ...
$^{106}$ Nikhef National Institute for Subatomic Physics and University of Amsterdam, Amsterdam, Netherlands\\
% ...
$^{112}$ Homer L. Dodge Department of Physics and Astronomy, University of Oklahoma, Norman OK, United States of America\\
$^{113}$ Department of Physics, Oklahoma State University, Stillwater OK, United States of America\\
% ...
$^{116}$ LAL, Universit{\'e} Paris-Sud and CNRS/IN2P3, Orsay, France\\
% ...
\end{flushleft}


\end{document}